\documentclass[10pt,journal,compsoc]{IEEEtran}
\usepackage{adjustbox}
\usepackage{lipsum}
\usepackage{dblfloatfix}
\usepackage[]{graphicx}
\usepackage{cite}
\usepackage{verbatim}
\usepackage{url}
\usepackage{amsmath}
\usepackage{amssymb}
\usepackage{breqn}
\usepackage{amsthm}
\usepackage{multirow}
\usepackage{lipsum}
\usepackage{bbm}
\usepackage[caption=false,font=footnotesize]{subfig}
\usepackage{caption}
\usepackage{algorithm}
\usepackage{algpseudocode}
\usepackage{mathtools}

\newtheorem{thm}{Theorem}
\newtheorem{lemma}{Lemma}

\newtheorem{defn}{Definition}

\DeclareMathOperator*{\argmax}{arg\,max}

\begin{document}

\title{Learning to Coordinate in a Decentralized Cognitive Radio Network in Presence of Jammers}
\author{Suneet~Sawant,
	         Rohit~Kumar,
		Manjesh K.~Hanawal,
       and~Sumit J.~Darak
\IEEEcompsocitemizethanks{\IEEEcompsocthanksitem Suneet Sawant is with the Department
of Electrical Engineering, IIT Bombay, India. E-mail: suneet.sawant0308@gmail.com. Rohit Kumar is with the Department of ECE, National Institute of Technology, Delhi, India. E-mail: rohitkumafr@nitdelhi.ac.in. Manjesh K. Hanawal is with IEOR, IIT Bombay, India. E-mail: mhanawal@iitb.ac.in.  Sumit J. Darak is with the Department
of ECE, Indraprastha Institute of Information Technology, Delhi, India. E-mail: sumit@iiitd.ac.in. 
}}


\IEEEtitleabstractindextext{
\begin{abstract}
Efficient utilization of licensed spectrum in the cognitive radio network is challenging due to lack of coordination among the Secondary Users (SUs). Distributed algorithms proposed in the literature aim to maximize the network throughput by ensuring orthogonal channel allocation for the SUs. However, these algorithms work under the assumption that all the SUs faithfully follow the algorithms which may not always hold due to the decentralized nature of the network. In this paper, we study distributed algorithms that are robust against malicious behavior (jamming attack). We consider both the cases of jammers launching coordinated and uncoordinated attacks. In the coordinated attack, the jammers select non-overlapping channels to attack in each time slot and can significantly increase the number of collisions for SUs. We setup the problem in each scenario as a multi-player bandit and develop algorithms. The analysis shows that when the SUs faithfully implement proposed algorithms, the regret is constant with high probability.  We validate our claims through exhaustive synthetic experiments and also through a realistic USRP based experiment.  

\end{abstract}
}

\maketitle
\section{Introduction}
The issue of spectrum scarcity has constantly driven researchers to look for an efficient utilization of available licensed spectrum. Among various envisioned paradigms such as device-to-device communications and LTE-unlicensed network, dynamic spectrum access (DSA) based cognitive radio network (CRN) seems to be a promising solution to deal with the problem of spectrum scarcity incurred due to static spectrum allocation policies \cite{p1,p2,p3}. DARPA's spectrum collaboration challenge 2016 and 2018 were significant steps to validate the feasibility of DSA in real radio environment\cite{darpa}. 

CRN consists of licensed or primary users (PUs) and unlicensed or secondary users (SUs). The PUs coordinate for channel assignments through central control. However, such coordination is not feasible for SUs in many scenarios. Such network is referred to as infrastructure-less CRN. To overcome lack of coordination among SUs, several distributed algorithms like $\rho^{rand}$ \cite{prand}, multi-user $\epsilon-$greedy collision avoiding (MEGA) \cite{MEGA}, musical chair (MC) \cite{MC} are proposed which guarantee orthogonal channel allocation if faithfully implemented by all the SUs. However, the decentralized network gives rise to the possibility that some of the SUs need not implement the algorithm faithfully, or even worse, launch Denial-of-Service attacks and degrade network performance. Such SUs are referred to as jammers or malicious users (MUs). It is thus important that the distributed algorithms for the legitimate SUs (henceforth simply referred as SUs) identify such malicious behavior and achieve best possible throughput in their presence. In this work, we develop distributed algorithms that are robust against various jamming attacks.
	
We consider an overlay CRN where SUs can transmit on a channel only if it is idle, i.e., no PU is active on that channel. For primary transmissions, we assume widely studied model that each channel being idle is an independent and identically distributed process\footnote{More realistic 
	Markovian channel behavior can also be studied using Multi-Armed Bandit for Markov Chains. We leave it as future work}. Transmissions of the SUs are time-slotted and packeted. At the end of each slot, the receiver acknowledges successful
reception of packets through ACK/NACK signals. A transmission by the SU fails if another SU or a jammer or both transmit on the same channel, and we refer to this event as a collision. Otherwise, transmission is considered to be successful. 

The aggregate network throughput for the SUs is highest if all of them select orthogonal channel from the \textquoteleft top\textquoteright  $N$ channels, where $N$ is the number of SUs. The existing distributed algorithms thus aim to learn the channel statistics as well as the number of SUs and then find orthogonal channel assignments in the top $N$ channels. In presence of jammers, the estimation of $N$ can be inaccurate and the SUs may end up settling on sub-optimal channels or on non-orthogonal channels resulting in sub-optimal network performance. We focus on learning methods that not only estimate the channel statistics and the number of SUs but also the number of jammers correctly so that SUs can find orthogonal channel assignment in the `appropriate' set of top channels. 
We assume that the jammers are another set of SUs and their behavior is similar to other (legitimate) SUs in all respects except that they do not strictly follow the common protocol (distributed algorithm). We allow the presence of multiple jammers who are either coordinated or can act independently; when they coordinate, they attack non-overlapping channels causing maximum damage. We consider the worst possible attack by the jammers. Specifically, we assume that once the jammers estimate $N$, they attack only top $N$-channels. Our algorithms perform only better against any weaker jamming attacks. 

Some part of our distributed algorithms is similar to the Musical Chairs (MC) algorithm\cite{MC}. MC is designed to work in a scenario where channels are always idle and there are no malicious users. We develop new algorithms that are robust to malicious jamming attacks and applicable to CRN where channels are not always idle. Also, our approach ensures fairness in channel assignment to SUs as we allow them to use the 'top' channels sequentially instead of locking each SU on non-overlapping channels (see sequential hopping later).  Our algorithms run in multiple phases; in the first phase, each SU randomly hops on all channels and learn the channel statistics and number of users (SUs and jammers) from the collision statistics. In the second phase, they find non-overlapping channels. When the jammers coordinate, they can estimate the number of SUs through collision information, but it is not possible for the SUs to estimate the number of SUs and the number of jammers through the collision information. We devise a new mechanism for SUs to get an estimate of the number of jammers and SUs and then find non-overlapping channels. Our algorithms minimize regret in a multi-player bandit where regret is defined as the difference between best aggregate throughput achievable when all the SUs cooperate with prior knowledge of network parameters (channel statistic, number of SUs and jammers) and aggregate throughput achieved when they do not have any prior knowledge. Our contributions can be summarized as follows:

\begin{itemize}
	\item When the jammers coordinate  and each SU can identify whether a collision is caused by the presence of other SUs or jammers, we propose algorithm CDJ (Coordination in
presence of Distinguishable Jammers) and show that it gives constant regret with high confidence. 
	\item When the jammers coordinate and SUs cannot identify whether a collision is caused by the presence of other SUs or jammers, we propose algorithm CNJ (Coordination in
presence of Non-distinguishable Jammers) which is also shown to give constant regret with high confidence.
	\item When the jammers do not coordinate, we show that a straightforward modification of CNJ, named CUJ (Co-ordination in presence of uncoordinated and non-distinguishable jammers) achieves constant regret with high confidence. 
	\item We validate our guarantees of the algorithms through extensive simulations. We give a realistic universal software radio peripheral (USRP) based experimental setup and demonstrate the effectiveness of our algorithms. 
\end{itemize}
 
The paper is organized as follows. The network model and setup is discussed in Section~\ref{networkmodel}. The proposed algorithms, CDJ, CNJ and CUJ, along with respective regret bounds are presented in Section~\ref{CJ-DC},~\ref{CJ-NDC} and \ref{UJ}
respectively. The simulation and experimental results on USRP testbed are discussed in Section~\ref{simulation} and Section~\ref{usrp} respectively. Section~\ref{conclusion} concludes the paper and discusses future directions. All the missing proof are given in the supplementary. \\

\noindent
{\bf Related Work}: In the last decade, significant research work has been observed in areas such as narrowband and wideband spectrum sensing \cite{p1,p2,p3,nus}, DSA \cite{p2,p3}. Among them, the DSA in a decentralized network is the most challenging and complex due to the lack of coordination among the SUs. Next, we discuss the papers most related to our work.
 
  Various algorithms \cite{MC,MEGA,quek,prand,rjain,seus,zhandi,zhandia} have been proposed to make the DSA feasible in the decentralized networks. The decentralized auction based algorithms in \cite{rjain,zhandia} allow SUs to transmit on their preferred frequency channel. However, for collision free transmissions, they either need a central controller or communication links between SUs which may not be feasible in the decentralized infrastructure-less CRN. The  $\rho^{rand}$ \cite{prand} employs multi-armed bandit algorithm based channel selection for estimation of the channel ranking and randomization based orthogonalization of SUs in the top $N$ channels and it is assumed that all SUs have prior knowledge of $N$. The $\rho^{rand}$ is extended in \cite{seus,zhandi} by replacing the randomization based SU orthogonalization with the learning based scheme leading to further improvement in regret. The channel hopping based algorithm in \cite{quek} allows SUs to orthogonalize into collision free hopping and hence does not need prior knowledge of $N$. However, SUs select all channels 
in sequential hopping manner leading to higher regret especially when some channels are significantly better than others.  In addition, the main drawback of the algorithms in \cite{prand,seus,rjain,zhandia,quek,zhandi} is the requirement of prior knowledge of number of SUs, $N$. 
  
  In the decentralized networks, it is difficult to have knowledge of the number of SUs, $N$. To the best of our knowledge, MEGA \cite{MEGA} and MC \cite{MC} are the only algorithms that allow SU to independently estimate $N$ in the decentralized network. It has been shown that MC outperforms MEGA and it is simpler to implement. All these algorithms guarantee good performance of the network only if they are implemented faithfully by all the SUs, otherwise the performance guarantee need not hold.
  
  Various algorithms in the literature have considered the DSA in the presence of jammers when some form of central control is available \cite{jams}. The common observation is that the conventional jammer avoidance strategies such as frequency hopping or direct sequence spread spectrum may not be efficient in CRN since they demand control channel link between SU transmitter and receiver. In addition, the channels over which SU transmits may change dynamically over time.
  The Jammer Inference-based Jamming Defense (jDefender) algorithm in \cite{zhu} identifies the SUs acting as jammers based on their channel access information. However, the channel allocation must be done using the central database and hence, it is not feasible in the decentralized network.
 
 
 In \cite{wang}, the PUs or central coordinator helps SUs by transmitting specific jamming signal in the vacant channels. The jammers will not jam such channels assuming the presence of PUs while the SUs will detect the channel as vacant using advanced sensing involving cancellation of jamming signal. In addition to the need of the controller, this algorithm demands advanced signal processing at the SU terminals and hence, may not be suitable for battery operated SU terminals. The algorithm in \cite{wang1} employs online learning algorithm for a transmitter receiver pair of the SU to rendezvous on a common channel. The algorithm incurs a significant number of collisions among SUs especially for large number of SUs. In \cite{xiao}, the anti-jamming power allocation strategy for SUs based on reinforcement learning
algorithms, including Q-learning and WoLF-PHC is proposed. It resists smart jamming in cooperative cognitive
radio networks but needs a high processing time. In \cite{minh}, a real-time Medium Access Control-based (MAC-based) jamming detection scheme is proposed
to meet the requirements of safety applications in vehicular networks (VNs). It reduces the false alarm rate and the time required to monitor VNs but no anti-jamming solution is provided in this work.

 To the best of our knowledge, the algorithms in the literature require some presence of central controller or prior knowledge of network parameters to improve performance in presence of jamming attack. This paper aims to develop algorithms that overcome these limitations. Further, we consider the worst possible attack from the jammers and hence our performance guarantees are pessimistic. They can be improved if a specific jammer strategy is known.

\section{Network Model}\label{networkmodel}
%

\begin{figure}[!h]
	\centering
	\subfloat[]{\includegraphics[scale=0.4]{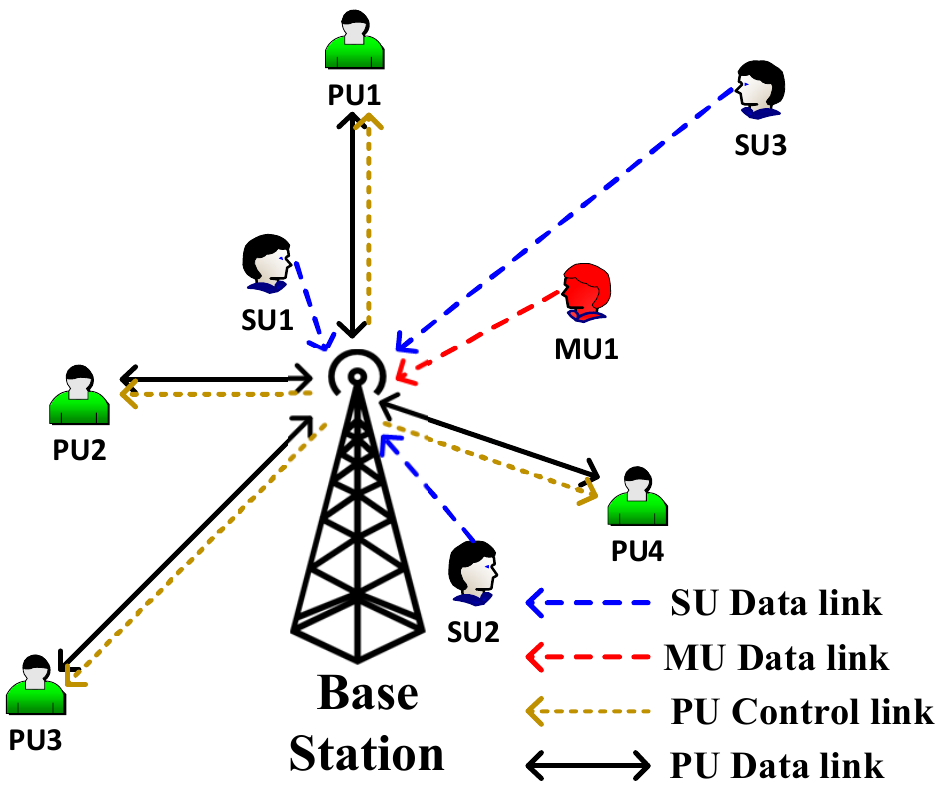}%
		\label{network}}
	\hspace{0.2cm}
	\subfloat[]{\includegraphics[scale=0.25]{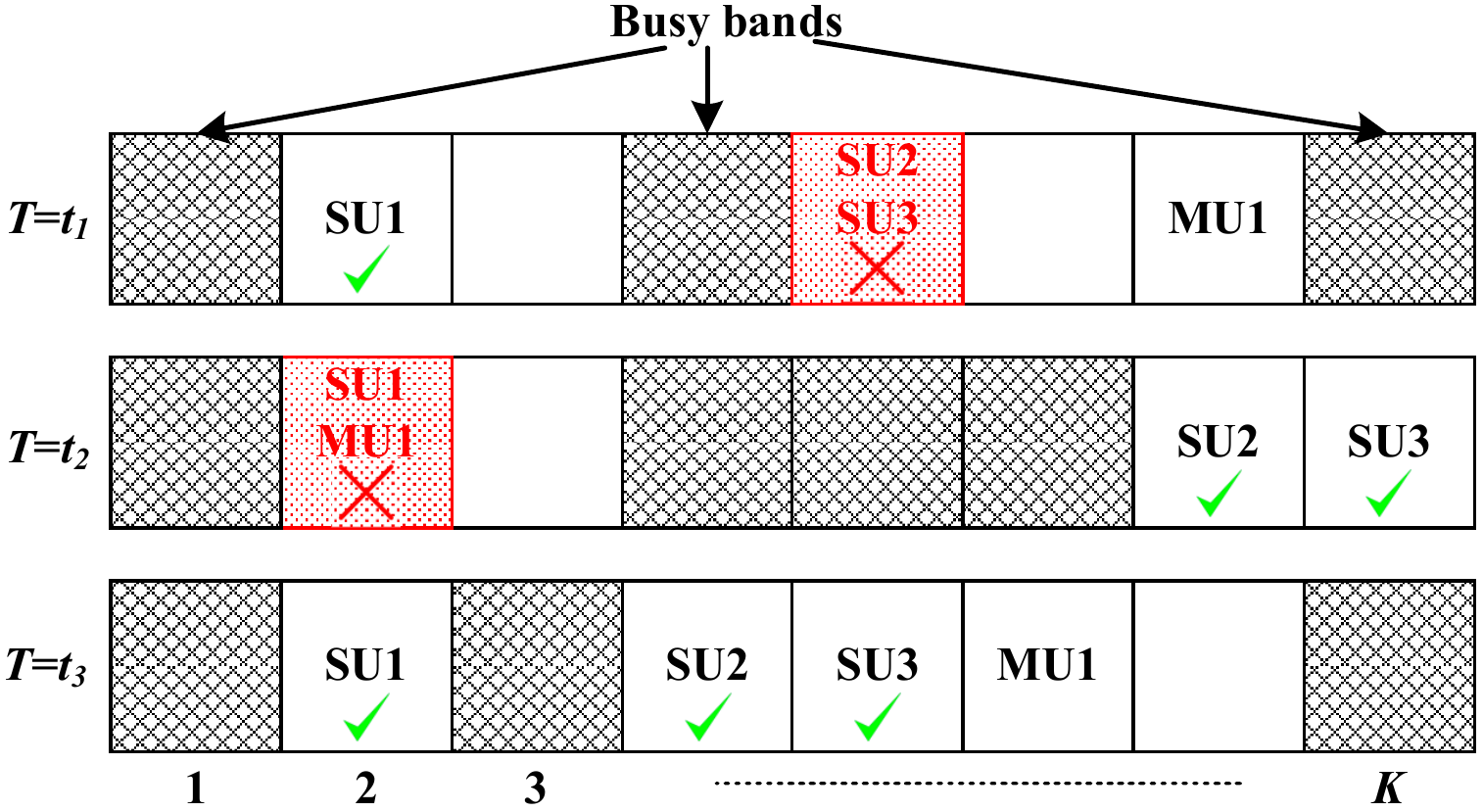}
		\label{band}}
		\caption{(a) The network model, and (b) Status of frequency bands at different time instant along with various collision events.}
	\label{fig1}
\end{figure}

Consider a decentralized network consisting of $N$ non-cooperative SUs competing for $K (\geq N)$ uniform bandwidth channels in the licensed spectrum. We assume time slotted communication, i.e., $t = 1,2,\cdots $, where each slot is further divided into two sub-slots. In the first sub-slot, each SU senses the channel for active PUs. In the second sub-slot, they transmit if the channel is vacant. We assume there are $J<N$ malicious users (or jammers) that have similar capability as that of SUs and follow identical process in each time slot. For illustration, consider the network shown in Fig.\ref{network}, consisting of the cellular base station (BS), PUs, SUs and malicious users (MUs) or jammers. The BS provides the data and control link only to the PUs. In case of packets received from SUs over any of the licensed channel, BS simply forwards the received packets to the destination based on the information in the header of the received packet. Such network is referred to as infrastructure-less CRN since BSs do not provide control link to SUs. As shown in Fig.\ref{band}, when two or more SUs or jammers transmit on the same vacant channel, collision occurs. In case of no collision, data transmission is assumed to be successful. For simplicity of analysis, we assume ideal detector, i.e. no sensing error. Also, SUs can sense only one channel in each time slot due to hardware and delay constraints.

The occupancy of channels by PUs in each slot is random. Let $p_i \in [0 \; 1], i \in [K]$ denote the probability that $i$-th channel is occupied by a PU (or busy) in a slot. This process is assumed to be independently and identically distributed across time slots and unknown to the SUs and the jammers. The same network model is considered for the study of non-cooperative CRN in several works including \cite{prand},\cite{quek}, \cite{zhandi} \cite{rjain}. Other important aspects of networking like, rendezvous, error-free sensing, time-synchronization are important but not considered here to keep the focus of the paper on the learning aspects as done in other papers \cite{prand},\cite{quek}, \cite{zhandi} \cite{rjain}. They can be addressed separately and out of scope of this work.

In the following we assume that $\sum_i p_i < K$, otherwise SUs cannot use the channels. This assumption also implies that there exists an $\theta>0$ such that $1-\sum_i {p_i/K} > \theta$. Let $\boldsymbol {\pi}$ denotes the vector of channel indices sorted according to increasing value of $\{p_i\}$ (with ties broken arbitrarily). We denote the $i^{th}$ component of $\boldsymbol {\pi}$ by $\boldsymbol {\pi}_i$. For later use, define $\Delta=p_{\boldsymbol {\pi}_{N}}-p_{\boldsymbol {\pi}_{N+1}}$ i.e., gap between the channel statistics of $N^{th}$ and $(N+1)^{th}$ best channels. We assume that $\Delta>0$. We use the notation round($a$) to denote that $a$ is rounded off to its nearest integer.

The SUs select a channel in each slot to transmit their packets. The channel selection is done by either hopping on to a channel randomly or sequentially. In random hopping, a channel is selected uniformly at random in each slot. In sequential hopping, SUs index the channels and select a new channel in each slot from the indices sequentially. When all the channels are selected once, they repeat the process. If SUs know the values of $N$,$J$ and $\{p_i\}$ and the jammers attack the top $N$ channels, then SUs can hop on top $N^*$=$N+m$, such that average throughput per channel is maximized for each SU. 
Note that by hopping on $N^*\geq N$ top channels the SUs may reduce the number of collisions with the jammer and get better throughput. For such case, $N^*$=$N+m$ is computed as follows:
\begin{multline} {\label{eq:optm}}
OP \left(N,J,\{p_i\}\right):\\
\hspace{-.2cm}  m = \argmax_w \left(\sum\limits_{i=1}^{N} \frac{1-p_{\boldsymbol {\pi}_i}}{N+w}\left (1-\frac{J}{N} \right) \hspace{-.1cm}+\hspace{-.3cm}\sum\limits_{i=N+1}^{N+w} \frac{1-p_{\boldsymbol {\pi}_i}}{N+w}\right).
\end{multline} 
The first term in the summation is the probability of a collision-free transmission on the top $N$ channels and the second term is the probability of a collision-free transmission on the next top $w$ channels. Note that SUs experience no collision from jammers on these $w$ channels.

The jammers are assumed to transmit enough power in each slot so that if the SU transmits on the same channel its transmission fails due to high interference. The jammers do not transmit on a channel if it is occupied by a PU, but always transmit power if the selected channel is idle. The jammers can be smart and detect transmissions from SUs on a vacant channel. In this case, they can continuously attack that channel till the SUs leave it. However, in our approach, SUs hop in every slot (either randomly or sequentially), hence sensing the presence of SUs is not useful to the jammers. If the jammers learn channel statistics, they can continuously attack the top $J$ channels. However this is not effective as the SUs can avoid the top $J$ channels (once they learn $J$), and they will never see jamming. So, the best strategy for the jammers is to attack the top $N$ channels randomly. In the following, we assume that the jammers attack $J$ channels in each slot selected uniformly at random from the top $N$ channels. Our method works without modification even if the jammers attack any $J$ channels in each slot selected uniformly at random.

We consider coordinated jammers in which jammers communicate with each other and attack non-overlapping (orthogonal) channels in each time slot, thus avoiding collisions among themselves. This case arises when there is a single jammer with multiple radios. The jammer can then attack distinct channels in each slot. For this case, we further consider two possible scenarios. In the first scenario, each SU can identify whether or not a jammer is involved in a collision experienced by it and in the second case, they cannot do so. We develop decentralized algorithms for both the scenarios. In the uncoordinated case where jammers cannot communicate, jammers operate independently and attack channels without the knowledge of the channels attacked by other jammers. 

We evaluate the performance of our algorithms in terms of regret. 
Regret is defined as the difference of expected optimal throughput and runtime average throughput defined as follows: 
\begin{align} \label{eq:regret} 
Regret &= R_{op}-E\left[\sum\limits_{t=1}^{T}\sum\limits_{n=1}^{N} r^n_t\right] \nonumber \\
 &= R_{op}-\sum\limits_{t=1}^{T}\sum\limits_{n=1}^{N} (1-p_{I_t^n})(1- E\left[\eta_{I_t}\right]).\end{align}
$R_{op}$ is the maximum total throughput achievable for SUs. It is achieved when each SU transmits on one of $N^*$ channels and do not collide with any other SU in each slot. $r^n_t$ is the reward at time $t$ for SU $n$,
$I_t^n$ denote the channel selected by SU $n$ at time $t$, $p_{I_t^n}$  and $\eta_{I_t^n}$ denote the busy probability and collision indicator on channel  $I_t^n$. If there is collision, collision indicator is set to $1$, otherwise it is $0$. The various notations and their definitions are summarized in Table \ref{gc2}.

\renewcommand{\arraystretch}{1.2}
\begin{table}[!h]
	\captionsetup{justification=raggedright,singlelinecheck=false}
	\caption{Notations and Definitions}
	\label{gc2}
	
	\begin{tabular}{|l|l|}
		\hline
		\textbf{Notations} & \textbf{Definitions} \\
		\hline
		$N$ (or $\hat{N}$) & Actual (or estimated) number of non-cooperative SUs \\
		\hline
		$K$ & No. of channels\\
		\hline
		$J$ (or $\hat{J}$) & Actual (or estimated) number of jammers\\
		\hline
		$\widehat{N+J}$ & Estimation of total number of SUs and jammers\\
		\hline
		$t$ & Current time slot\\
		\hline
		$T$ & Length of the time horizon\\
			\hline			
		$T_C$ & Time required for estimating correct channel ranking, \\ & number of the SUs and Jammers.\\
		\hline
		$p_i$ (or $\hat{p_i}$) & Actual (or estimated) probability that $i^{th}$ channel \\ & is occupied by a PU  in a slot \\
		\hline
		$N^*$ & Optimum number for channels for maximizing SU \\ &throughput after learning.\\
		\hline
		$\boldsymbol {\pi}$ & Vector of channel indices sorted according to \\ & increasing value of $p_i$ \\
		\hline
		$\Delta$ & Gap between $N^{th}$ and $(N+1)^{th}$ best channel \\
		\hline
		$R_{op}$ & Maximum total throughput achievable            		for SUs \\
		\hline
		$r_t^n$ & Reward at time $t$ for SU $n$ \\		
		\hline		
	 $I_t^n$ & Channel selected by SU $n$ at time $t$ \\
	   \hline	
	   $\eta_{I_t^n}$ & Collision indicator on channel  $I_t^n$ \\ 
	   	\hline
	   $i_0$ & Index of selected channel at the end of OR phase\\
	   \hline	
	   $p_{I_t^n}$ & Busy probability on channel  $I_t^n$ \\ 
	    \hline	
	   $C$ & Total number of collisions\\
	   \hline
	   $C_J$ &  Number of collisions from jammers \\		
		\hline	
	   $O_{i}$ & Number of times $i^{th}$ channel is selected \\ 
\hline	
	   $B_{i}$ & Number of times $i^{th}$ channel is sensed as busy \\ 
\hline	
	   $F$ & Number of times channel is sensed as vacant\\
	  
	   \hline	
	   $l$ & Indicator of successful transmission on a vacant \\ & channel\\
			
				\hline
		$x$ & Index of selected channel in $\pi$\\
\hline		
$p_{nc}$ & Probability of no collision\\
\hline		
$p_{c}$ & Collision probability of a SU at any time slot t\\
\hline		
$T_{O}$ & Time required for orthogonalization of all SUs on $K$ \\&  channels with high probability\\
\hline
$T_{J}$ & Time required for estimation of $J$ \\
\hline
$T_{L}$ & Sum of $T_C$, $T_O$ and $T_J$\\
\hline
$T_{R}$ & Time required for $\epsilon$-correct ranking of channels \\
\hline
$T_{NE}$ & Time required for estimation of $N$ \\
\hline
$T_{E}$ & Time required for estimation of number of active users \\
		\hline
		
	\end{tabular}
\end{table}	


\section{Distinguishable Jammers}\label{CJ-DC}

In this section, we assume that the SUs can identify if a collision is due to the presence of a jammer. We say that collisions, in this case, are distinguishable. For illustration, we discuss two scenarios where SU might be able to detect a collision due to a jammer:
1)~When the SUs are equipped with in-band full-duplex radios: In this case, the SUs can transmit and receive on the same channel. As a protocol, all the SUs stop to transmit (back-off) if they experience a collision.  But, the jammers
need not back-off and continue to transmit. After backing-off, if the SU observes activity on the channel, it must be the case that a jammer is present on the channel, otherwise, collision is due to the other SUs.
2)~Through acknowledgments: At the end of each time-slot, each SU receives an ACK/NACK feedback. If a collision happens due to other SUs but not the jammers, each SU receives a NACK signal. However, if the collision is due to the jammers, NACK signal is also jammed and hence, SUs do not receive it. 

Our proposed algorithm, named CDJ (Secondary User Coordination in Network with Jammers), is run at each SU terminal. The algorithm has two phases namely, Channel Ranking (CR) and Orthogonalization (OR). In the CR phase, the primary goal of each SU is to identify the channel ranking and number of jammers ($J$) and SUs ($N$). Since each user can identify the presence of a jammer in a collision, the total number of collisions caused by the jammers can be known. Using this information and the total number of collisions, $N$ and $J$ are estimated. In the OR phase, SUs hop randomly till they get a collision-free transmission on a channel and hop sequentially thereafter. As the jammers can cause maximum damage when they attack the top $N$ channels, they also learn the channel ranking and the number of SUs following the CR phase. They attack $J$ randomly selected channels from the top $N$ channels in each slot thereafter.
\begin{algorithm}[!h]
	\caption{CDJ} \label{Algo1}
	\begin{algorithmic}
	\State	Input: {$T_C,T,K$}
		\State	$(\hat{N},\hat{J}, \{\hat{p}_i\} )= CR1(T_C, K)$
		\State	 $N^*= OP(\hat{N},\hat{J}, \{\hat{p}_i\})$
		\State $ OR (\boldsymbol {\pi}, N^*, T)$ 
	\end{algorithmic}
\end{algorithm}
\begin{algorithm}[!t]
\caption*{\textbf{Subroutine 1:} $CR1$}
\begin{algorithmic}[1]
\State Input: $\ T_C,\ K$
\State Set $\ C_J= 0 ,\ C = 0, \ O_i = 0 ,\ B_i = 0 \ \forall i \in 1\dots K $ 
	\For{$t=1 \dots T_C$}
	\State Select $I_t \sim U(1\dots K)$\; 
	\State $O_{I_t} \leftarrow O_{I_t} + 1$ 
    \If{selected channel $I_t$ busy}
    \State $B_{I_t} \leftarrow B_{I_t} + 1, \ r_t \leftarrow 0 $
    \Else 
    \State Transmit on $I_t$,  $F \leftarrow F + 1$ 
    \If{collision}
    \State $C \leftarrow C + 1,  \ r_t \leftarrow 0$
	\If{collision from Jammer}
	\State	$C_J \leftarrow C_J + 1$
	\EndIf	
	\Else	
	\State	$r_t \leftarrow 1$
	
	\EndIf
	\EndIf	
	\EndFor	
	\State set $\hat{p}_i = \frac{B_i}{O_i}  \ \forall i \in 1\dots K \mbox{ and } \hat{J} = round\left(KC_J/F)\right)$ 
	\State $ \hat{N} =round\left( 1+ \frac{\log \left(1-\frac{C}{F}\right)-\log\left(1-\frac{\hat{J}}{K}\right)}{\log\left(1-\frac{1}{K}\right)}\right)$
		\end{algorithmic}
\end{algorithm}

\subsection{SU Algorithm: CDJ}
CDJ described in Algorithm \ref{Algo1} consists of two subroutines that run sequentially. The first subroutine named CR1 runs for $T_C$ (see (\ref{eq:TCLengh})) time slots. In each slot, the SU selects a channel ($I_t$) uniformly at random.
If $I_t$ is idle then the SU transmits on $I_t$ and checks for collision. If a collision occurs, the SU looks for the presence of a jammer and, if detected, updates the count of the number of collisions from jammer ($C_J$). If no collision occurs, transmission is treated as successful and reward ($r_t$) of $1$ unit is assigned in that slot. 
The algorithm keeps track of number of times each channel is selected ($O_{I_t}$), how many times selected channels are found busy ($B_{I_t}$),  total number of vacant slots ($F$), total number of collisions $(C)$, and use it to estimate $N, J$ and $\{p_i\}$.
These estimates are then used to find $N^*$ from (\ref{eq:optm}). 
The second subroutine OR takes an array of channel indices ($\pi$) sorted in increasing order of the estimated channel busy probabilities, $N^*$, and time horizon $T$ as inputs.  In this subroutine, each SU randomly hops on top $N^*$ channels till its transmission succeeds on a vacant channel ($l=1$), and once successful on a channel, they hop sequentially from that channel following indices from $\boldsymbol {\pi}$.


\vspace{-.1cm}

\subsection{Jammer Algorithm}
The algorithm for the jammer is described in Algorithm \ref{Algo2}, which is similar to the SUs till $T_C$. After $T_C$, 
the jammer deviates and attack any $J$ channels from top $N$ channels in each slot. This results in maximum throughput loss for SUs. 

\begin{algorithm}[!t]
\caption*{\textbf{Subroutine 2:} $OR$}
\begin{algorithmic}[1]
\State Input: $\boldsymbol {\pi}, N^*, \ T$ 
\State Set $l = 0 , x = 0$ \; 
	\For{$t=T_C+1, T_C+2 \ldots T $}
    \If{$l=1$}
    \State $x  \leftarrow (x+1)  \mbox{ modulo } N^* $ and transmit on $ {\pi}_x$
    \If{channel $ {\pi}_x$ idle \textbf{and} no collision}
    \State $ r_t \leftarrow  1 $ \;
    \EndIf
    \Else 
    \State select $I_t \sim U({\pi}_1 \dots  {\pi}_{N^*}	)$ and transmit on $I_t$	  
    \If{channel $ {\pi}_{I_t}$ idle \textbf{and} no collision}
    \State $l  \leftarrow 1, x \leftarrow index \ of \ I_t \ in \ \boldsymbol {\pi}, r_t \leftarrow  1 $ \;
    \EndIf
	\EndIf	
	\EndFor	
		\end{algorithmic}
\end{algorithm}

\begin{algorithm}[!h]
	\caption{Jammer Algorithm} \label{Algo2}
	\begin{algorithmic}[1]
	\State	Input: {$\ T_C,\ T$}
	\State	Set $\ C=0, \ B=0, \ O_i = 0 ,\ B_i = 0 \ \forall i \in 1\dots K $ 
	\For{$t=1 \dots T_C$}
	\State	 Select $J$ orthogonal channels from [K] randomly
	\State $O_{I_t} \leftarrow O_{I_t} + 1$ for all the selected channels 
	\If{any selected channel $I_t$ is busy}
	\State $B_{I_t} \leftarrow B_{I_t} + 1, \ B\leftarrow B+1 $
	\Else
	\State Attack selected $J$ channels
	\If{ $x$ collision are observed}
	\State $C \leftarrow C+ x$
	\EndIf
	\EndIf	
	\EndFor	
	\State $\hat{p}_i = \frac{B_i}{O_i}  \ \forall i \in 1\dots K \mbox{ and } \hat{N} = round \left(\frac {\log \left( 1-\frac{C}{JT_C - B} \right) } {\log \left(1-\frac{1}{K} \right)} \right) $
\For{$t=T_C+1 \cdots T$}
\State Select $J$ orthogonal channels from top $\hat{N}$ randomly\;
	\EndFor	
	\end{algorithmic}
\end{algorithm}

\vspace{-.1cm}
\subsection{\textbf{Analysis of CDJ Algorithm}} 
In this subsection, we bound the regret of CDJ. We begin with the following definition given in \cite{MC,MEGA}.
\begin{defn}
An $\epsilon$-correct ranking of $K$ channels is a sorted list of empirical mean values of channel occupancy probabilities such that for all $i,j$, $\hat{p}_i$ is listed before $\hat{p}_j$ if $p_j - p_i >  \epsilon$.
\end{defn}
\noindent

\begin{lemma} \label{thm:le1}
By the end of CR1 phase, all SUs have $\epsilon$-correct ranking of the channels, number of SUs ($N$) and number of Jammers ($J$) with probability at least $1-\delta$.
\end{lemma}
\noindent \textbf{Proof}: The proof steps are similar to \cite{MC}[Lemma 1].
If for any user $n$, it is true that $\forall k \in 1 \cdots K$ $\left | \hat{p}_k-p_k \right| \leq\frac{\epsilon}{2} $, then the 
user has an  $\epsilon-$ correct ranking. We will upper bound the probability that no user has $\epsilon-$correct ranking given the user has $S_1$ observations of each 
channel. 
We define the following events : \\

\noindent$J_n$ - event that an user $n$ observed each channel $j\geq0$ times.\\
$A$ - event that all users have an $\epsilon$-correct ranking. \\
$A_n$ - event that an user $n$ has $\epsilon$- correct ranking. \\
$B$ - event that all users have at least $S_1$ observations of each channel. \\
$B_n$ - event that an user $n$ has at least $S_1$ observations of each channel. Let $\overline{X}$ denote complement of any event $X$\\

We first upper bound $Pr(\overline{A}_n|B_n)$. We have, 
\begin{align*}
Pr(\overline{A}_n|B_n) &\leq Pr\left ( \exists k \in 1 \cdots K \ \ s.t | \hat{p}_k-p_k | >\frac{\epsilon}{2} | \ B_n\right) \\
 \tag{By Union Bound}      &\leq \sum_{k=1}^{K} Pr\left ( | \hat{p}_k-p_k | >\frac{\epsilon}{2} | \ B_n\right) \\ 
       			&= \sum_{k=1}^{K} \sum_{j=C}^{\infty} Pr\left (| \hat{p}_k-p_k | >\frac{\epsilon}{2} | \ 
J_n  \right) Pr \left( J_n | \ B_n \right) \\ 
 \tag {By Hoeffding's Inequality} &\leq \sum_{k=1}^{K} \sum_{j=C}^{\infty} 2  \exp \left(\frac{-j  \epsilon^2}{2} \right) Pr \left( J_n | \ B_n \right) \\ 
	&\leq \sum_{k=1}^{K} 2  \exp \left(\frac{-S_1  \epsilon^2}{2} \right) \sum_{j=C}^{\infty} Pr \left( J_n | \ B_n \right) \\ 
	 &\leq \sum_{k=1}^{K} 2 \exp \left(\frac{-S_1 \epsilon^2}{2} \right) \leq 2 K  \exp \left(\frac{-S_1 \epsilon^2}{2} \right)
\end{align*}
We can apply Hoeffding's inequality since each observation of the channel is independent of the number of times we observe that
channel. This is true since each SU and jammer is randomly selecting channels independent of the
previous rounds in each round. In order for this to be $< \frac{\delta}{4K}$, 
	 \[2 K  \exp \left(\frac{-S_1  \epsilon^2}{2} \right) < \frac{\delta}{4K} \newline
		\implies  S_1 >   \ln \left ( \frac{8  K^2}{\delta}   \right) \frac{2}{\epsilon^2}
\]

Now to show that if all users have at least $ S_1 > \ln \left ( \frac{8  K^2}{\delta}   \right) \frac{2}{\epsilon^2}$ observations of each channel, then all users have an $\epsilon-$correct ranking of channels with probability at least $1-\frac{\delta}{4}$. 

\begin{align*} 
     Pr(A|B) &\geq 1-Pr\left( \vee_n Pr(\overline{A}_n | B_n) \right) \\
            &\geq 1-\sum_{n=1}^{N} Pr(\overline{A}_n | B_n)    \tag{By Union Bound}\\
	    &\geq 1-\frac{\delta}{4K}  (N) \geq 1-\frac{\delta}{4K}  K= 1-\frac{\delta}{4}
\end{align*}
We want to show that there exists a $T_{S1}$ large enough such that all SUs have greater than $S_1$ observations of each channel.  
Let $B_{n,k}(t)$ be a random variable which indicates that channel $k$ is selected by SU $n$. 
\begin{equation*}\label{eq:ct}
				B_{n,k}(t)= \left\{ \begin{array}{cl}
				1 &\ w.p \ \frac{1}{K} \\
				0 &\ w.p \ \frac{1}{K}. \end{array} \right .
                \end{equation*}
 We have   
\begin{align*}
Pr &\left(n^{th} \ SU \ has < \frac{1}{2} T_{S_1}\ E[B_{n,k}(t)] \ observations\right)  \\ 
 &= Pr\left(\sum\limits_{t=1}^{T_{S_1}}B_{n,k}(t) < \frac{1}{2}T_{S_1}\ E[B_{n,k}(t)]\right)\\
 &\leq \exp \left(-\frac{1}{8} T_{S_1}\ E[B_{n,k}(t)] \right) \tag{ Chernoff's Bound}
\end{align*}
We can apply Chernoff bound here since for any $n, k, B_{n,k}$ is i.i.d across $t$ and all SUs employ random sampling in CR phase.
We want,
\begin{align*}
&Pr \left(\exists n,k \ s.t.\sum\limits_{t=1}^{T_{S_1}} B_{n,k}(t) < \frac{1}{2} T_{S_1} E[B_{n,k}(t)]\right) <\frac{\delta}{4} \\
&\implies N\exp \left(-\frac{1}{8} T_{S_1} E[B_{n,k}(t)] \right) < \frac{\delta}{4K} \\
&\implies T_{S_1} > 8  \ln \left(\frac{4KN}{\delta}\right)  \frac{1}{E[B_{n,k}(t)] } \\
&\implies T_{S_1} > 8  \ln \left(\frac{4K^2}{\delta}\right)  \frac{1}{E[B_{n,k}(t)] } \\
&\implies T_{S_1} > 8K\ln \left(4K^2/\delta\right)
\end{align*}
The last inequality follows using $E\left[B_{n,k}(t)\right] = 1/K$. We have shown that for $T_{S_1} >8K\ln \left(4K^2/\delta\right)$ , the number of observations of selected channels for all SUs, $\sum\limits_{t=1}^{T_{S_1}} B_{n,k}(t)$ is greater than $ \frac{1}{2} T_{S_1} E[B_{n,k}(t)]$ with probability at least
$1-~\frac{\delta}{4} $. We also want that total number of observations of each channel be at least $S_1$. Therefore,
\begin{align*}
&\sum\limits_{t=1}^{T_{S_1}} B_{n,k}(t) \geq \frac{1}{2} T_{S_1} E[B_{n,k}(t)] \geq S_1\\
&\implies \frac{1}{2}T_{S_1} E[B_{n,k}(t)] \geq \frac{2}{\epsilon^2} \ln \left ( \frac{8  K^2}{\delta}\right) \\
&\implies  T_{S_1} \geq \frac{4}{\epsilon^2 E[B_{n,k}(t)]}\ln \left ( \frac{8  K^2}{\delta} \right)    \\
&\implies  T_{S_1} \geq \frac{4K}{\epsilon^2 }\ln \left ( \frac{8  K^2}{\delta} \right)   
\end{align*}
Combining all, the probability that all the users will have $\epsilon$-correct ranking is given by
\begin{align*}
Pr(A) &= 1 - Pr(\overline{A}) \\
      &= 1 - Pr(\overline{A}/B)Pr(B)-Pr(\overline{A}/\overline{B})Pr(\overline{B}) \\
      &\geq 1 - Pr(\overline{A}/B) - Pr(\overline{B})\geq 1- \frac{\delta}{4} - \frac{\delta}{4} \geq 1- \frac{\delta}{2}.
\end{align*}
Thus, if $ T_{R} = \max \left(8K\ln \left(\frac{4K^2}{\delta}\right),\frac{4K}{\epsilon^2}\ln \left ( \frac{8  K^2}{\delta} \right) \right)$, then for any $t\geq T_{R}$ all SUs will have $\epsilon$-correct ranking of channels with probability at least 
 $1- \frac{\delta}{2}.$ \\


\noindent \textbf{Estimation of $N$}: Now, we compute number of slots required to estimate $N$ and $J$ with high probability.
We first compute slots required to estimate $N$ given $J$. We define the following events : \\

\noindent$W_k$ - event that channel $k$ is selected by a SU \\
$X_k$ - event that none of the other SUs/Jammer selects $k$ \\
$U_k$ - event that channel $k$ is free. \\

\noindent
Let $S_2$ be the number of times a selected channel is found free. 
Probability of collision, denoted $p_c$, for any SU conditioned on the events that the channel selected is free and the number of Jammers is $J$ is given by
\begin{align*}
Pr(collision) &= \sum\limits_{k=1}^{K}  Pr(W_k/U_k)Pr(X_k/U_k)  \\
p_c	&= \sum\limits_{k=1}^{K} \frac{1}{K} \left(1-\left(1-\frac{J}{K}\right)\left(1-\frac{1}{K}\right)^{N-1}  \right)\\
p_c	&= 1-\left(1-\frac{J}{K}\right)\left(1-\frac{1}{K}\right)^{N-1}   \\
\end{align*}
Solving for N, we get
\[N = 1 + \frac{\log \left (1-p_c\right)-\log \left ( 1-\frac{J}{K} \right )} {\log \left ( 1-\frac{1}{K} \right )}\] 
Let $\hat{p_c}$  be the estimate of $p_c$. 
\[\hat{N} = 1 + \frac{\log \left (1-\hat{p}_c\right)-\log \left ( 1-\frac{J}{K} \right )} {\log \left ( 1-\frac{1}{K} \right )}\] 
We want $\hat{N}$ to be a good estimate of $N$. Therefore, for some $\gamma < \frac{1}{2}$, we set $ \left | \hat{N} - N \right| < \gamma$
  
\[\left| \frac{\log \left (1-\hat{p}_c\right)} {\log \left ( 1-\frac{1}{K} \right )}-\frac{\log \left (1-p_c\right)} {\log \left ( 1-\frac{1}{K} \right )} \right|  < \gamma  \iff \left| \frac{\log \left(\frac{1-\hat{p}_c}{1-p_c}\right)}{\log \left ( 1-\frac{1}{K} \right )}\right| < \gamma   \]
Let $\alpha$ be the difference between $\hat{p}_c$ and $p_c$. Substituting $\hat{p}_c = p_c +\alpha $.  

\[-\gamma\log \left ( 1-\frac{1}{K} \right )\leq \log \left(\frac{1-p_c - \alpha}{ 1-p_c}  \right) \leq  \gamma\log \left ( 1-\frac{1}{K} \right )  \]
\[ \log \left ( 1-\frac{1}{K} \right )^{-\gamma}\leq \log \left(\frac{1-p_c - \alpha}{ 1-p_c}  \right) \leq  \log \left ( 1-\frac{1}{K} \right )^\gamma \]
\[ \left ( 1-\frac{1}{K} \right )^{\gamma}\leq  \left(\frac{1-p_c - \alpha}{ 1-p_c}  \right) \leq   \left ( 1-\frac{1}{K} \right )^{-\gamma} \]
\[ \left ( 1-\frac{1}{K} \right )^{\gamma} \left(1-p_c \right)\leq  \left(1-p_c - \alpha  \right) \leq   \left ( 1-\frac{1}{K} \right )^{-\gamma} \left(1-p_c \right)  \]
\[ \left(1-p_c \right)\left(1-\left ( 1-\frac{1}{K} \right )^{-\gamma} \right) \leq \alpha \leq \left(1-p_c \right)\left(1-\left ( 1-\frac{1}{K} \right )^{\gamma} \right). \]
If we ensure $|\hat{p}_c - p_c| < \epsilon_1$ where
\begin{multline}\epsilon_1 = \min \left \{\left| \left(1-p_c \right)\left(1-\left ( 1-\frac{1}{K} \right )^{-\gamma} \right)\right| \right.,
\\ \left. \left| \left(1-p_c \right)\left(1-\left ( 1-\frac{1}{K} \right )^{\gamma} \right)\right|\right \}\end{multline}
for some $\gamma<1/2$ then we can have $\hat{N} = N$. The number of observations $S_2$ required to compute $\hat{p_c}$
such that $|\hat{p}_c - p_c| \leq \epsilon_1$  for all SUs is given by Hoeffding's inequality. 
 \[Pr(|\hat{p}_c - p_c| >\epsilon_1) \leq 2\exp(-2 S_2\epsilon^2_1) < \frac{\delta}{6K} \]
\[\implies S_2 > \frac{1}{2\epsilon^2_1}\ln \left(\frac{12K}{\delta}\right) \]
Following steps in the proof of \cite{MC}, we can further simplify the expression for $\epsilon_1$. Substituting the value of $p_c$, we have 
 \begin{multline}\epsilon_1 = \min \left \{\left|\left(1-\frac{1}{K}\right)^{N-1} \left(1-\frac{J}{K} \right) \left(1-\left ( 1-\frac{1}{K} \right )^{-\gamma} \right)\right| \right., \\ \left.
                    \left| \left(1-\frac{1}{K}\right)^{N-1} \left(1-\frac{J}{K} \right) \left(1-\left ( 1-\frac{1}{K} \right )^{\gamma} \right)\right|\right \}\end{multline}
Using $\left(1-\frac{J}{K} \right)\geq \frac{1}{2}$  as $K>N>J$, we have 
\[\left|\left(1-\frac{1}{K}\right)^{N-1} \left(1-\frac{J}{K} \right)\left(1-\left ( 1-\frac{1}{K} \right )^{-\gamma} \right)\right| \geq \frac{\gamma}{2\exp(1)K}\]
\[\left|\left(1-\frac{1}{K}\right)^{N-1}\left(1-\frac{J}{K} \right) \left(1-\left ( 1-\frac{1}{K} \right )^{\gamma} \right)\right| \geq \frac{\gamma}{2\exp(1)K}\]
Combining above two, we have 
\[\epsilon_1 \geq \frac{\gamma}{2\exp(1)K}\]
If all SUs have $S_2>\frac{1}{2\epsilon^2_1}\ln \left(\frac{12K}{\delta}\right)$ observations of channel when they are free, where $\epsilon_1 =\frac{\gamma/K}{2\exp(1)}$, then with 
probability at least $1-\frac{\delta}{6}$, we have  $\hat{N} = N$ for all SUs given $J$.\\

\noindent
\textbf{Estimation of $J$:} Next we compute number of slots required to estimate $J$ correctly by all SUs with high probability. 
We compute collision probability with jammers when the channel is free.
Let $S_3$ be the number of times we observe that selected channel is free.
Given $S_3$, we want to ensure correct estimation of $J$ with high probability.
\begin{align*} {\label{eq:Pc4}}
Pr(jammer \; from\; collision) = p_{cj }&= \sum\limits_{i=1}^{K} \frac{1}{K} \left(\frac{J}{K}\right) \\
\implies J = Kp_{cj}
\end{align*}
Let $\hat{p}_{cj}$ be the estimate of $p_{cj}$. Then the estimate of $J$ is given by $\hat{J}=K\hat{p}_{cj}$.
We want to ensure $\hat{J}$ is close to $J$. Therefore, for some $\gamma < 1/2$ we set
\begin{align*} 
\left| \hat{J}- J \right| \leq \gamma 
\iff \left| \hat{p}_{cj} - p_{cj} \right| \leq \frac{\gamma}{K} \\
\end{align*}
The number of observations $S_3$ required to compute $\hat{p_{cj}}$
such that $|\hat{p}_{cj} - p_{cj}| \leq \epsilon_2$, where $\epsilon_2= \frac{\gamma}{K}$ for all SUs is given by applying the Hoeffding's inequality. 
 \[Pr(|\hat{p}_{cj }- p_{cj}| >\epsilon_2) \leq 2\exp(-2 C_3\epsilon^2_2) < \frac{\delta}{12K} \]
\[\implies S_3 > \frac{1}{2\epsilon^2_2}\ln \left(\frac{24K}{\delta}\right) \]
Next, we find the number of time slots ($T_F$) required to get the desired number of observations in which selected channel were 
free.
Let $A_{I_t}(t)$ be a random variable which indicates if selected channel $k$ is free at round $t$. 
\begin{equation}\label{eq:ct}
				A_{I_t}(t)= \left\{ \begin{array}{cl}
				1& \mbox{ if  $I_t$  is free}\\
				0&   \mbox{otherwise} \end{array} \right . 
                \end{equation}
We have
\begin{align*}
Pr &\left(a \ SU \ has < \frac{1}{2}\cdot T_{F}\cdot E[A_{I_t}(t)] observations\right)  \\ 
 &= Pr\left(\sum\limits_{t=1}^{T_{F}} A_{I_t}(t) < \frac{1}{2}\cdot T_{F}\cdot E[A_{I_t}(t)]\right)\\
 &\leq \exp \left(-\frac{1}{8}\cdot T_{F}\cdot E[A_{I_t}(t)] \right) \tag{ Chernoff's Bound}
\end{align*}
We set,
\begin{align*}
&Pr \left(\mbox{SUs do not have} > \frac{1}{2} T_{F}E[A_{I_t}(t)] \mbox{observations}\right)< \frac{\delta}{18} \\
&\implies \exp \left(-\frac{1}{8} T_{F} E[A_{I_t}(t)] \right) < \frac{\delta}{18K} \\
&\implies T_{F} > 8  \ln \left(\frac{18K}{\delta}\right)  \frac{1}{E[A_{I_t}(t)] } \\ 
&\implies T_{F} >\frac{8}{\theta}  \ln \left(\frac{18K}{\delta}\right),  
\end{align*}
In the last if-and-only statement we used the relation
\begin{align*}
E[A_{I_t}(t)] &= \sum\limits_{k=1}^{K}Pr(I_t=k) Pr(A_{I_t}(t)=1/I_t = k) \\
&= \sum\limits_{k=1}^{K}\frac{1}{K} (1-p_k)\geq \theta. 
\end{align*}
We have shown that for $T_{F} > \frac{8}{\theta} \cdot \ln \left(\frac{18K}{\delta}\right)  $ , the number of observations of 
selected channels being free for all SUs, $\sum\limits_{t=1}^{T_F} A_{I_t}(t)$ is greater than $ \frac{1}{2}\cdot T_{F}\cdot E[A_{I_t}(t)]$ 
with probability at least $1-\frac{\delta}{18} $

We also want that total number of observations of selected channels being free to be at least $\max(S_2,S_3)$.
Therefore, \\ 
\[ \sum\limits_{t=1}^{T_{S_2}} A_{I_t}(t) \geq \frac{1}{2}\cdot T_{S_2}\cdot E[A_{I_t}(t)] \geq S_2\]
\[\implies \frac{1}{2}\cdot T_{S_2}\cdot E[A_{I_t}(t)] \geq \frac{1}{2\epsilon^2_1}\ln \left(\frac{12K}{\delta}\right)\]
\[\implies  T_{S_2} \geq \frac{1}{\epsilon^2_1 E[A_{I_t}(t)]}\ln \left(\frac{12K}{\delta}\right) \]
\[\implies  T_{S_2} \geq \frac{1}{\epsilon^2_1 \theta}\ln \left(\frac{12K}{\delta}\right) \]

and 

\[ \sum\limits_{t=1}^{T_{S_3}} A_{I_t}(t) \geq \frac{1}{2}\cdot T_{S_3}\cdot E[A_{I_t}(t)] \geq S_3\]
\[\implies \frac{1}{2}\cdot T_{S_3}\cdot E[A_{I_t}(t)] \geq \frac{1}{2\epsilon^2_2}\ln \left(\frac{24K}{\delta}\right)  \]
\[\implies  T_{S_3} \geq \frac{1}{\epsilon^2_2 E[A_{I_t}(t)]}\ln \left(\frac{24K}{\delta}\right) \]
\[\implies  T_{S_3} \geq \frac{1}{\epsilon^2_2 \theta}\ln \left(\frac{24K}{\delta}\right) \]

\[T_{{E}} = max\left(\frac{8}{\theta} \cdot \ln \left(\frac{18K}{\delta}\right),\frac{1}{\epsilon^2_1 \theta}\ln \left(\frac{12K}{\delta}\right),\frac{1}{\epsilon^2_2 \theta}\ln \left(\frac{24K}{\delta}\right)\right)\]
Let us define the following events:
$E$ - event that all SUs have observed at least $S_2$ times the selected channel was free \\ 
$F$ - event that all SUs have correct estimate of $J$.\\
$D$ - event that all SUs have correct estimate of $N$.  \\
Then, we have
\begin{align*}
Pr(DF) \geq & 1 - Pr(\overline{D})-Pr(\overline{F})   \\
     &= 1 - Pr(\overline{D}|EF)Pr(EF) - Pr(\overline{D}|\overline{EF})Pr(\overline{EF})-\\&Pr(\overline{F}) \\ 
     &\geq 1-Pr(\overline{D}|EF)-Pr(\overline{EF})-Pr(\overline{F}) \\
     &\geq 1-Pr(\overline{D}|EF)-Pr(\overline{E})-2Pr(\overline{F})\\
     &\geq 1-Pr(\overline{D}|EF)-Pr(\overline{E})-2Pr(\overline{F}|E)Pr(E)-\\&2Pr(\overline{F}|\overline{E})Pr(\overline{E})\\
     &\geq 1-Pr(\overline{D}|EF)-Pr(\overline{E})-2Pr(\overline{F}|E)-2Pr(\overline{E})\\
     &\geq 1-Pr(\overline{D}|EF)-2Pr(\overline{F}|E)-3Pr(\overline{E})\\
     &\geq 1 - \frac{\delta}{6}-\frac{2\delta}{12}-\frac{3\delta}{18}\geq 1 - \frac{\delta}{2} \\
\end{align*}
Finally, if $T_{{E}} = \max\left(\frac{8}{\theta}  \ln \left(\frac{18K}{\delta}\right),\frac{1}{\epsilon^2_1 \theta}\ln \left(\frac{12K}
{\delta}\right),\frac{1}{\epsilon^2_2 \theta}\ln \left(\frac{24K}{\delta}\right)\right)$ then for $t \geq T_{{E}}$, all SUs would have 
correct estimate of $N$ and $J$ with probability atleast $1-\frac{\delta}{2}$. 
\newline
For $A$,$D$ and $F$ be as defined previously,
\begin{align*}
Pr(ADF)&= 1-Pr(\overline{ADF}) \\
      &\geq 1 - Pr(\overline{A}) - Pr(\overline{DF})\\
      &\geq 1 -  \frac{\delta}{2}-\frac{\delta}{2}= 1-\delta 
\end{align*}
Hence if $T_{C} = max (T_{R},T_{E})$, then with probability $1-\delta$, all SUs would have $\epsilon$-correct channel ranking as well as correct estimate of $N$ given $J$.


The following theorem states a high confidence bound on the expected regret of CDJ.  The expectation is over the algorithm's randomness. 
\begin{thm}\label{thm:t1}
For all $\Delta > \epsilon, \gamma \in (0, 0.5)$ and $\delta \in [0,1]$, with probability at least $1-\delta$, the expected regret of $N$ SUs using CDJ in the 
	presence of $J$ jammers with $T_C$ slots of learning is upper bounded by $NT_C + \frac{N^2exp(1)}{\theta}$, where $T_C$ is 
\begin{eqnarray}
\label{eq:TCLengh}
&& \mbox{round} \bigg(max\bigg(\frac{8}{\theta} \cdot \ln \left(\frac{18K}{\delta}\right),\frac{1}{\epsilon^2_1 \theta}\ln \left(\frac{12K}{\delta}\right), \\
\nonumber
&& \hspace{.5cm} \frac{1}{\epsilon^2_2 \theta}\ln \left(\frac{24K}{\delta}\right), 8K \ln \left(\frac{4K^2}{\delta}\right),\frac{4K}{\epsilon^2}\ln \left( \frac{8 K^2}{\delta} \right) \bigg)\bigg)
\end{eqnarray}	
with $\epsilon_1=\gamma/2K\exp(1)$ and $\epsilon_2= \gamma/K$.
\end{thm}
\noindent
The first term ($NT_C$) is the total regret incurred by all users in the CR1 phase. This phase runs for $T_C$ number of time slots, required for estimating correct channel ranking, number of the SUs and Jammers. The second part ($N^2exp(1)$) corresponds to regret incurred in the final OR phase. \\
\noindent \textbf{Proof:} After estimating $N,J$ and channel ranking from Lemma \ref{thm:le1}, all SUs run the OR subroutine for the second time. We will compute the maximum slots required such that all SUs get orthogonalized and start sequential hopping. Our approach to bound regret is similar to approach in \cite{MC}. Probability of collision for any SU at any round $t$ is given by: 
\begin{align*}
p_{nc} &= \sum\limits_{k=1}^{N} \frac{1-p_k}{N+m} \left(1-\frac{J}{N}\right)\left(1-\frac{1}{N+m}\right)^{N-1}  \\
	&+ \sum\limits_{k=N+1}^{N+m} \frac{1-p_k}{N+m} \left(1-\frac{1}{N+m}\right)^{N-1}+ \sum\limits_{k=1}^{N+m} \frac{p_k}{N+m}\\		
 &\geq \sum\limits_{k=1}^{N+m} \frac{1-p_k}{N+m} \left(1-\frac{J}{N}\right)\left(1-\frac{1}{N+m}\right)^{N-1}\\								
 &\geq \left(1-\frac{J}{N}\right)\left(1-\frac{1}{N+m}\right)^{N-1} \sum\limits_{k=1}^{N+m} \frac{1-p_k}{N+m}\\	
 &\geq \left(\frac{N-J}{N}\right)\left(1-\frac{1}{N+m}\right)^{N-1} \theta \\
 &\geq \left(\frac{1}{N}\right)\left(1-\frac{1}{N+m}\right)^{N+m-1} \theta \geq \frac{\theta}{Nexp(1)}\\
\end{align*}
Since the number of slots required to find a collision free transmission and start sequential hopping is geometric distributed, the expected number of slots for first success, denoted by $T_{SH}$, is bounded by :
\[T_{SH} < \frac{1}{p_{nc}} \implies T_{{SH}} < \frac{Nexp(1)}{\theta} .\]  
We will now bound the expected regret due to users when running the OR subroutine for second time.
\begin{eqnarray*}
 \lefteqn{E\left[\sum\limits_{t=T_C+1}^{T} \sum\limits_{i=1}^{N}r_{t}^n \right]}\\
&\leq& \hspace{-.2cm} E\left[\sum\limits_{t=T_C+1}^{T} \sum\limits_{i=1}^{N}\mathbbm{1}\{\mbox{SUs not  orthogonalized in round $t$}\} \right]\\
&\leq&  \hspace{-.2cm}  E\left[\sum\limits_{t=T_C+1}^{T} N\mathbbm{1}\{\mbox{SUs not  orthogonalized in round $t$}\} \right] \\
 &\leq & NT_{SH} \leq \frac{N^2exp(1)}{\theta}.
\end{eqnarray*}
Using Lemma \ref{thm:le1} and the above we prove that channel ranking, number of SUs and jammers can be estimated with probability $1-\delta$
and total regret is bounded by $NT_C + \frac{N^2exp(1)}{\theta}$.
Hence completing proof of Theorem \ref{thm:t1}.


\section{Non-Distinguishable Jammers}\label{CJ-NDC}

In this section, we remove the requirement that the SUs can identify whether or not a collision happened due to the presence of a jammer using either hardware capability or a signaling scheme. We say collisions, in this case, are indistinguishable. 

The main difficulty in this scenario is one cannot estimate $N$ and $J$ using the collision information. To see this, notice that the expression for collision probability when all SUs and jammers hop randomly and conditioned on the event that channels are vacant, given by 
\begin{equation} \label{eq:PC}
Pr(collision)	= \left(1-\left(1-J/K\right)\left(1-1/K \right)^{N-1}  \right)
\end{equation} 
cannot be inverted to get the value of $N$ or $J$ without knowing the value of other. But both the values are unknown a priori. So, the empirical estimates of collisions are not useful to get estimates of $J$ and $N$ as in the previous section where collisions are distinguishable. Our next algorithm CNJ (Secondary User Coordination in a Network with Jammers 2) given in Algorithm \ref{Algo3} overcomes this difficulty by first allowing the SUs to orthogonalize and then count the collisions. After orthogonalization, collisions are only due to the jammers as there are no collisions among SUs. From the information on collision caused by the jammers the SUs can estimate $J$, and use it in turn to find an estimate for $N$ using ($\ref{eq:PC}$) and the estimate of collision probability obtained before orthogonalization where collisions are due to both SUs and jammers. 

The CNJ algorithm has three phases. The two phases CR and OR are as in the previous section. The third phase is named Jammer Estimation (JE) phase. The OR phase occurs two times, one before JE phase and one after it. The OR phase before the JE phase allows SUs to find non-overlapping channels. Then, any collisions seen in the JE phase are only due to the jammers. In the JE phase, the SUs use collision count from the jammers to estimate their number.  Once $J$ is estimated with good accuracy, OR phase is again applied so that the SUs orthogonalize on the top $N^*$ channels.

\subsection{SU Algorithm: CNJ }
\begin{figure}[!h]
	\vspace{-0.5cm}
	\centering
	\includegraphics[scale=0.4]{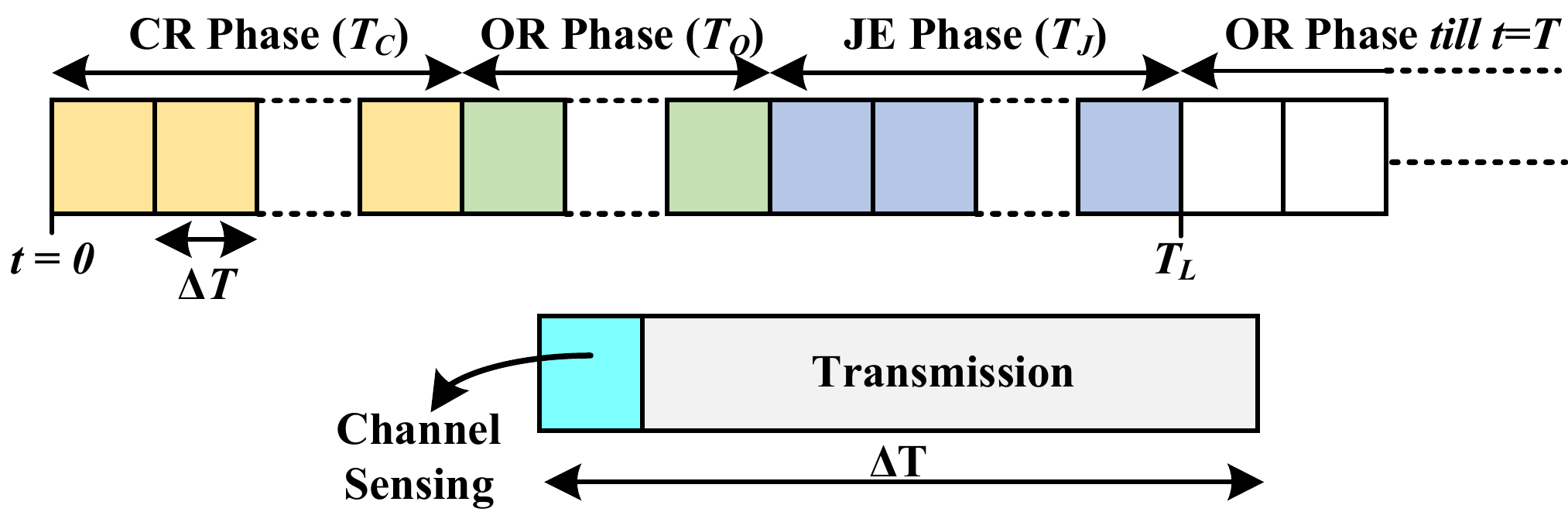}
	\caption{Phases in CNJ algorithm at different instants of horizon.}
	\label{Model2}
	
\end{figure}

The CNJ algorithm consists of four subroutines that run sequentially as shown in Fig.~\ref{Model2}. The first subroutine named CR2 runs for $T_C$ number of time slots (given in (\ref{eq:TC2})). It is similar to CR1 subroutine in CDJ except that SUs can only keep a track of the total number of collisions ($C$) but cannot identify in how many of these collisions jammers are involved.
At the end of CR2, SUs get an estimate of $\{p_i\}$. The second subroutine named OR runs for $T_O$ (specified in (\ref{eq:TO2})) time slots and end at slot $T_C+T_O$. This subroutine is the same OR subroutine in CDJ except that it ends after $T_O$ time slots by the end of which the SUs find non-overlapping channels from $K$ channels. The third subroutine named JE (Jammer Estimation) runs for $T_J$ slots (given in (\ref{eq:TJ2})) and keeps count of collisions from jammers $(C_{\pi_i})$ on each channel. The jammers could be attacking any subset of $K$ channels, hence it is important here that we count the collisions on each channel separately\footnote{If the jammers are attacking a subset of channels (selected uniformly at random), collisions occur only on these channels in JE phase. Hence the collision information we get across channels is asymmetric.} to get a correct estimate of $J$.

   \begin{algorithm}[!h]
	\caption{CNJ} \label{Algo3}
	\begin{algorithmic}
	\State	Input: {$T_C,T_O, T_J, T,K$ }
		\State	$(C/F,\boldsymbol {\pi},\{\hat{p}_i\} )= CR2(T_C, K)$ \;
		\State	$OR(\boldsymbol {\pi}, K, T_C+T_O)$
		\State $ (\hat{N},\hat{J} )= JE(\boldsymbol {\pi},T_J,C/F)$
		 \State $N^*= OP(\hat{N},\hat{J}, \{\hat{p}_i\})$
		 \State $OR(\boldsymbol {\pi},N^*,T)$
	\end{algorithmic}
\end{algorithm}

\begin{algorithm}[!t]
\caption*{\textbf{Subroutine 3:} $CR2$}
\begin{algorithmic}[1]
\State Input: $ T_C,K$
\State Set	$r_t = 0,\ F=0,\ C=0,\ O_i = 0 ,\ B_i = 0 \ \forall i \in 1..K$\;
\For{$t=1 \dots T_C$}
\State Select $I_t \sim U(1\dots K)$ and  sense \; 
\State $O_{I_t} \leftarrow O_{I_t} + 1$ \;
\If{selected channel $I_t$ busy}
\State $B_{I_t} \leftarrow B_{I_t} + 1, r_t \leftarrow 0 $
\Else
\State Transmit on $I_t$, $F \leftarrow  F + 1$ \;
\If{collision }
\State $C \leftarrow C + 1,  r_t \leftarrow 0$
\Else
\State $r_t \leftarrow 1$
\EndIf
\EndIf	
\EndFor	
\State Set	$\hat{p}_i = \frac{B_i}{O_i} \forall i \in 1\cdots K $
\State $\pi \leftarrow \ indices \ in \ [K] \ sorted \ according \ to \   \hat{p}_i$
	\end{algorithmic}
\end{algorithm}
  The initial value of $i_0$ in the Subroutine 4: JE denotes the channel on which it is operated at the end of previous OR phase. Note that only one  SU is operating on $i_0$ as all SUs orthogonalized (with high probability) at the end of the previous OR subroutine. At the end, JE subroutine gives an estimate of $J$. Using this value and the collision information from subroutine CR2, an estimate for $N$ is obtained. Now OP (\ref{eq:optm}) is used to find $N^*$ using all the estimates of $N, J$ and $\{p_i\}$ as in CDJ. The last subroutine is again OR, but this time it finds non-overlapping channels on the top $N^*$ channel and runs until the end of horizon $T$. 

\begin{algorithm}[!t]
	\caption*{\textbf{Subroutine 4:} $JE$}
	\begin{algorithmic}[1]
		\State Input: $\boldsymbol {\pi}, T_J,C/F$
		\State $F_i=0,C_i=0,O_i = 0, i=i_0  \ \forall \ i \in 1\cdots K$  \; 
		\For{$t=T_C+T_0+1, \ldots,  T_C+T_0+T_J$}
		\State $i = (i+1) \ modulo \ K$ and transmit on $ {\pi}_i$
		\State $O_{{\pi}_i} = O_{ {\pi}_i} + 1$
		\If{selected channel ${\pi_{i}}$ free}
		\State $F_{{\pi}_{i}} = F_{ {\pi}_{i}} + 1$   \;
		\If{collision from Jammer}
		\State $C_{ {\pi}_{i}} = C_{{\pi}_{i}} + 1$
		\Else
		\State $r_t \leftarrow  1$
		\EndIf
		\EndIf	
		\EndFor	
		\State $ \hat{J} = round \left ( \sum_{i=1}^{K} K\frac{O_i}{T_J}\frac{C_{ {\pi}_i}}{F_{ {\pi}_{i}}}\right)$
		\State $\hat{N} = round\left( 1+ \frac{\log \left (1-\frac{C}{F}\right)-\log \left ( 1-\frac{\hat{J}}{K} \right )} {\log \left ( 1-\frac{1}{K} \right )}\right)$
		
	\end{algorithmic}
\end{algorithm}
\subsection{Jammer Algorithm}
The jammers use Algorithm \ref{Algo2} as in the previous section and estimates the number of SUs after $T_C$ time slots. Then, it continuously attacks the top $N$ channel. We note that our method to estimate number of the jammers works as long as they are hopping randomly on any number of channels, not necessarily only top $N$ channels. So our algorithm need not know which channels the jammers are attacking. Since we consider a worst possible attack, we consider that the jammers also learn $N$ and channel ranking so that it can attack the top $N$ channels to cause maximum damage.

\subsection{\textbf{Analysis of CNJ}}
\begin{lemma} \label{thm:le2}
For any $\delta>0$, all SUs will have $\epsilon$-correct ranking of channels and   $\hat{N}=N$ given $J$ with probability at least $1-\delta/3$ if CR2 subroutine is run for $T_C$ number of time slots.
\end{lemma}
\noindent
The proof of this lemma is similar to that of Lemma \ref{thm:le1} and can be found in the supplementary.
\begin{lemma} \label{thm:le3}
For any $\delta>0$, all SUs will find non-overlapping channels with probability at least $1-\delta/3$ if OR subroutine is run for $T_O$ number of time slots.
\end{lemma}
\noindent \textbf{Proof:} We want to compute $T_O$ such that all SUs are orthogonalized on $K$ channels with high probability within $T_O$.
If $p_c$ is the collision probability of a SU at any time slot $t$, and if none of the other SUs are orthogonalized (worst-case) 
then probability that a SUs will find a an orthogonal channel within $T_O$ is given by 
   \begin{equation*} \label{eq:2} \sum\limits_{t=1}^{T_O} p_c^{t-1}\cdot (1-p_c). \end{equation*}
We want this probability to be at least $1-\frac{\delta_2}{K}$ for every SU to ensure all SUs are orthogonalized within $T_O$ with probability $1-\delta_2$.

\[\sum\limits_{t=1}^{T_O} p_c^{t-1}\cdot (1-p_c) \geq 1-\frac{\delta_2}{K} \]
Simplifying the summations, we get the following relations
\begin{align*}
 1-p_c^{T_O} &\geq 1-\frac{\delta_2}{K}
\implies	T_O \log p_c \leq \log \left(\frac{\delta_2}{K}\right) 
\end{align*}
yielding $T_O \geq \frac{\log \left(\frac{\delta_2}{K}\right)}{\log p_c}$. 
We next obtain an upper bound on $p_c$. For notational convenience define the following: 
$p_{ns}$ denote the probability of no collision due to non-settled SUs \\
$p_{s}$= probability of no collision due to settled SUs.
Probability of collision can be bounded as follows:
\begin{align*}
&p_{nc}=Pr(\mbox{no collision}) \\
	&= \sum\limits_{k=1}^{N}\frac{1}{K} (1-p_k) \left(p_{ns}+p_{s}\right)\left(1-\frac{J}{N}\right) \\ 
  &+ \sum\limits_{k=N+1}^{K}\frac{1}{K} (1-p_k) \left(p_{ns}+p_{s}\right) + \sum\limits_{k=1}^{K}\frac{p_k}{K}\\
 &\geq \sum\limits_{k=1}^{K}\frac{1}{K} (1-p_k) \left(p_{ns}+p_{s}\right)\left(1-\frac{J}{N}\right)\\
    &\geq \sum\limits_{k=1}^{K}\frac{1}{K} (1-p_k) p_{ns}\left(1-\frac{J}{N}\right)\\
    &\geq \sum\limits_{k=1}^{K}\frac{1}{K} (1-p_k) \left(1-\frac{1}{K} \right)^{N-1}\left(1-\frac{J}{N}\right) \\
    &\geq \left(1-\frac{1}{K} \right)^{N-1}\left(\frac{N-J}{N}\right) \theta\\  
  &\geq \left(1-\frac{1}{K} \right)^{K-1}\left(\frac{1}{N}\right) \theta\geq \left(1-\frac{1}{K} \right)^{K-1}\left(\frac{\theta}{K}\right) 
\end{align*}
In the second last inequality we used  the relations $\sum\limits_{K=1}^{K} \frac{1}{K}(1-p_k) \geq \theta$ and $N>J$.
Note that when we lower bound $p_{nc}$, we consider the worst case that all SUs are not settled. We get the following upper bound on $p_c$
\begin{align*}
p_c &= 1-p_{nc} \\
p_c &\leq 1-\frac{\theta}{K}\left(1-\frac{1}{K} \right)^{K-1} .
\end{align*}
Using this bound we get $T_O$, we get that  
\[T_O \geq \frac{\log \left(\frac{\delta_2}{K}\right)}{\log \left(1-\frac{\theta}{K}\left(1-\frac{1}{K} \right)^{K-1}\right)}.\]
This implies that if OR phase is run for at least $\frac{\log \left(\frac{\delta_2}{K}\right)}{\log \left(1-\frac{\theta}{K}\left(1-\frac{1}{K} \right)^{K-1}\right)}$ number of time slots, a SU will find a collision free slot and orthogonalize with probability at least $1-\delta_2$.
\begin{lemma} \label{thm:le4}
For any $\delta>0$, all SUs will estimate number of jammers correctly ($\hat{J}=J$) with probability at least $1-\delta/3$ if JE subroutine is run for $T_J$ number of time slots.\end{lemma}
The proof of this lemma is again based on careful applications of Hoeffding's inequality and Chernoff bound and appropriately bounding the collision probabilities as done in the proof of Lemma \ref{thm:le1}. Detailed proof is given in the supplementary. 

The following theorem states the expected regret of the CNJ against the jammers who employs Algorithm \ref{Algo2}. Again the expectation is over the randomness of the algorithm.
\begin{thm}\label{thm:t2}
For all 
	$\Delta > \epsilon, \gamma \in (0,0.5)$ and $0 < \delta\leq 1$, with probability at least $1-\delta$, the expected regret of SUs using CNJ in the presence of $J$ jammers after $T_L=T_C+T_O+T_J$ slots of learning is upper bounded by $NT_L + N^2exp(1)$, where
	\begin{eqnarray}
	\nonumber
	T_{C}&=&\mbox{round} \bigg (\max\bigg(\frac{8}{\theta}\ln \left(\frac{12K}{\delta}\right),\frac{1}{\epsilon^2_1 \theta}\ln \left(\frac{24K}{\delta}\right) ,\\
	\label{eq:TC2}
	&& \hspace{1cm}8K \ln \left(\frac{12K^2}{\delta}\right),
	\frac{4K}{\epsilon^2}\ln \left ( \frac{24K^2}{\delta} \right) \bigg)\bigg)\\
		\label{eq:TO2}
	T_O &=&\mbox{round}\bigg(\frac{\log \left(\frac{\delta}{3K}\right)}{\log \left(1-\frac{\theta}{K}\left(1-\frac{1}{K} \right) ^	{K-1} \right)} \bigg )\\
		\label{eq:TJ2}
	T_J &=& \hspace{-.2cm}\mbox{round} \bigg (max\left(\frac{8}{\theta }  \ln \left(\frac{6K}{\delta}\right),\frac{1}{\epsilon^2_2 \theta}\ln \left(\frac{12K}{\delta}\right)\right)\bigg)
	\end{eqnarray}
with $\epsilon_1=\frac{\gamma}{2\exp(1).K}$ and $\epsilon_2= \frac
{\gamma}{K}$.
\end{thm}
\noindent
The terms $NT_C, NT_O, NT_J$ correspond to the total regret incurred by all the users in the first three subroutines, i.e., CR2, OR and JE respectively. The last term corresponds to the OR subroutine that runs till the end of time horizon. The proof of this theorem is similar to proof of theorem \ref{thm:t1} and it is given in the supplementary.

\section{Uncoordinated and Non-distinguishable Jammers}\label{UJ}
In this section, we consider the case where jammers cannot communicate with each other and attack channels independently. This attack is more practical as the jammers can simply use multiple SU terminals and tune them to behave maliciously. However, the downside is that multiple jammers may attack the same channel in a slot, thus reducing their effectiveness. We consider indistinguishable collisions here, if they are distinguishable we can directly apply CDJ.

For the case of uncoordinated jammers where the collisions cannot be distinguished, it turns out that one can directly estimate the total number of users in the network, i.e., $N+J$ from an estimate of collision probability. To see this, note that  collision probability conditioned on the event that channels are idle is given by
 \begin{equation} {\label{eq:PC1}}
  Pr(collision) = \left(1-\left(1-1/K \right) ^ {N+J-1} \right).
  \end{equation}
 One can get an estimate for $N+J$ from an estimate of collision probability from the above relation. Though the SUs still cannot know the values of $N$ and $J$, but knowing $N+J$, they can find non-overlapping channels on the top $N+J$ channels and then estimate $J$ from the collision information. Note that in the previous case where jammers coordinated, the SUs cannot know $N+J$, so we took a pessimistic approach and orthogonalized SUs over $K$ channels. Once the SUs orthogonalize on the $N+J$ they can estimate $J$ and subsequently get an estimate of $N$. We next give  CUJ (Co-ordination in presence of uncoordinated and non-distinguishable jammers) given in Algorithm \ref{Algo4} which is a modification of CNJ algorithm for the case of uncoordinated jamming attack.

 \begin{algorithm}[!h]
	\caption{CUJ} \label{Algo4}
	\begin{algorithmic}
	\State	Input: {$T_C,T_O, T_J, T,K$}
		\State	$(\widehat{N+J},\ \{\hat{p}_i\} )= CR2(T_C, K)$;
		\State	$OR(\boldsymbol {\pi}, \widehat{N+J}, T_O)$
		\State $ (\hat{J},\hat{N}) = JE(\boldsymbol {\pi},T_J)$ 
		\State $N^*=OP1(\hat{N},\hat{J}, \{\hat{p}_i\})$
		\State $OP(\boldsymbol {\pi}, N^*,T)$ 
	\end{algorithmic}
\end{algorithm}
%

 \subsection{SU Algorithm: CUJ}
CUJ algorithm also has four subroutines as in CNJ and they run sequentially. The CR2, OR and JE subroutines are the same as in the CNJ algorithm. At the end of CR2 subroutine, using values of $C$ and $F$ each SU finds an estimate for $N+J$ as $\widehat{N+J}= round \left( 
1 + \frac {\log \left( 1-\frac{C}{F}  \right)} {\log \left(1-\frac{1}{K} \right)}
\right) $. The SUs then find non-overlapping channels on the top $\widehat{N+J}$ channels using the OR subroutine. In the JE subroutine, any collision SUs observe are necessarily from the jammers. JE subroutine is run for  $T_J$ number of time slots and collisions observed in this period is used to estimate $J$ using relation  $\hat{J} = round \left ( \sum_{i=1}^{K} K\frac{O_i}{T_J}\frac{C_{ {\pi}_i}}{F_{{\pi}_{i}}}\right)$(see Line 15 of subroutine 4:~JE). Then an estimate for $N$ is obtained from $\hat{N}=\widehat{N+J}-\hat{J}$. Using estimates of $N,J$ and $\{p_i\}$, SUs compute $N^*$=$N+m$ using the program $OP1(N,J, \{p_i\}):$
\begin{equation} \label{eq:optm2} 
			m = \argmax_w\left\{ \begin{array}{cl}
				\hspace{-.3cm} \sum\limits_{i=1}^{N+w}\frac{1-p_{ {\pi}_i}}{N+w}  \left(1-\frac{1}{N+J-1} \right)^J  if \ w \leq J-1\\
				
				 \hspace{-.3cm}\sum\limits_{i=1}^{N+J-1}\frac{1-p_{ {\pi}_i}}{N+w}\left(1-\frac{1}{N+J-1} \right)^J \hspace{-.2cm}+ \hspace{-.2cm} \sum\limits_{i=N+J}^{N+w}\frac{1-p_{ {\pi}_i}}{N+w} \\ if \ K-N \geq w>J-1. \end{array} \right . 
\end{equation}
Note that (\ref{eq:optm2}) is different from (\ref{eq:optm}). In the case of coordinated jammers, jammers know how many jammers are present ($J$), whereas it is unknown to them when they are  not coordinating. Hence in the latter case after estimating $N+J$, each jammer attacks $N+J-1$ channels. (\ref{eq:optm2}) gives $m$ for which average  throughput is maximum for a SU with jammers attacking as in this case. Finally, OR subroutine is again used to find orthogonal channel assignment from top $N^*$ channels. SUs then continue to hop sequentially according to the indices in ${\pi}$ till the end.

Each Jammer independently runs the CR2 module to get an estimate of $N$ and $J$ and attacks the top $N+J-1$ channels continuously after that.
\subsection{\textbf{Analysis of CUJ}}
The regret of the CUJ algorithm is same as the CNJ algorithm. Please refer to supplementary for proof of the Theorem 3.

\begin{thm}\label{thm:t3}
For all  $\Delta > \epsilon$ and $0 \leq\delta\leq 1$, with probability at least $1-\delta$, the expected regret of $N$ SUs using the CUJ algorithm in the presence of $J$ Jammers after $T_L=T_C+T_O+T_J$ time slots of learning is upper bounded by $NT_L + N^2exp(1)$, where the values of $T_C, T_O,$ and $T_J$ are given in  (\ref{eq:TC2}), (\ref{eq:TO2}), and (\ref{eq:TJ2}), respectively with $\epsilon_1=\epsilon_2=\frac{\gamma}{\exp(1).K}$.
\end{thm}

\noindent
{\bf Discussion:}We only considered the static case in this paper, i.e., the number of SUs and the jammers remains fixed throughout. However, in practice the network may be dynamic, where the number of SUs may changes as SUs can enter and leave the network at any time. Our work can be extended for the dynamic case as done in the Dynamic Musical Chairs (DMC) in \cite{MC}. Specifically, if all the SUs share a global clock then they can restart the algorithms for their static counterparts periodically in run it in epochs. One can show that the regret with this approach results in regret of at most $\mathcal{O}(\sqrt{T})$ with high probability by carefully tuning the length of epochs. Due to lack to space we skip these details.\\


\section{Simulations and Experimental Results} 
In this section, the functionality of the proposed algorithms is validated using synthetic experiments in MATLAB and USRP based experiments in real radio environment. Please refer to supplementary for additional results.

\subsection{Synthetic Experiment} \label{simulation}
For synthetic experiments, we consider three algorithms, CDJ, CNJ and CUJ, separately. For each algorithm, the comparison with the respective myopic algorithm, $\rho^{rand}$ \cite{prand} and MC \cite{MC} algorithm in terms of regret and throughput is shown. Then, the effect of $N$, $K$ and $J$ on the regret and throughput of the proposed algorithms is analyzed. The channel busy probability for each $K$ is given by $p_{\lceil \frac{K}{2}\rceil} = 0.5$ and for $i > {\frac{K}{2}}$ and $i < {\frac{K}{2}}$, the gap between the channel busy probability of the $i^{th}$ and $(i+1)^{th}$ channel should be at least 0.06. For instance, for $K=16$, we set $p_i~=~\{0.08,0.14,0.20,0.26,0.32,0.38,0.44,0.5,0.56,0.62,\\ 0.68,0.74,0.80,0.86,0.92,0.98\}$. For these $p_i$, the values of $T_J$ and $T_O$ are fixed and equal to 1000 and 50, respectively. The value of the $T_C$ for CDJ algorithm is 13000 for $K=16$ and is incremented by 2000 for each increase in $K$. The value of $T_C$ for CNJ and CUJ algorithms are 10000 and the values for $T_C+T_O+T_J$ and $T_C+T_O$ is 13000 for CNJ and CUJ algorithm, respectively. The length of learning stage in MC \cite{MC} is 3000. Each numerical result shown here is the average of the values obtained over 50 independent experiments and each experiment considers a time horizon of 30000 slots. Note that the solid line in all the figures correspond to the average regret while the dashed lines correspond to the average throughput.

\begin{figure*}[!b]
	\centering
	\subfloat[]{\includegraphics[scale=0.35]{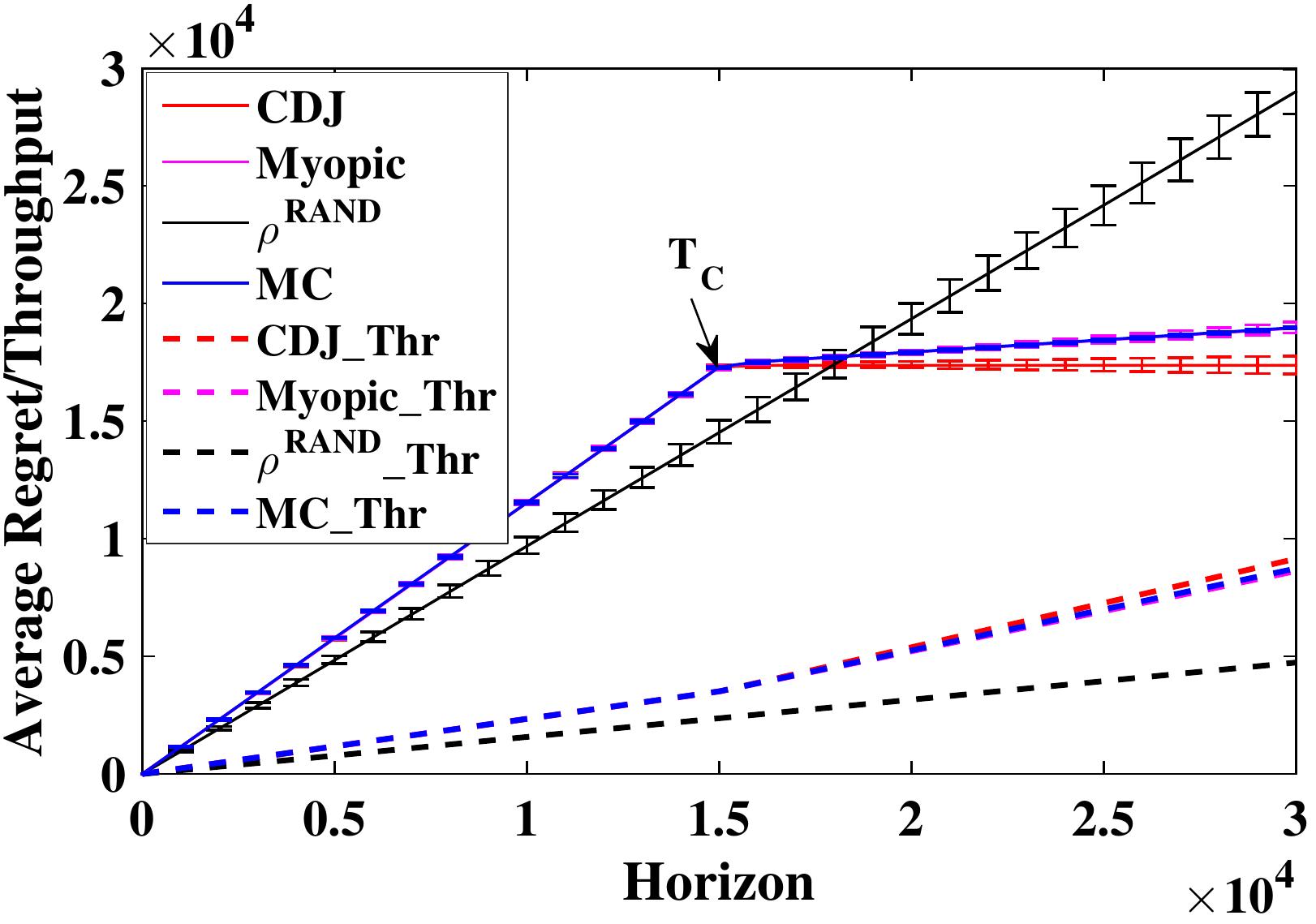}%
		\label{fixnkj1}}
	\hspace{0.1cm}
	\subfloat[]{\includegraphics[scale=0.35]{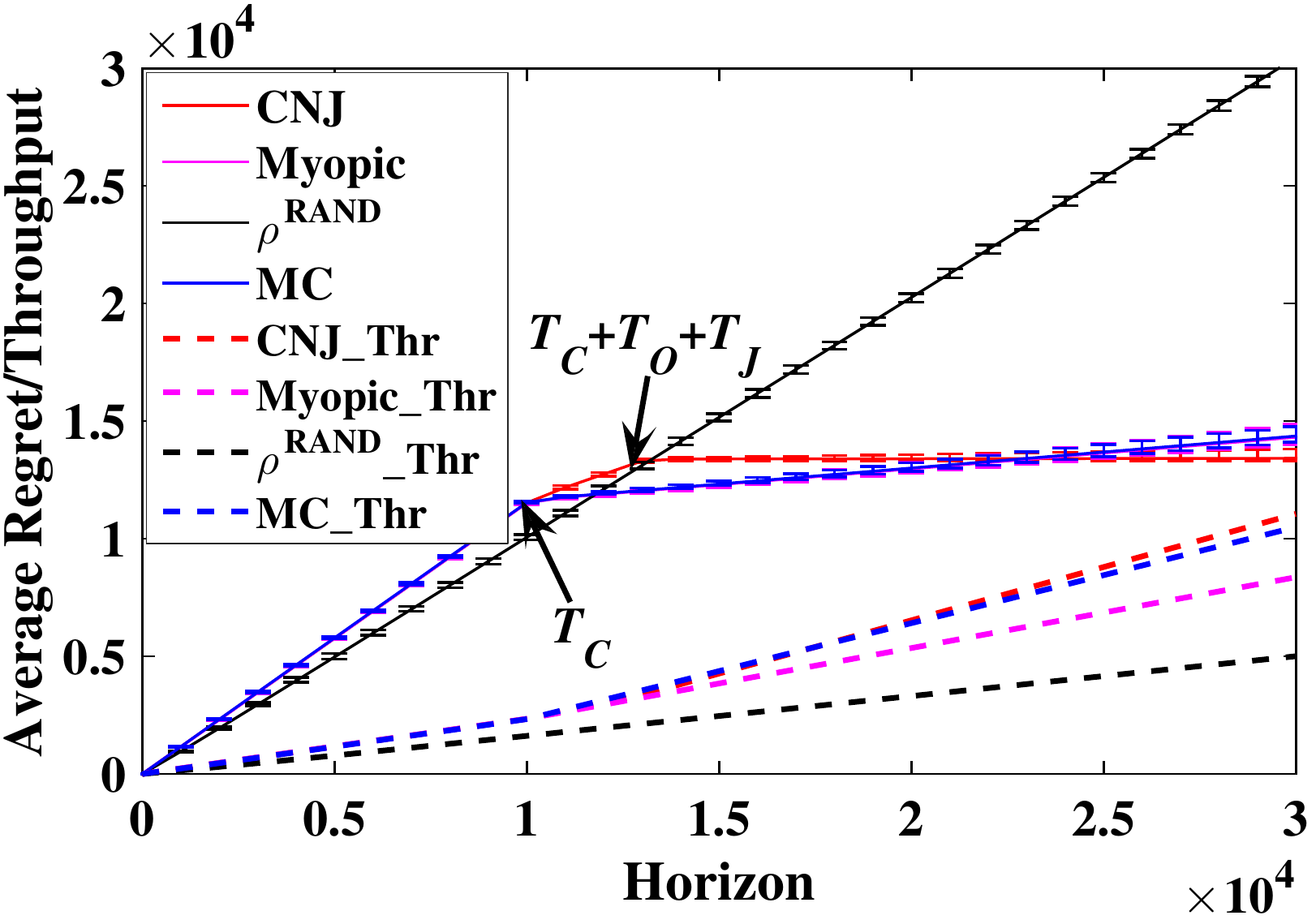}%
		\label{fixnkj2}}
	\hspace{0.1cm}
	\subfloat[]{\includegraphics[scale=0.35]{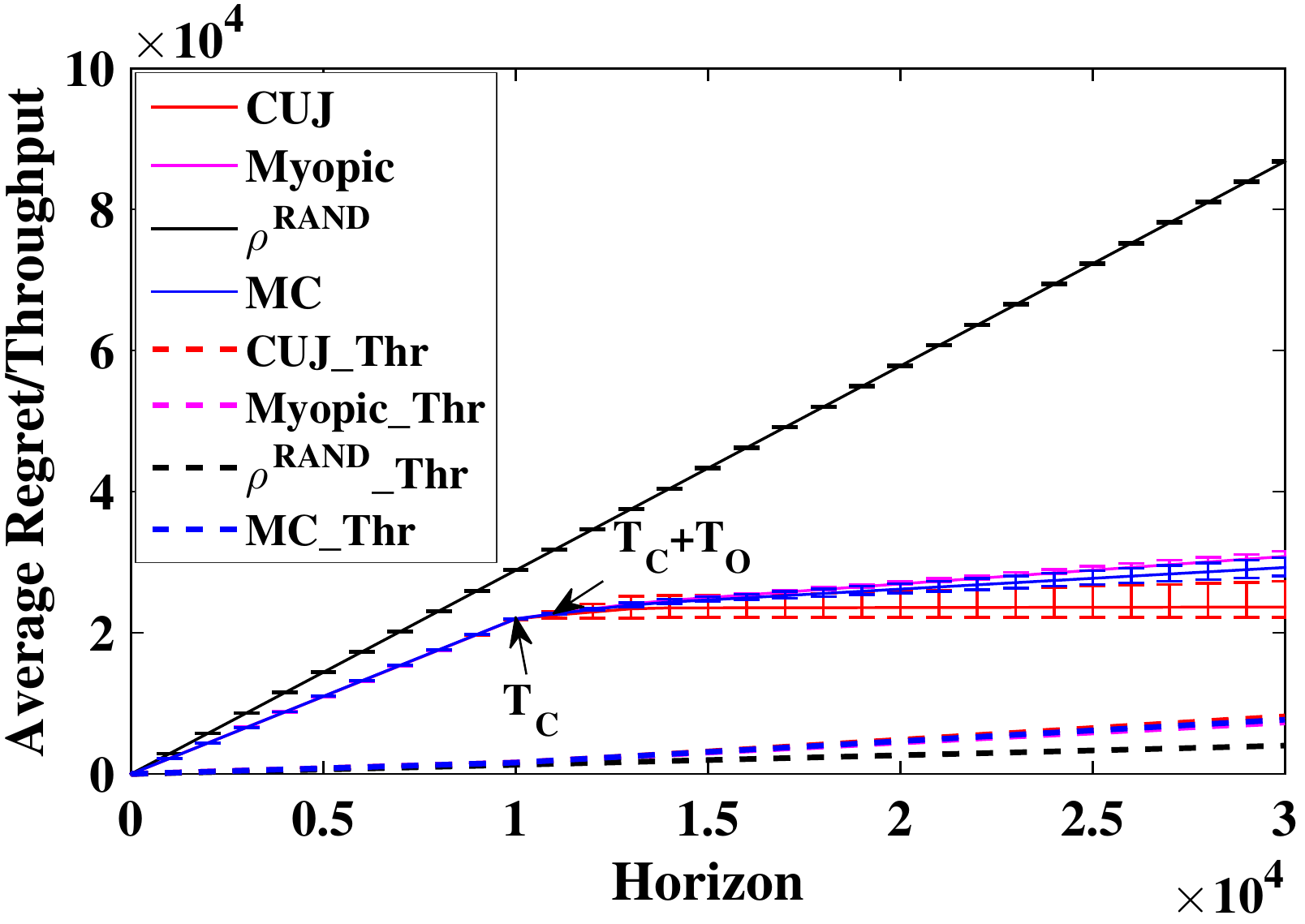}%
		\label{fixnkj3}}	
	\caption {The comparison of average regret and average throughput of the Myopic, $\rho^{rand}$ and MC algorithms with (a) CDJ algorithm, (b) CNJ algorithm and (c) CUJ algorithm at different instants of the horizon. Here, we fix  $N=8$, $K=16$, $J=4$.}
	\label{fixnkj}
	
\end{figure*}

\subsubsection{Average Regret \& Throughput for fixed $N$, $K$ and $J$}
We compare the average regret and throughput of the proposed algorithms with that of the myopic, $\rho^{rand}$ \cite{prand} and MC \cite{MC} algorithm. The corresponding plots for CDJ algorithm, CNJ algorithm and CUJ algorithm are shown in Fig.~\ref{fixnkj} (a), (b) and (c), respectively. In Fig.~\ref{fixnkj1}, we consider the CDJ algorithm designed for the scenario where the SUs can distinguish the collisions with other SUs from the collisions with jammers. In case of myopic algorithm, the SUs select the channel via random hopping till $T_C$ followed by orthogonalization and sequential hopping among top $N$ channels till the end of the horizon. In the $\rho^{rand}$ \cite{prand} algorithm which assumes that all SUs have prior knowledge of $N$, each SU is randomly assigned a rank, $R \in {1,..., N+J}$. In the subsequent time slots, SU with rank $R$ selects the channel with the $R^{th}$ best quality index based on the characterization by the UCB algorithm. When SUs collide, the rank is randomly and independently recalculated at the colliding SUs. The MC algorithm \cite{MC} divides the time horizon into two stages: 1) Learning stage, 2) MC stage. In the learning stage,
the SUs select the channel via random hopping till $T_C$ followed by orthogonalization on one of the top $N$ channels in the MC stage and then get locked on it till the end of the horizon. Jammer algorithm is same in all algorithms except $\rho^{rand}$ where jammers have prior knowledge of $N$ and hence, they attack best $J$ channels as per $\rho^{rand}$ algorithm.

As shown in from Fig. \ref{fixnkj1}, the average regret incurred by CDJ, myopic and MC algorithms is identical till $T_C$ due to identical channel selection approach whereas regret of the $\rho^{rand}$ algorithm is lower in this duration due to learning based channel selection. After $T_C$ time slots, the SUs in the CDJ algorithm sequentially hops among top $N^*$ channels leading to fewer number of collisions with jammers and hence, lower average regret, than the myopic algorithm, $\rho^{rand}$ and MC algorithm where SUs have considerable number of collisions with jammer and thus, higher regret.

Next, we consider the CNJ algorithm designed for the scenario where the SUs can not distinguish between the collisions with other SUs and the collisions with jammers. Also, it considers coordinating jammers. The Fig. \ref{fixnkj2} compares the regret of the CNJ algorithm with myopic, $\rho^{rand}$ and MC algorithm.
It can be observed that these algorithms incur the same regret till $T_C$ due to random hopping. After $T_C$, the SUs in myopic algorithm orthogonalize on the top $N+J$ channels followed by sequential hopping. Whereas in the MC algorithm, the SUs orthogonalize on the top $N+J$ channels and get locked on it. In the CNJ algorithm, the SUs orthogonalizes in the $K$ channels followed by sequential hopping till $T_C+T_O+T_J$ time slots. Thus, the myopic and MC algorithm incurs lower regret than the CNJ algorithm during this phase. After $T_C+T_O+T_J$ time slots, the SUs in CNJ algorithm sequentially hops in the top $N^*$ channels and hence, does not incur any regret leading to the constant average regret. Its regret is lower since $N^* \leq N+J$ and it guarantees the same value of $N^*$ at each SUs. This is not true for myopic and MC algorithm leading to non-zero regret even after $T_C+T_O+T_J$ slots. Whereas $\rho^{rand}$ algorithm performs similar to Fig. \ref{fixnkj1}.

At the end, we consider the CUJ algorithm designed for the scenario where the SUs can not distinguish collisions with other SUs from the collisions with jammers and non-coordinating jammers. As shown in Fig.~\ref{fixnkj3}, the regret plots are similar to the plots in Fig.~\ref{fixnkj2} except $T_J$ phase where these algorithms have identical regret. This is because the SUs in the CUJ algorithm sequentially hops in the top $N+J$ channels compared to $K$ channels in the CNJ algorithm leading to lower regret in former. 

For all plots, the upper and lower bounds on the regret are tight and do not overlap with that of the other algorithms. This validates the superiority of the proposed algorithms.


\begin{figure*}[!t]
	\centering
	\subfloat[]{\includegraphics[scale=0.34]{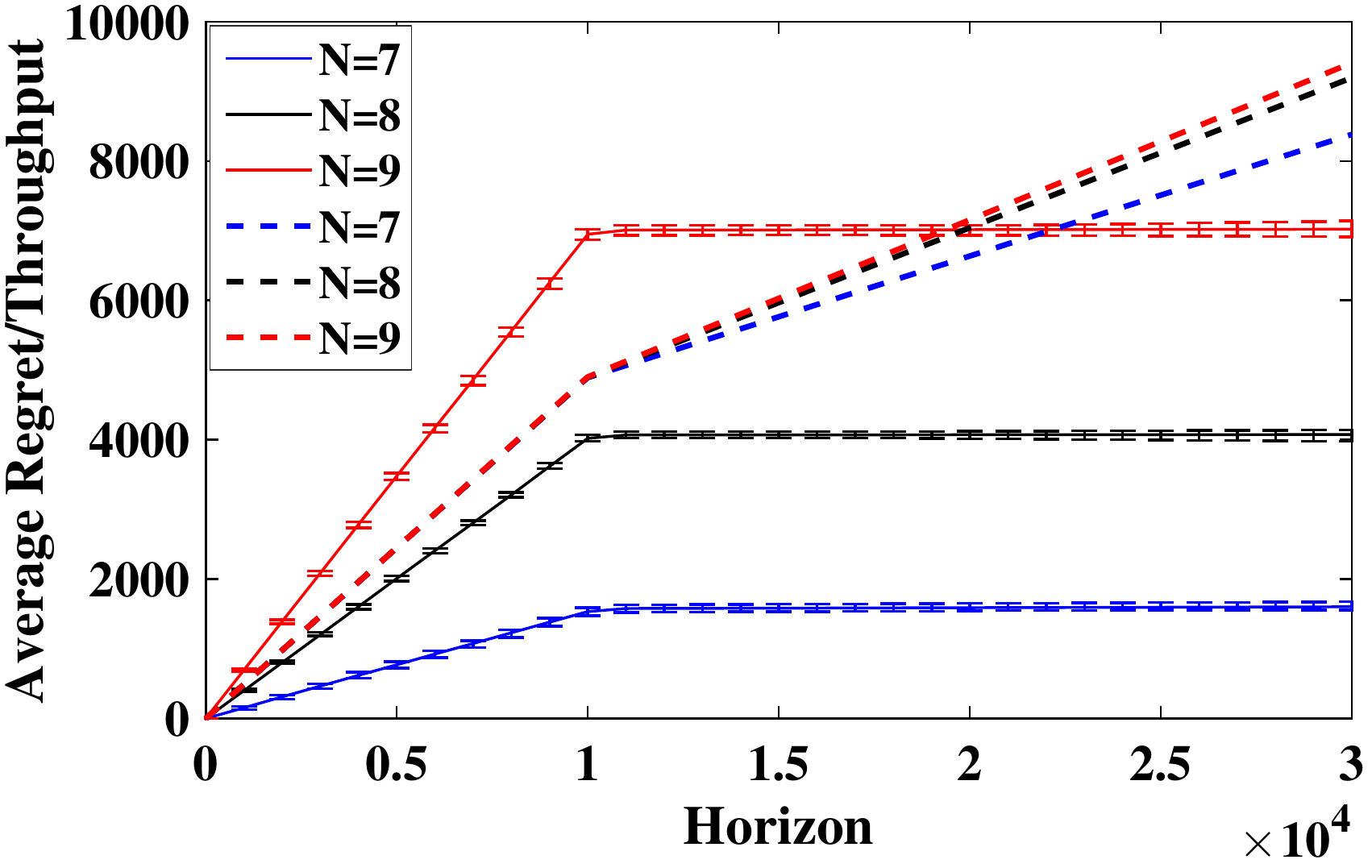}%
		\label{varN1}}
		\hspace{0.1cm}
	\subfloat[]{\includegraphics[scale=0.34]{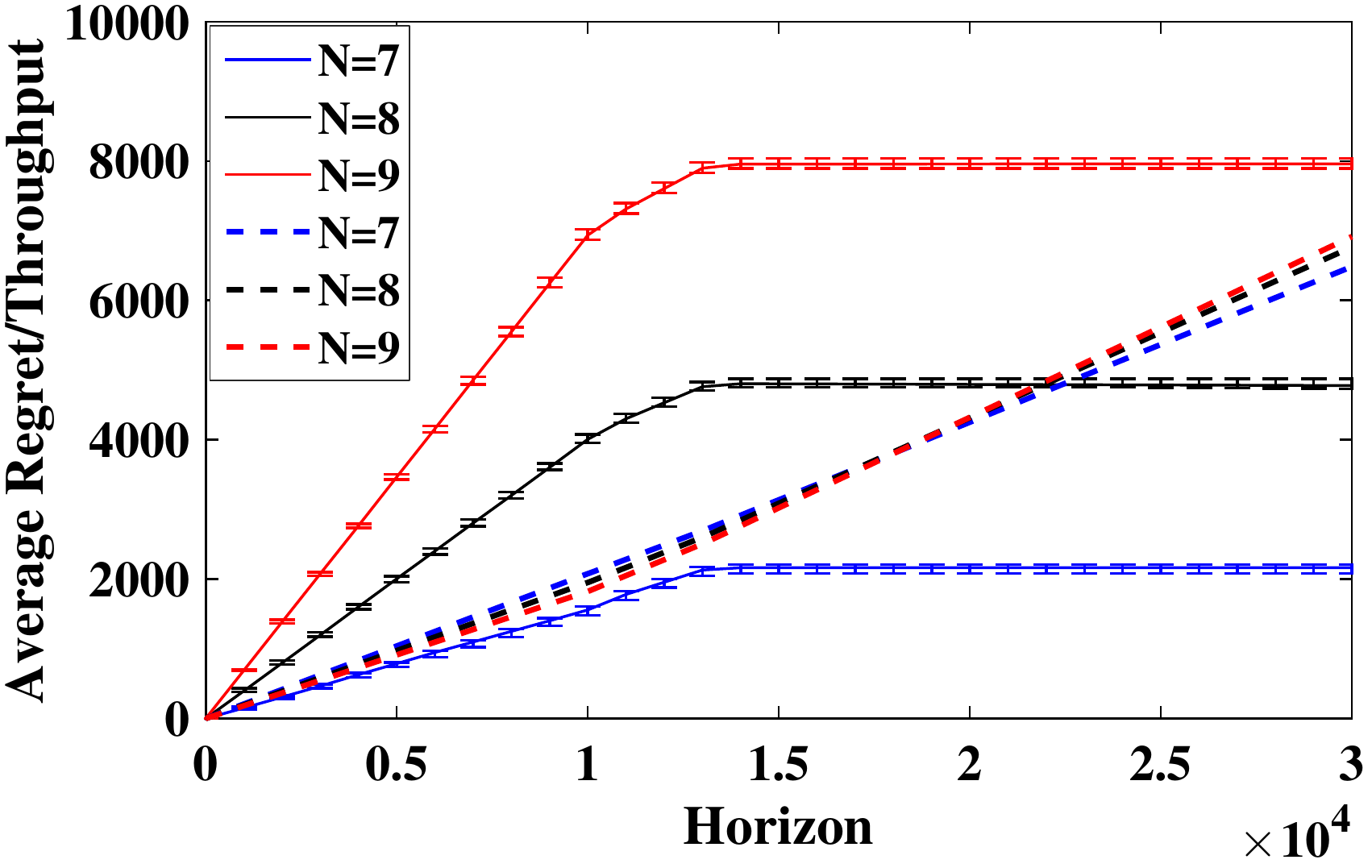}%
		\label{varN2}}
	\hspace{0.1cm}
	\subfloat[]{\includegraphics[scale=0.34]{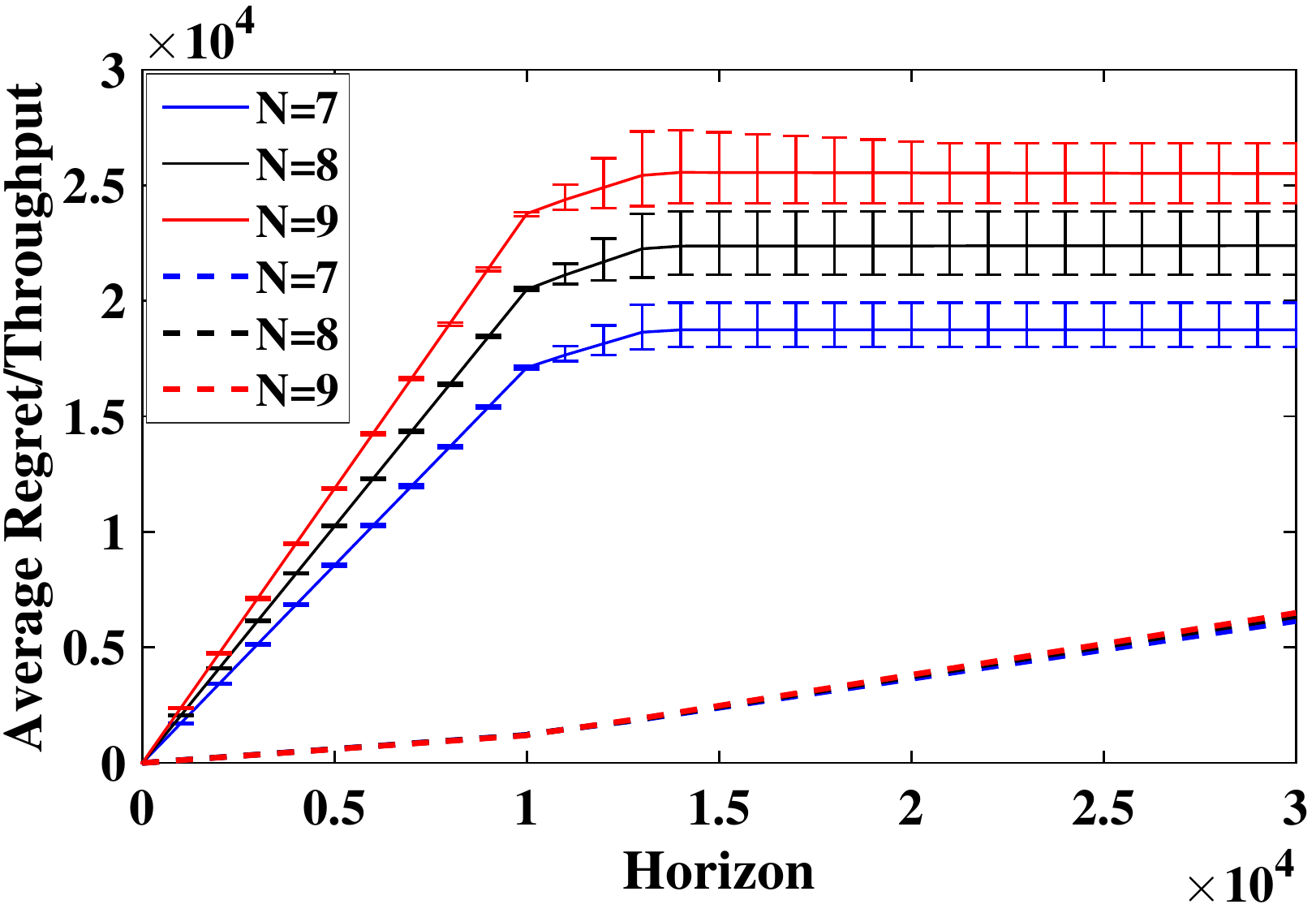}%
		\label{varN3}}	
	\caption {The plots showing the effect of $N=\{7,8,9\}$ on the average regret and average throughput of (a) CDJ algorithm, (b) CNJ algorithm, and (c) CUJ algorithm, at different instants of the horizon. Here, we fix $K=16$ and $J=6$.}
	\label{varN}
\end{figure*}
 
 \begin{figure*}[!t]
 	\centering
 	\subfloat[]{\includegraphics[scale=0.34]{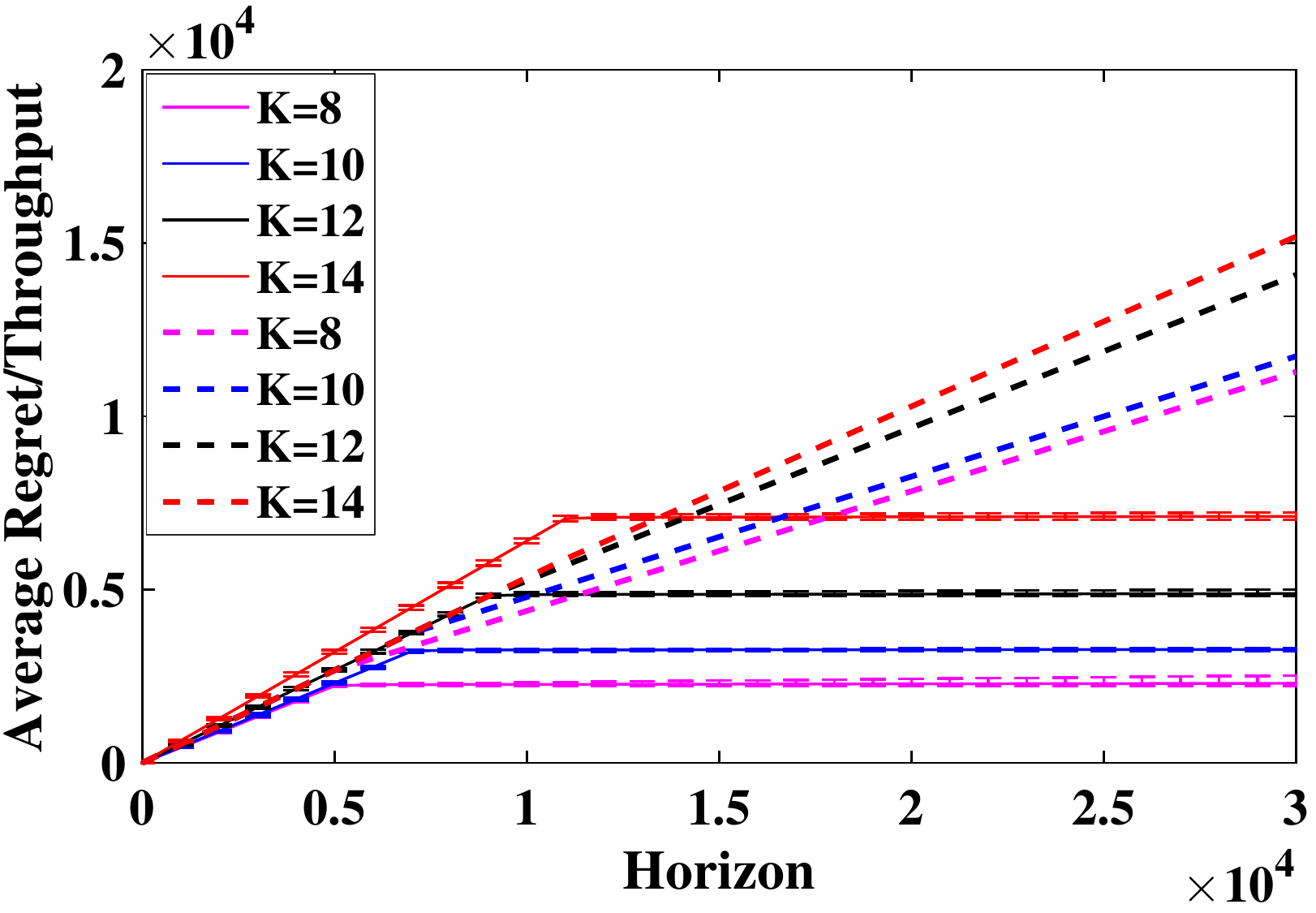}%
 		\label{varK1}}
 	\hspace{0.08cm}
 	\subfloat[]{\includegraphics[scale=0.34]{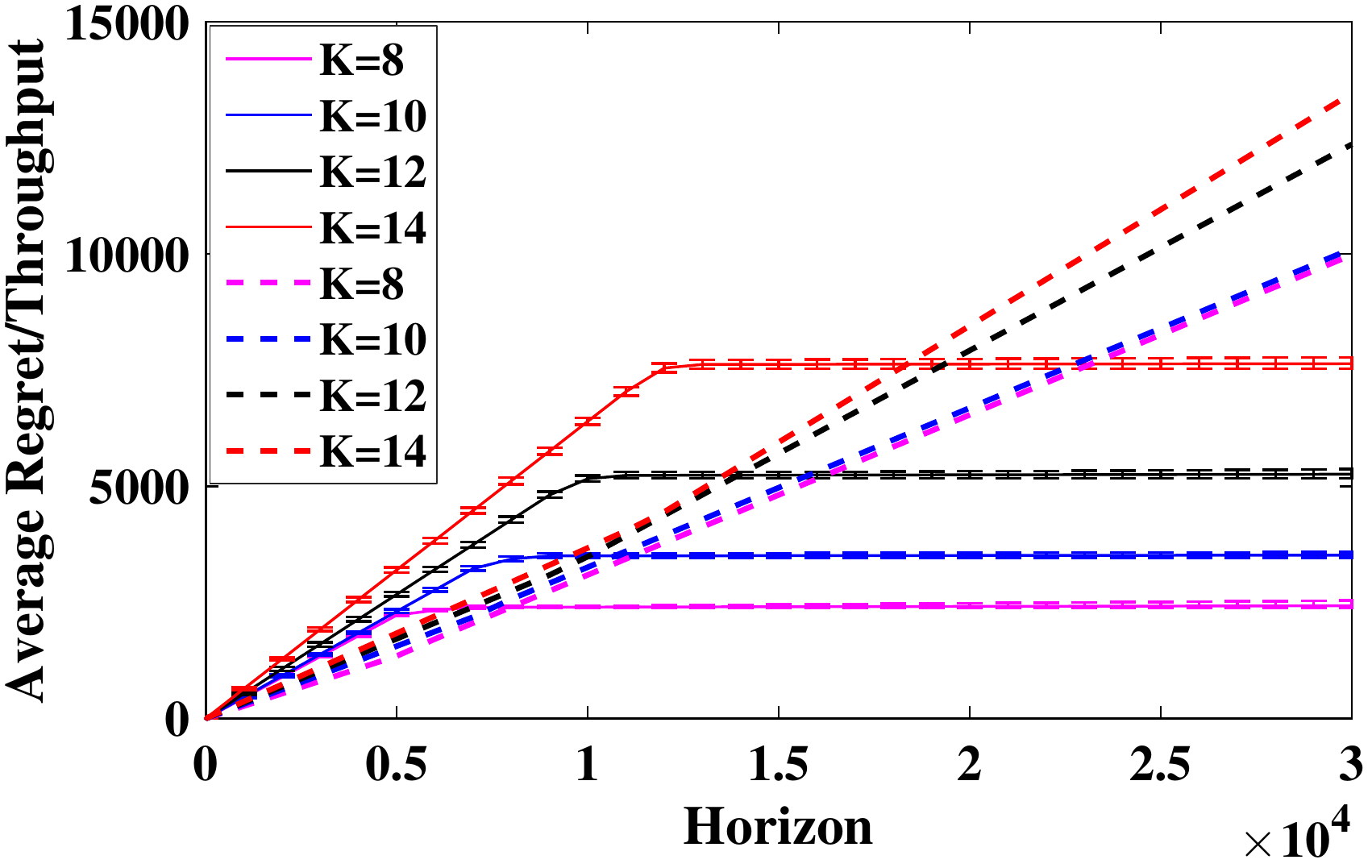}%
 		\label{varK2}}
 	\hspace{0.08cm}
 \subfloat[]{\includegraphics[scale=0.34]{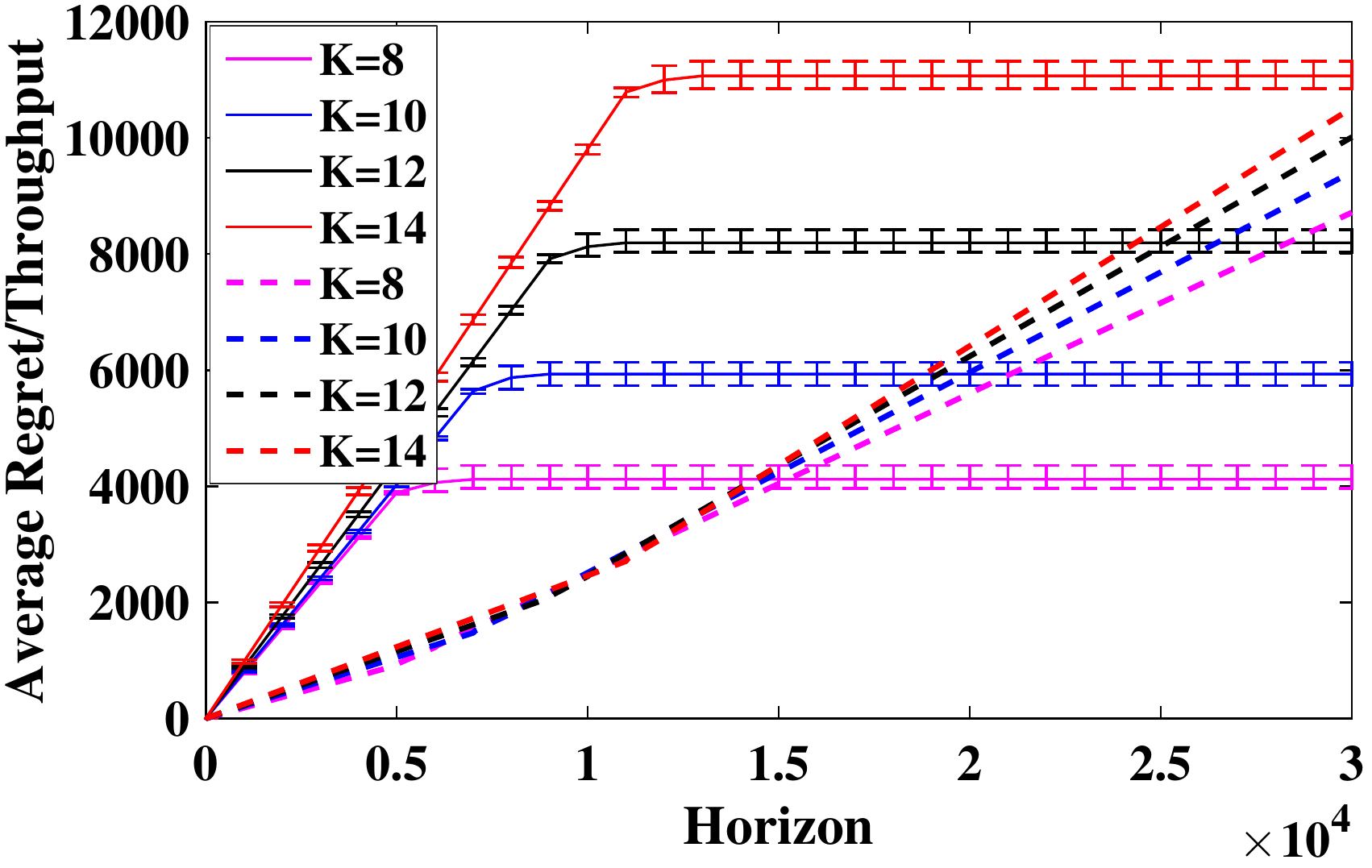}%
 		\label{varK3}}	
 		\caption {The plots showing the effect of $K=\{8,10,12,14\}$ on the average regret and average throughput of (a) CDJ algorithm, (b) CNJ algorithm, and (c) CUJ algorithm, at different instants of the horizon. Here, we fix $N=4$ and $J=2$.}
 	\label{varK}

 \end{figure*}
 
 \begin{figure*}[!t]
 	\centering
 	\subfloat[]{\includegraphics[scale=0.34]{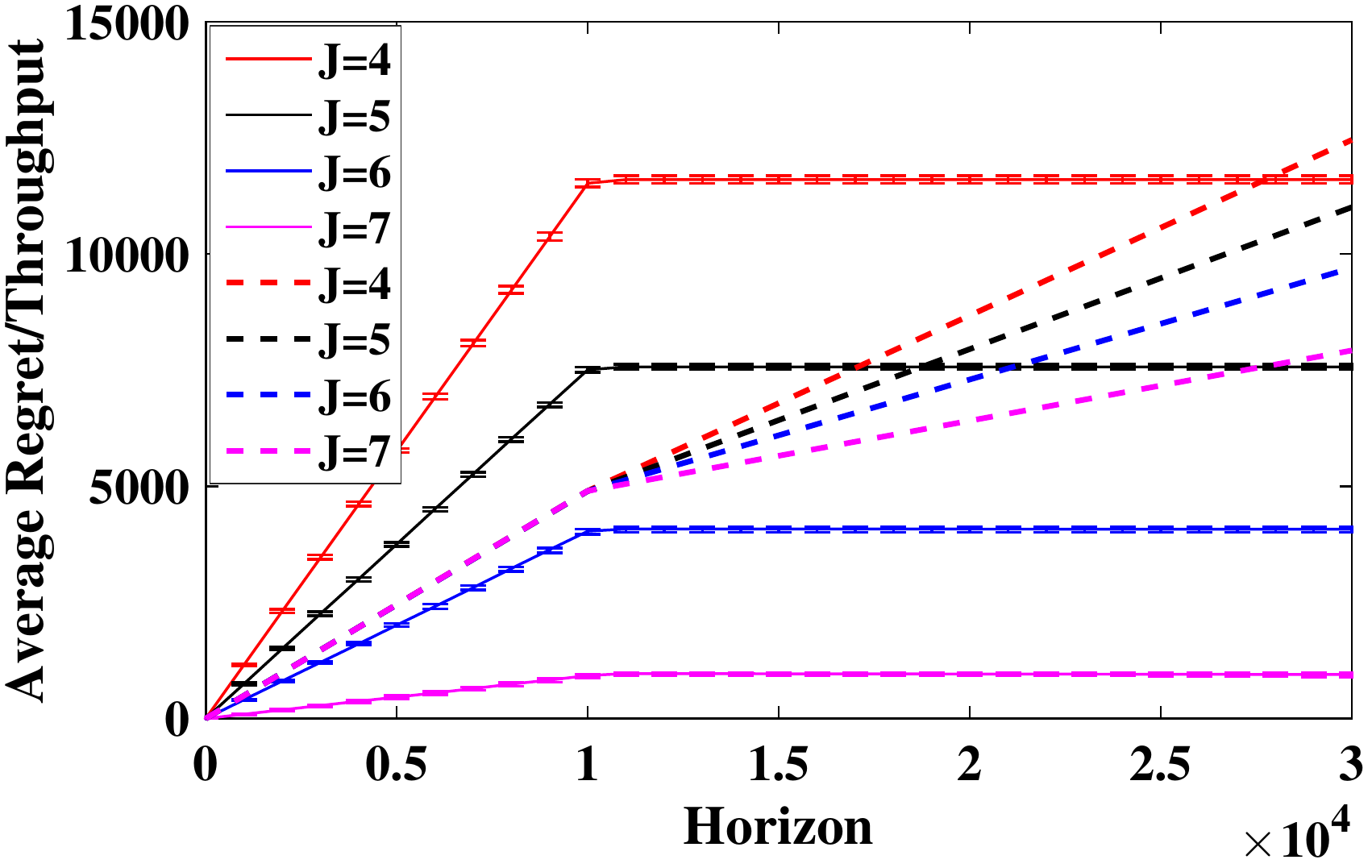}%
 		\label{varJ1}}
 	\hspace{0.1cm}
 	\subfloat[]{\includegraphics[scale=0.34]{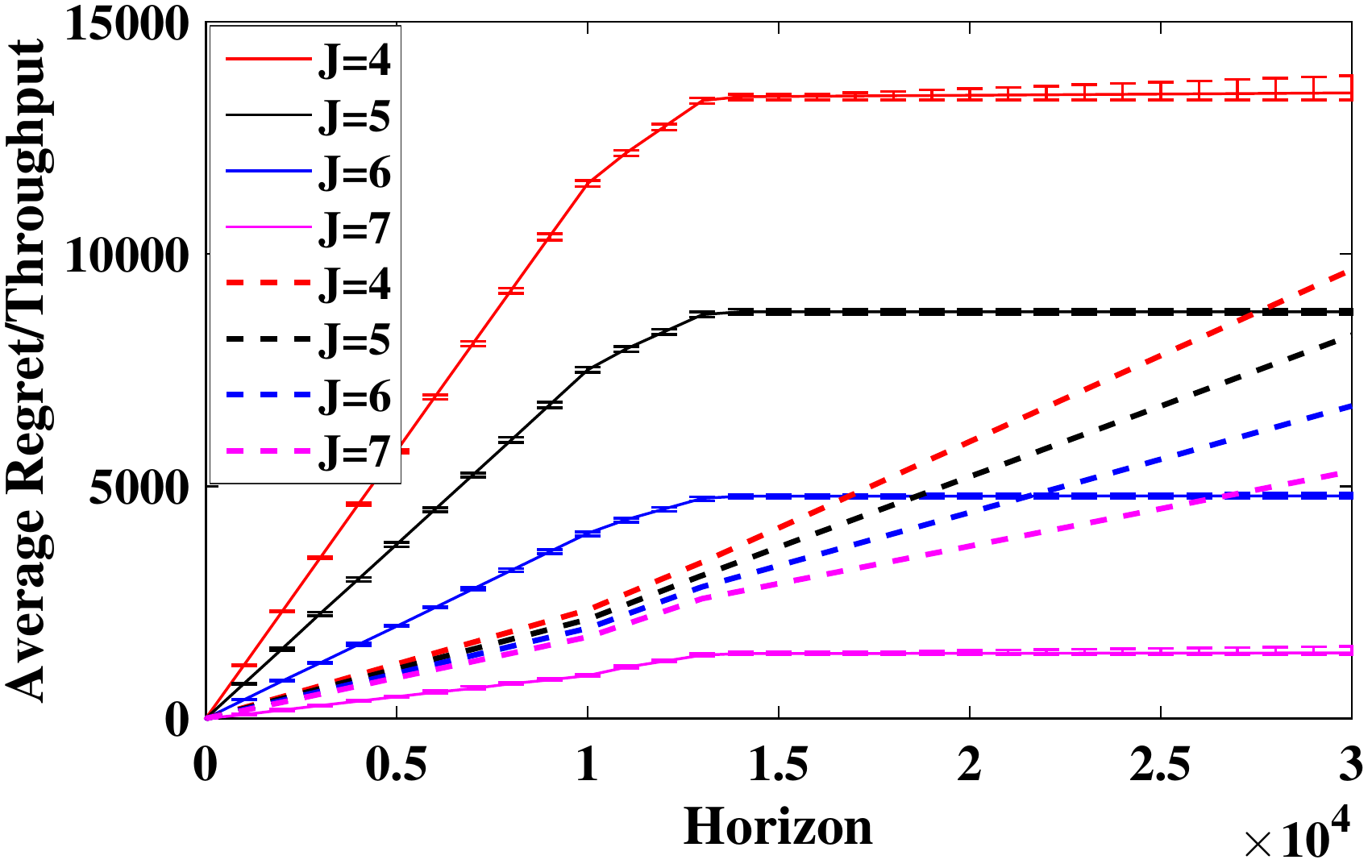}%
 		\label{varJ2}}
 	\hspace{0.1cm}
 \subfloat[]{\includegraphics[scale=0.34]{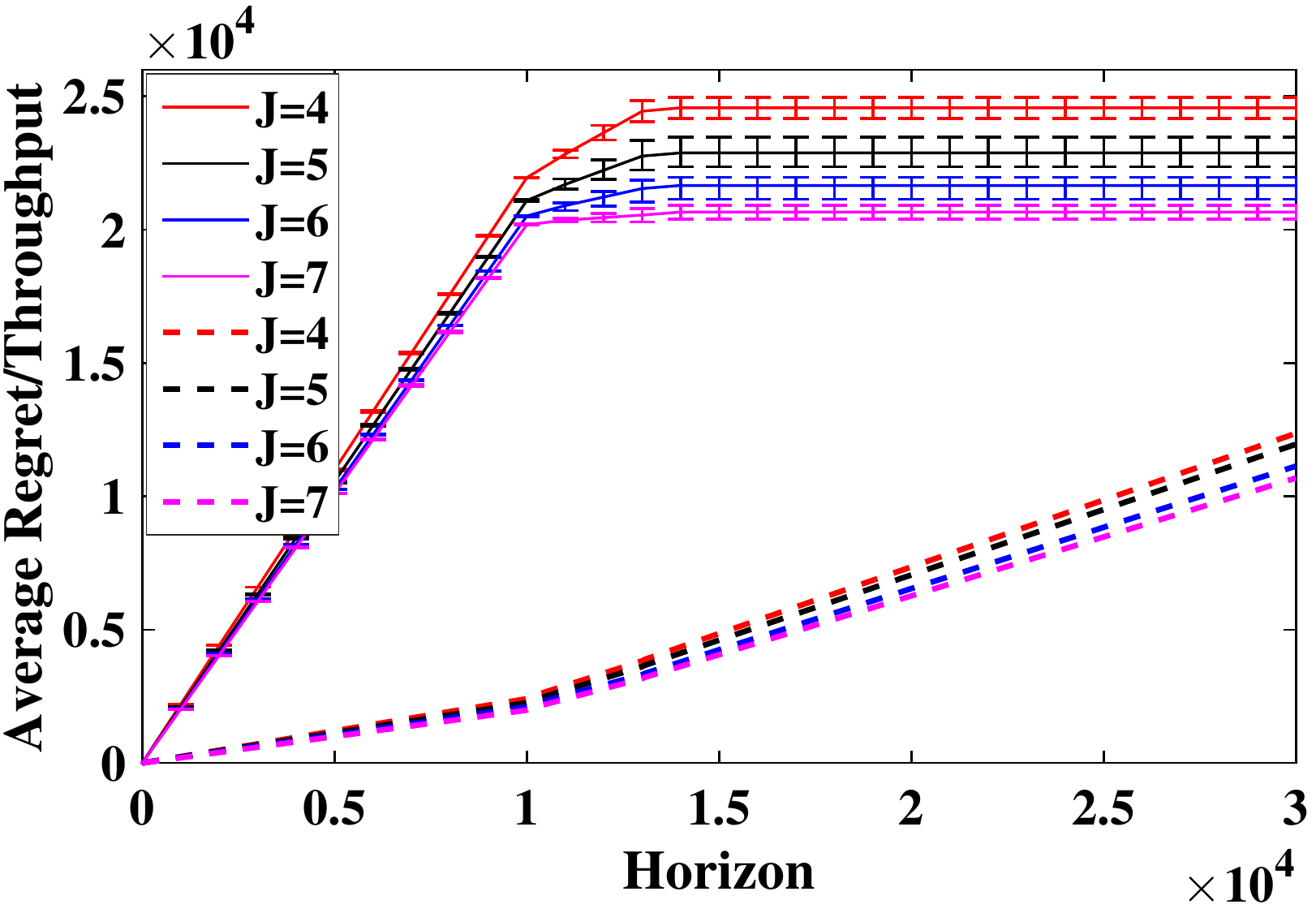}%
 		\label{varJ3}}	
 	\caption {The plots showing the effect of $J=\{4,5,6,7\}$ on the average regret and average throughput of (a) CDJ algorithm, (b) CNJ algorithm, and (c) CUJ algorithm, at different instants of the horizon. Here, we fix $N=8$ and $K=16$.}
 	\label{varJ}

 \end{figure*}


\subsubsection{Average Regret for variable $N$ with fixed $K$ and $J$}
In Fig. \ref{varN}, we compare the average regret and the throughput of the proposed algorithms for $N=\{7,8,9\}$ with $K=16$ and $J=6$. Since the probability of collision in random hopping increases with the increase in $N$, the average regret also increases with increase in $N$. After orthogonalization in top $N^*$ channels, proposed algorithms do not incur any regret leading to the constant average regret. As expected, the average regret of the CUJ algorithm is highest followed by CNJ algorithm with the CDJ algorithm having the lowest regret. This is because expected throughput in case of non-coordinating jammers increases due to collisions among jammers but expected throughput of the CUJ algorithm is same as that of CDJ and CNJ algorithms due to identical random hopping approach. The CDJ algorithm has lowest regret due to the capability to distinguish between the collisions from other SUs from the collisions with jammers. With fixed $J$, jammers can attack only $J$ out of the top $N$ channels. Therefore with increase in value of $N$, the remaining $N+m-J$ channels are always available for transmission leading to increase in the system throughput. Thus, the average throughput increases with increase in $N$.

\subsubsection{Average Regret for variable $K$ with fixed $N$ and $J$}
In Fig. \ref{varK}, we compare the average regret and the throughput of the proposed algorithms for $K=\{8,10,12,14\}$ with $N=4$ and $J=2$. Higher the value of $K$, higher is the orthogonalization time and hence, higher is the average regret. Thus, the average regret increases till orthogonalization in top $N^*$ channels after which there is no further increase in regret. The average throughput also increases with the increase in value of $K$ due to high transmission opportunities on the optimal channels. 

\subsubsection{Average Regret for variable $J$ with fixed $N$ and $K$}
In Fig. \ref{varJ}, we compare the average regret and the throughput of the proposed algorithms for $J=\{4,5,6,7\}$ with $N=8$ and $K=16$. The average regret decreases with the increase in the value of $J$. This is because, the number of sub-optimal channels decreases as $J$ increases. Fewer the number of sub-optimal channels, lower is the regret. The average throughput also decreases with the increase in value of $J$ due to lesser opportunities of transmission with increase in $J$. In case of CUJ algorithm, the expected throughput also increases as the number of jammers increases due to non-coordinating jammers. Hence, the decrease in the average regret and throughput as the value of $J$ increases is not substantial. Next, experimental results on USRP testbed are discussed.

 \par

\subsection{USRP Experiment}

The USRP based testbed has been developed to validate the functionality of the proposed work in real radio environment and compare it with the myopic algorithm. Due to limited page constraints, we restrict the discussion to the CNJ algorithm and corresponding myopic algorithm since they consider the most challenging scenario of coordinating jammers with the SUs not capable of distinguishing between the collisions from other SUs and collisions from jammers.

	\begin{figure}[!h]

	\centering
	\includegraphics[scale=0.07]{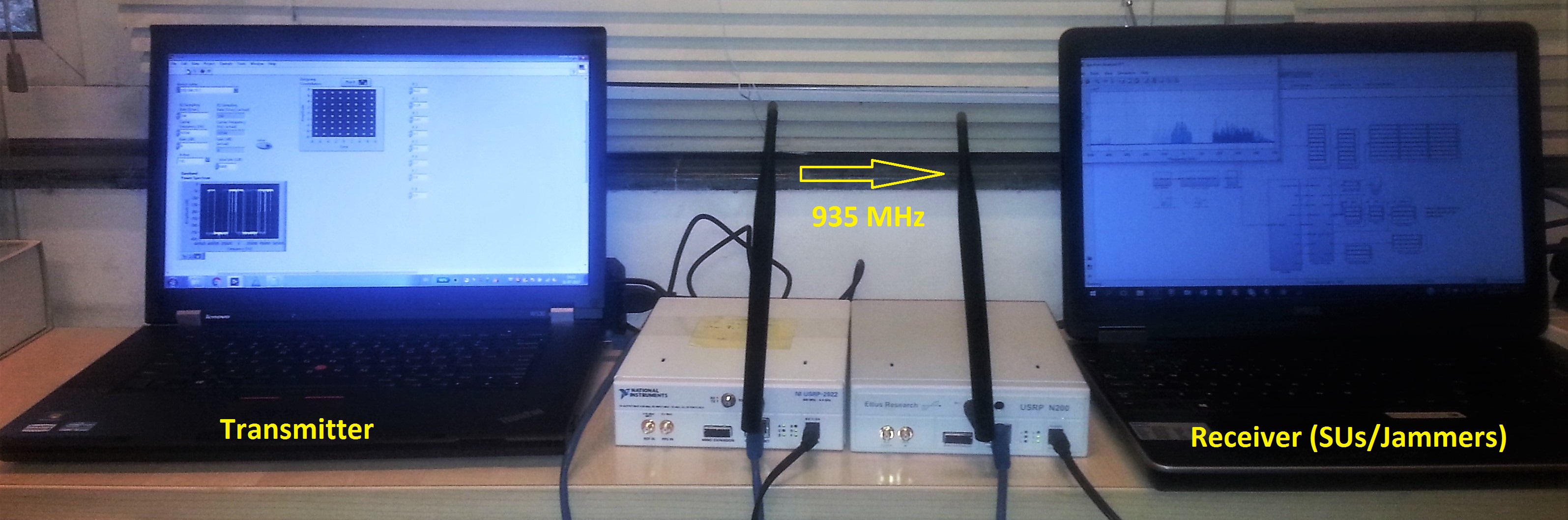}
	\caption{Proposed USRP based experimental testbed.}
	\label{usrp1}

\end{figure}

\begin{figure*}[!h]
	\centering
	\subfloat[]{\includegraphics[scale=0.275]{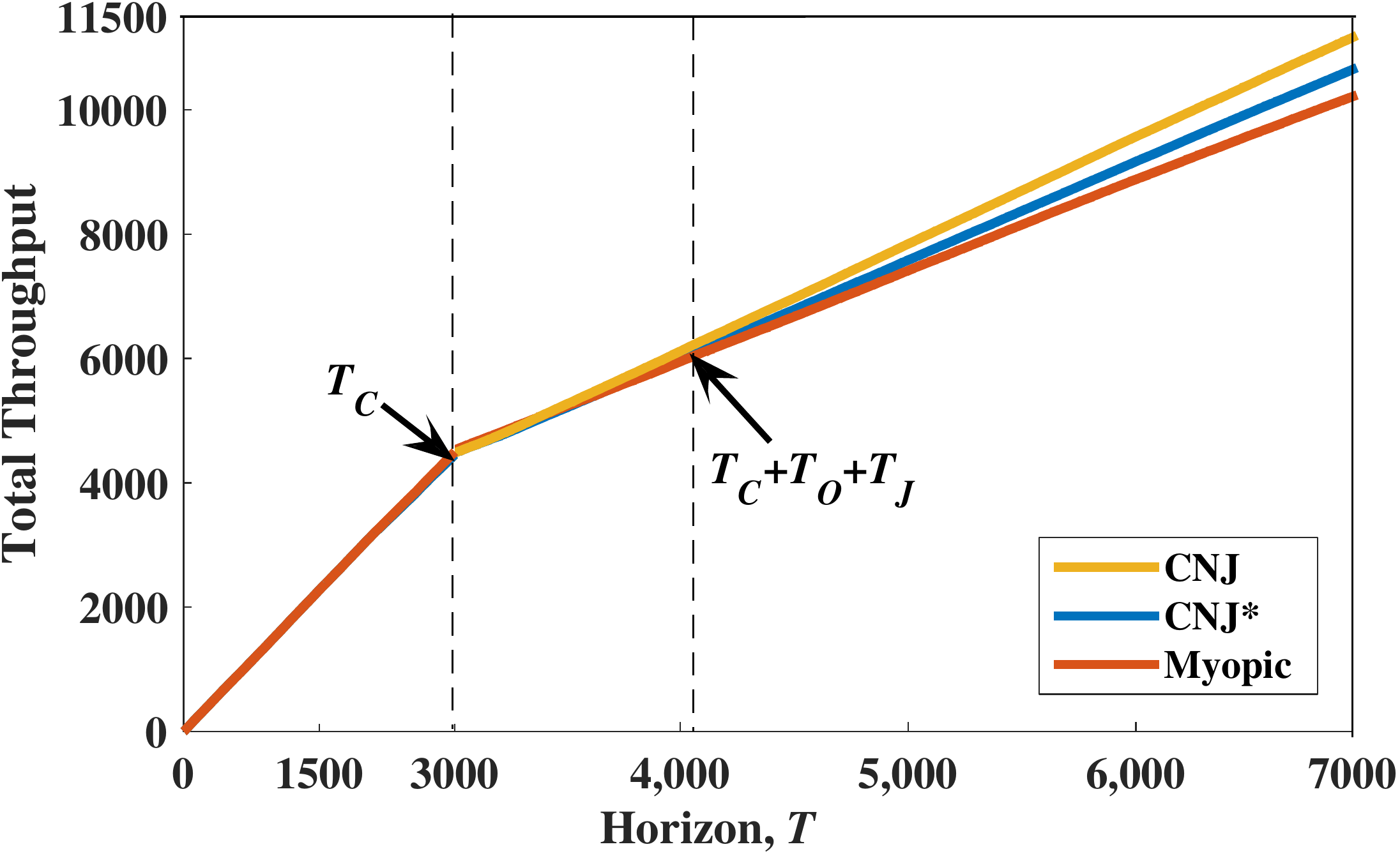}%
		\label{d1}}
	\hspace{0.5mm}
	\subfloat[]{\includegraphics[scale=0.275]{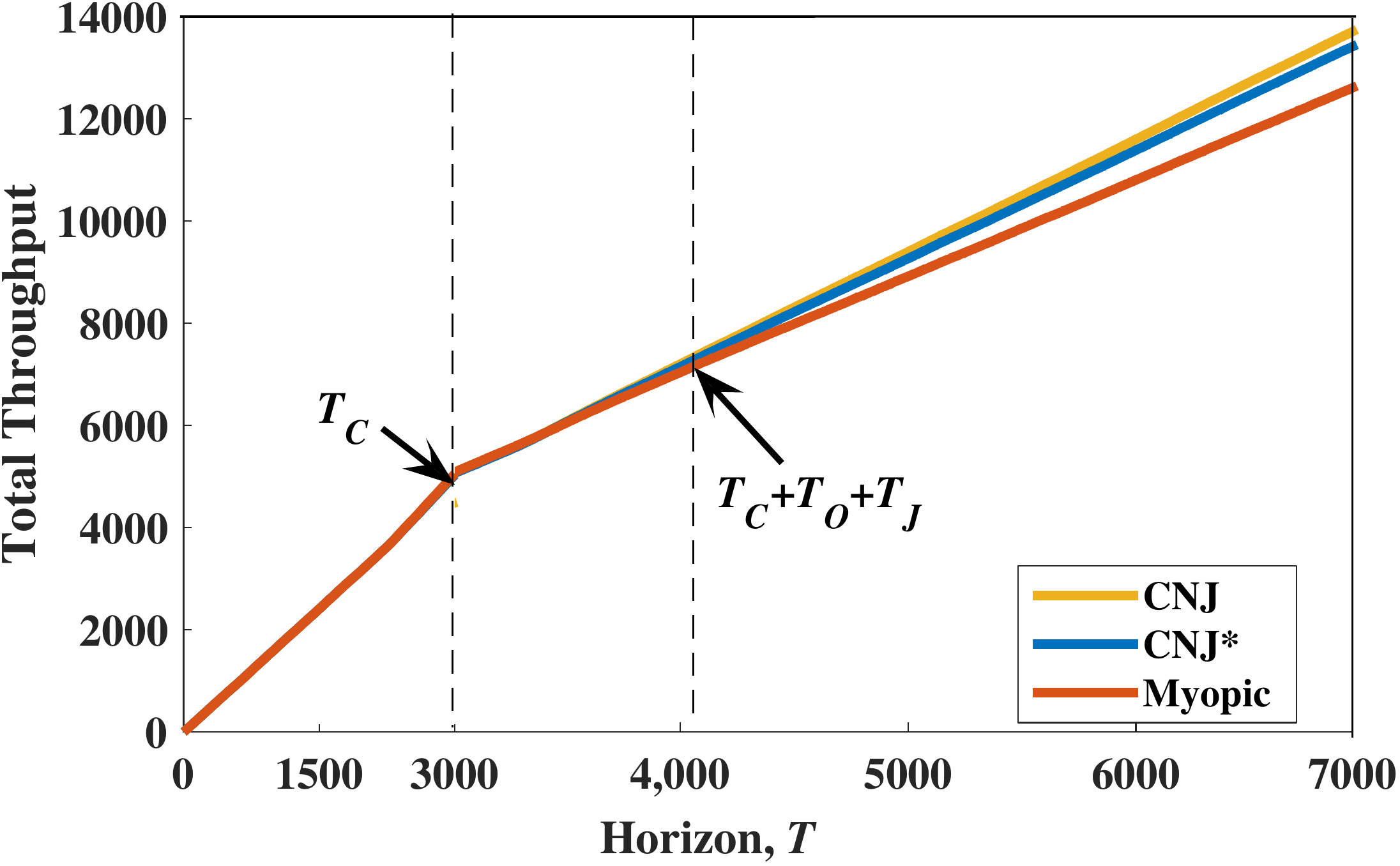}%
		\label{d2}}
	\hspace{0.5 mm}
	\subfloat[]{\includegraphics[scale=0.275]{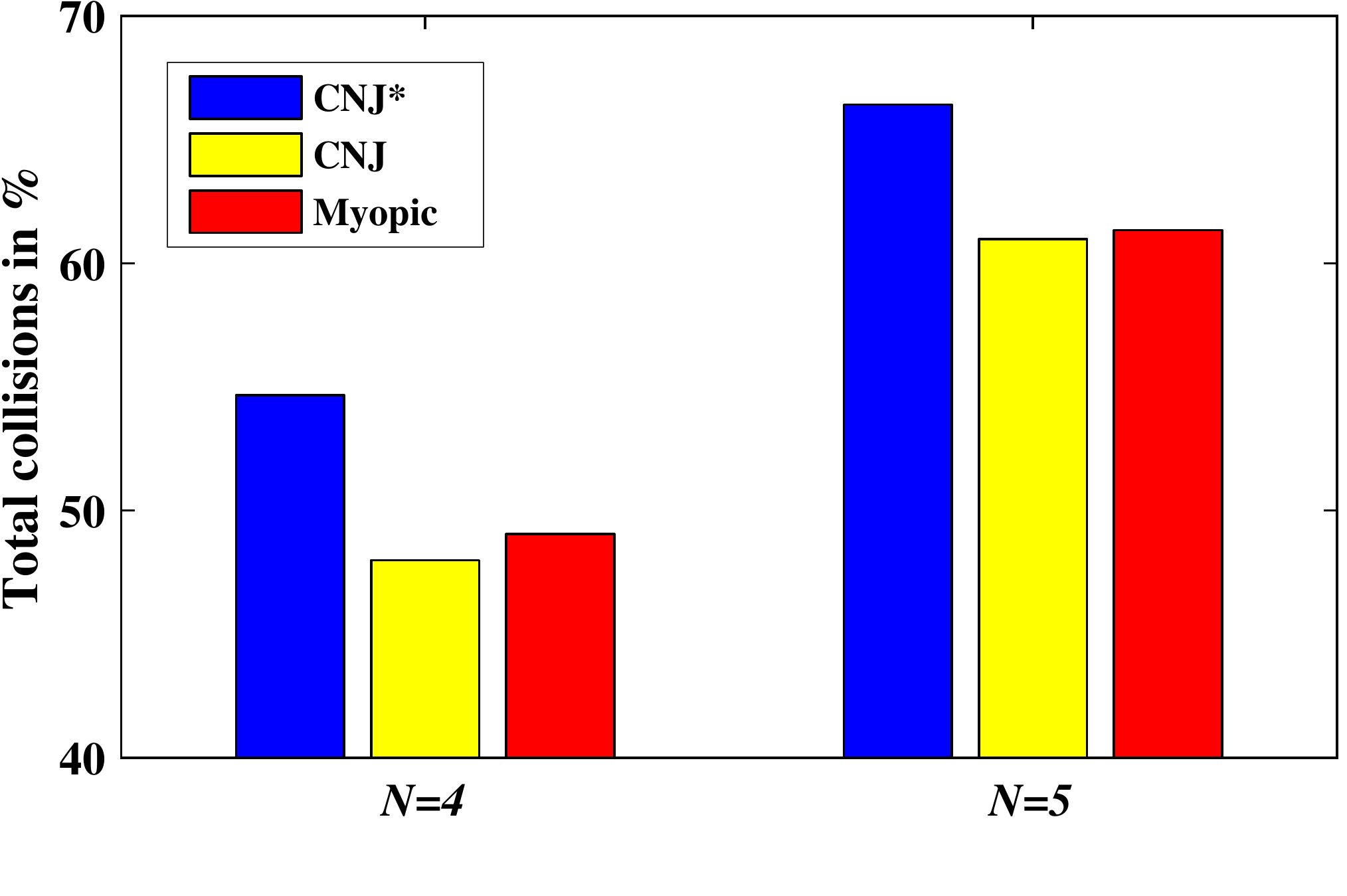}%
		\label{d3}}	
	\caption{Experimental results on USRP testbed for $J$=2 and $K$=8: a) Total throughput of SUs for $N$=4, b) Total throughput of SUs for $N$=5, and c) Total number of times in \% the SUs face collision with other SUs and jammers for $N$=4 and $N$=5.}

	\label{dthr}
	
\end{figure*}

The testbed consists of primary user traffic generator designed using OFDM based transmitter realized in LabView and USRP-2922 from National Instruments for over-the-air transmission as shown in Fig.~\ref{usrp1}. It transmits the signal in one or multiple channels based on their statistics. The first channel is dedicated for synchronization and hence, it is not used by the PUs/SUs/jammers for the transmission. The synchronization has been achieved by switching the corresponding channel from occupied to vacant states or vice-versa in each time slot. Each slot duration ($\Delta t$) is 0.1 second so that channel status can be followed by human eye. For experiments, the transmission parameters such as the number of OFDM sub-carriers, number of channels, center frequency, and bandwidth are 1024, 8, 935 MHz and 2 MHz, respectively. At the receiver side, SUs and jammers are implemented using MATLAB and USRP N200 from Ettus Research. At each SU or jammer, the channel selected by the underlining algorithm is passed through non-ideal energy detector to check whether it is vacant or occupied. When the channel is vacant and it is not selected by other SUs and jammers, it is assumed that the SU transmits over the channel and transmission is successful. The size of the horizon, $T$, is limited to 7000 due to large $\Delta t$ and each result is averaged over 15 independent experiments.

Consider $N=\{4,5\}$, $J=2$ and $p_i=\{0.2, 0.3, 0.4, 0.5, 0.6, 0.7, 0.8, 0.9\}$. The values of $T_C$, $T_O$ and $T_J$ are 3000, 50 and 1000, respectively. The Fig.~\ref{dthr} (a) and (b) show the comparison of total throughput, i.e. total number of successful transmissions, at different instants of horizon for $N$=4 and $N$=5, respectively. As expected, all algorithms have same throughput till $T_C$ slots. From $T_C$ till $T_C+T_O+T_J$ slots, the myopic algorithm may offer higher throughput than the proposed algorithm as discussed above. Thereafter, the proposed algorithms outperform myopic algorithm. Here, CNJ$^*$ refers to the proposed algorithm without the optimization function in (1) and hence, the SUs sequentially hops among top $N$ channels after $T_C+T_O+T_J$ time slots. The difference in the throughput of the CNJ and CNJ$^*$ algorithms show that identifying the subset size of optimum channels is as important as accurate estimation of $N$, $J$ and $p_i$ to achieve higher throughput. As evident from the slope of the plots after $T_C+T_O+T_J$ slots, the difference in the throughput between the proposed and myopic algorithms increases with time. Similar behavior can also be seen in the simulation results presented in previous section.
 
 The Fig.~\ref{dthr} (c) shows the total number of collisions faced by the SUs at the end of horizon. As expected, the SUs using CNJ$^*$ algorithm suffer highest number of collisions due to sequential hopping in top $N$ channels out of which $J$ channels are jammed by jammers. The number of collisions in myopic algorithm are slightly more than the CNJ algorithm though the SUs in latter sequentially hops in the top $N+J$ channels compared to the CNJ algorithm where the SUs sequentially hops in top $N+m$ channels, where $m\leq J$. This is because, $N+J$ estimation in the myopic algorithm may not be same at all SUs leading to frequent collisions among the SUs. The proposed algorithm guarantee accurate estimation of $N$ and $J$ with a high probability and hence, offers superior performance in terms of throughput as well as number of collisions (hence, longer battery life at SUs). 


%

%
%
%
%
\label{usrp}

\section{Conclusions and Future Directions} \label{conclusion}
In this paper, we presented three distributed algorithms to aid secondary users (SUs) for dynamic spectrum access in infrastructure-less cognitive radio network where some SUs can be malicious (i.e., jammers). The novel contribution of the proposed algorithms is to enable coordination of SUs in the network of unknown number of SUs and jammers without the need of a common communication link or a central controller. The proposed algorithms achieve optimal throughput in the secondary network after a bounded number of time slots with high probability when all the legitimate SUs implement them faithfully. We have validated the functionality of the proposed algorithm via extensive synthetic experiments and USRP experiments in the real radio environment. The comparisons with Myopic algorithms show that the proposed algorithm offers superior performance in terms of total throughput (and regret) as well as the number of collisions. The higher throughput leads to better spectrum utilization while fewer collisions leads to longer battery life at SU terminals. Future works include analysis and validation for the case where channels busy periods follow Markovian property and dynamic networks where SUs can enter or leave the network anytime.

\begin{IEEEbiography}{Sunnet Sawant}
received his B.Tech and M.Tech. dual degree in Electrical
Engineering from IIT Bombay in 2017. His research interests include communication networks

\end{IEEEbiography}

\begin{IEEEbiography}{Rohit Kumar}
	received his
	B.Tech. in Electronics and
	Communication Engineering in
	2010 and M.Tech. in Digital
	Communication in 2012, both
	from Guru Gobind Singh
	Indraprastha University, Delhi,
	India. He is currently pursuing
	the Ph.D. degree program in
	Electronics and Communication
	Enginnering at National Institute of Technology Delhi. 
\end{IEEEbiography}

\begin{IEEEbiography}{Manjesh K. Hanawal}
received the M.S. degree in ECE from the Indian Institute of Science, Bangalore, India, in 2009,
  and the Ph.D. degree from INRIA, Sophia Antipolis, France, and the University of Avignon, Avignon, France, in 2013. After spending two
  years as a postdoctoral associate at Boston University, he is now an Assistant Professor in Industrial Engineering and Operations Research
  at the Indian Institute of Technology Bombay, Mumbai, India. His research interests include communication networks, machine learning
  and network economics.
\end{IEEEbiography}

\begin{IEEEbiography}{Sumit J. Darak}
received his B.E. degree in Electronics and Telecommunications
Engineering from Pune University, India in 2007, and PhD degree from the
School of Computer Engineering, Nanyang Technological University (NTU),
Singapore in 2013. He is currently an Assistant Professor at Indraprastha Institute of Information Technology, Delhi (IIIT-Delhi), India. Dr. Sumit has been awarded India Government's “DST Inspire Faculty Award”
which is a prestigious award for young researchers under 32 years age. He has
also received “Best Demo Award” at CROWNCOM 2016, "Young Scientist Paper" award at URSI 2014and 2017, "Best Student Paper Award" at DASC 2017. His current research interests include the reinforcement learning algorithms and reconfigurable architectures for applications such as wireless communications, energy harvesting, smart grids etc.

\end{IEEEbiography}

\newpage
\onecolumn

  \appendices

\section*{\huge Supplementary for Learning to Coordinate in a Decentralized Cognitive Radio Network in Presence of Jammers}\vspace{1cm}
\section{Analysis of CDJ Algorithm}\label{CJ-DC}

In this section, we bound the regret of CDJ. We begin with the following definition given in \cite{MC,MEGA}.
\begin{defn}
An $\epsilon$-correct ranking of $K$ channels is a sorted list of empirical mean values of channel occupancy probabilities such that for all $i,j$, $\hat{p}_i$ is listed before $\hat{p}_j$ if $p_j - p_i >  \epsilon$.
\end{defn}
\noindent

\begin{lemma} \label{thm:le1}
By the end of CR1 phase, all SUs have $\epsilon$-correct ranking of the channels, number of SUs (N) and Jammer $J$ with probability at least $1-\delta$.
\end{lemma}
\begin{thm}\label{thm:t1}
For all $\Delta > \epsilon, \gamma \in (0, 0.5)$ and $\delta \in [0,1]$, with probability at least $1-\delta$, the expected regret of $N$ SUs using CDJ in the 
	presence of $J$ jammers with $T_C$ slots of learning is upper bounded by $NT_C + N^2exp(1)$, where $T_C$ is 
\begin{equation}
\label{eq:TCLengh}
 \mbox{round} \bigg(max\bigg(\frac{8}{\theta} \cdot \ln \left(\frac{18K}{\delta}\right),\frac{1}{\epsilon^2_1 \theta}\ln \left(\frac{12K}{\delta}\right),
 \frac{1}{\epsilon^2_2 \theta}\ln \left(\frac{24K}{\delta}\right), 8K \ln \left(\frac{4K^2}{\delta}\right),\frac{4K}{\epsilon^2}\ln \left( \frac{8 K^2}{\delta} \right) \bigg)\bigg)
\end{equation}	
with $\epsilon_1=\gamma/2K\exp(1)$ and $\epsilon_2= \gamma/K$.
\end{thm}

\section{Analysis of CNJ Algorithm}\label{CJ-NDC}
\begin{lemma} \label{thm:le2}
For any $\delta>0$, all SUs will have $\epsilon$-correct ranking of channels and   $\hat{N}=N$ given $J$ with probability at least $1-\delta/3$ if CR2 subroutine is run for $T_C$ number of time slots.
\end{lemma}
\noindent \textbf{Proof:} \textbf{Channel Ranking Estimation}
If for any user $n$ it is true that $\forall k \in 1 \cdots K$ $\left | \hat{p}_k-p_k \right| \leq\frac{\epsilon}{2} $ then the 
user has an  $\epsilon-$correct ranking.

We will upper bound the probability that no user has $\epsilon-$correct ranking given the user has $S_{1}$ observations of each 
channel. 
We define the following events : \\

\noindent$J_n$ - event that an user $n$ has observed each channel $j\geq0$ times.\\
$A$ - event that all users have an $\epsilon$-correct ranking. \\
$A_n$ - event that an user $n$ has $\epsilon$-correct ranking. \\
$B$ - event that all users have at least $S_1$ observations of each channel. \\
$B_n$ - event that an user $n$ has at least $S_1$ observations of each channel. \\ 

Note*: $\overline{X}$ denotes complement of any event $X$\\

We want to compute,
\[Pr(\overline{A}_n|B_n) < \frac{\delta_1}{4K} \]

Also, we know , 
\begin{align*}
Pr(\overline{A}_n|B_n) &\leq Pr\left ( \exists k \in 1 \cdots K \ \ s.t | \hat{p}_k-p_k | >\frac{\epsilon}{2} | \ B_n\right) \\
 \tag{By Union Bound}      &\leq \sum_{k=1}^{K} Pr\left ( | \hat{p}_k-p_k | >\frac{\epsilon}{2} | \ B_n\right) \\ 
       			&= \sum_{k=1}^{K} \sum_{j=C}^{\infty} Pr\left (| \hat{p}_k-p_k | >\frac{\epsilon}{2} | \ 
J_n  \right) Pr \left( J_n | \ B_n \right) \\ 
 \tag {By Hoeffding's Inequality} &\leq \sum_{k=1}^{K} \sum_{j=C}^{\infty} 2 \cdot \exp \left(\frac{-j \cdot \epsilon^2}{2} \right) Pr \left( J_n | \ B_n \right) \\ 
	&\leq \sum_{k=1}^{K} 2 \cdot \exp \left(\frac{-S_1 \cdot \epsilon^2}{2} \right) \sum_{j=C}^{\infty} Pr \left( J_n | \ B_n \right) \\ 
	 &\leq \sum_{k=1}^{K} 2 \cdot \exp \left(\frac{-S_1 \cdot \epsilon^2}{2} \right)  \\
	 &\leq K \cdot 2 \cdot \exp \left(\frac{-S_1 \cdot \epsilon^2}{2} \right)
\end{align*}

We can apply Hoeffding's Inequality since each observation of the channel is independent of the number of times we observe that
channel. This is true since each SU is randomly selecting channels in CR phase, which is independent of the
previous rounds. In order for this to be $< \frac{\delta_1}{4K}$, 
	 \[ K \cdot 2 \cdot \exp \left(\frac{-S_1 \cdot \epsilon^2}{2} \right) < \frac{\delta_1}{4K} \newline
		\implies  S_1 >   \ln \left ( \frac{8 \cdot K^2}{\delta_1} \cdot  \right) \frac{2}{\epsilon^2}
\]


Now, to show that if all users have at least $ S_1 > \ln \left ( \frac{8 \cdot K^2}{\delta_1} \cdot  \right) \frac{2}{\epsilon^2}$ observations of each channel, then all users have an $\epsilon-$correct ranking of channels with probability at least $1-\frac{\delta_1}{4}$. 

\begin{align*} 
     Pr(A|B) &\geq 1-Pr\left( \vee_n Pr(\overline{A}_n | B_n) \right) \\
            &\geq 1-\sum_{n=1}^{N} Pr(\overline{A}_n | B_n)    \tag{By Union Bound}\\
	    &\geq 1-\frac{\delta_1}{4K} \cdot (N) \\
            &\geq 1-\frac{\delta_1}{4K} \cdot K\\
	    &= 1-\frac{\delta_1}{4}
\end{align*}

We want to show that there exists a $T_{S_1}$ large enough such that all SUs have greater than $S_1$ observations of each channel.  
Let $B_{n,k}(t)$ be a random variable which indicates that channel $k$ is selected by an user $n$. 
\begin{equation*}\label{eq:ct}
				B_{n,k}(t)= \left\{ \begin{array}{cl}
				1 &\ w.p \ \frac{1}{K} \\
				0 &\ w.p \ \frac{1}{K} \end{array} \right . 
                \end{equation*}
\[E\left[B_{n,k}(t)\right] = \frac{1}{K} \]

\begin{align*}
Pr &\left(A \ SU \ has < \frac{1}{2}\cdot T_{S_1}\cdot E[B_{n,k}(t) observations\right)  \\ 
 &= Pr\left(\sum\limits_{t=1}^{T_{S_1}} B_{n,k}(t) < \frac{1}{2}\cdot T_{S_1}\cdot E[B_{n,k}(t)]\right)\\
 &\leq \exp \left(-\frac{1}{8}\cdot T_{S_1}\cdot B_{n,k}(t)] \right) \tag{Using Chernoff's Bound}
\end{align*}

We want,
\begin{align*}
Pr \left(\exists n,k \ s.t.\sum\limits_{t=1}^{T_{S_1}} B_{n,k}(t) < \frac{1}{2}\cdot T_{S_1}\cdot E[B_{n,k}(t)]\right)  <\frac{\delta_1}{4} \\
\implies N\exp \left(-\frac{1}{8}\cdot T_{S_1}\cdot E[B_{n,k}(t)] \right) < \frac{\delta_1}{4K} \\
\implies T_{S_1} > 8 \cdot \ln \left(\frac{4K^2}{\delta_1}\right) \cdot \frac{1}{E[B_{n,k}(t)] } \\
\implies T_{S_1} > 8K\ln \left(\frac{4K^2}{\delta_1}\right)
\end{align*}

We have shown that for $T_{S_1} >8K\ln \left(\frac{4K^2}{\delta_1}\right)$ , the number of observations of selected channels for all SUs, $\sum\limits_{t=1}^{T_c} B_{n,k}(t)$ is greater than $ \frac{1}{2}\cdot T_{S_1}\cdot E[B_{n,k}(t)]$ with probability at least
$1-\frac{\delta_1}{4} $

We also want that total number of observations of each channel be at least $S_1$.
Therefore, \\ 
\[ \sum\limits_{t=1}^{T_{S_1}} B_{n,k}(t) \geq \frac{1}{2}\cdot T_{S_1}\cdot E[B_{n,k}(t)] \geq S_1\]
\[\implies \frac{1}{2}\cdot T_{S_1}\cdot E[B_{n,k}(t)] \geq \ln \left ( \frac{8 \cdot K^2}{\delta_1}\right) \frac{2}{\epsilon^2} \]
\[\implies  T_{S_1} \geq \frac{4}{\epsilon^2 E[B_{n,k}(t)]}\ln \left ( \frac{8 \cdot K^2}{\delta_1} \right)    \]
\[\implies  T_{S_1} \geq \frac{4K}{\epsilon^2 }\ln \left ( \frac{8 \cdot K^2}{\delta_1} \right)    \]

\begin{align*}
Pr(A) &= 1 - Pr(\overline{A}) \\
      &= 1 - Pr(\overline{A}/B)Pr(B)-Pr(\overline{A}/\overline{B})Pr(\overline{B}) \\
      &\geq 1 - Pr(\overline{A}/B) - Pr(\overline{B})\\
      &\geq 1- \frac{\delta_1}{4} - \frac{\delta_1}{4} \\
      &\geq 1- \frac{\delta_1}{2}
\end{align*}
If $ T_{R} = max \left(8K\ln \left(\frac{4K^2}{\delta_1}\right),\frac{4K}{\epsilon^2}\ln \left ( \frac{8 \cdot K^2}{\delta_1} \right) \right)$, then for $t\geq T_{R}$ all SUs will have $\epsilon$-correct ranking of channels with probability at least 
 $1- \frac{\delta_1}{2}.$ 


\noindent \textbf{Estimation of N:} Now, we compute number of slots required to estimate $N$ with high probability given $J$. We define the following events: \\
$W_k$ = event that channel $k$ is selected by a SU \\
$X_k$ = event that none of the other SUs/Jammer selects $k$ \\
$U_k$ = event that channel $k$ is free \\
Let $S_2$ be the number of times a selected channel is found free. 
Probability of collision, denoted $p_c$, for any SU conditioned on the events that the channel selected is free and the number of Jammers is $J$ is given by

\begin{align*}
Pr(collision) &= \sum\limits_{k=1}^{K}  Pr(W_k/U_k)Pr(X_k/U_k)  \\
p_c	&= \sum\limits_{k=1}^{K} \frac{1}{K} \left(1-\left(1-\frac{J}{K}\right)\left(1-\frac{1}{K}\right)^{N-1}  \right)\\
p_c	&= 1-\left(1-\frac{J}{K}\right)\left(1-\frac{1}{K}\right)^{N-1}   \\
\end{align*}

Solving for $N$, we get 

\[N = 1 + \frac{\log \left (1-p_c\right)-\log \left ( 1-\frac{J}{K} \right )} {\log \left ( 1-\frac{1}{K} \right )}\] 
Let $\hat{p_c}$  be the estimate of $p_c$. 
\[\hat{N} = 1 + \frac{\log \left (1-\hat{p}_c\right)-\log \left ( 1-\frac{J}{K} \right )} {\log \left ( 1-\frac{1}{K} \right )}\]

We want $\hat{N}$ to be a good estimate of $N$. Therefore, for some $\gamma < \frac{1}{2}$, we set $ \left | \hat{N} - N \right| < \gamma$

\begin{align*}
\left| \frac{\log \left (1-\hat{p}_c\right)} {\log \left ( 1-\frac{1}{K} \right )}-\frac{\log \left (1-p_c\right)} {\log \left ( 1-\frac{1}{K} \right )} \right|  &< \gamma  \\
\left| \frac{\log \left(\frac{1-\hat{p}_c}{1-p_c}\right)}{\log \left ( 1-\frac{1}{K} \right )}\right| &< \gamma  \\
\end{align*}

Let $\alpha$ be the difference between $\hat{p}_c$ and $p_c$. Substituting,
 $\hat{p}_c = p_c +\alpha $.  

\begin{align*}
\left| \frac{\log \left(\frac{1- p_c -\alpha}{1-p_c}\right)}{\log \left ( 1-\frac{1}{K} \right )}\right| &< \gamma  \\
 -\gamma\log \left ( 1-\frac{1}{K} \right )&\leq \log \left(\frac{1-p_c - \alpha}{ 1-p_c}  \right) \leq  \gamma\log \left ( 1-\frac{1}{K} \right ) \\
\log \left ( 1-\frac{1}{K} \right )^{-\gamma}&\leq \log \left(\frac{1-p_c - \alpha}{ 1-p_c}  \right) \leq  \log \left ( 1-\frac{1}{K} \right )^\gamma \\
 \left ( 1-\frac{1}{K} \right )^{\gamma}&\leq  \left(\frac{1-p_c - \alpha}{ 1-p_c}  \right) \leq   \left ( 1-\frac{1}{K} \right )^{-\gamma} \\
 \left ( 1-\frac{1}{K} \right )^{\gamma} \left(1-p_c \right)&\leq  \left(1-p_c - \alpha  \right) \leq   \left ( 1-\frac{1}{K} \right )^{-\gamma} \left(1-p_c \right) \\
\left(1-p_c \right)\left(1-\left ( 1-\frac{1}{K} \right )^{-\gamma} \right) &\leq \alpha \leq \left(1-p_c \right)\left(1-\left ( 1-\frac{1}{K} \right )^{\gamma} \right)
\end{align*}

If we ensure $|\hat{p}_c - p_c| < \epsilon_1$ where
\begin{equation}\epsilon_1 = min \left \{\left| \left(1-p_c \right)\left(1-\left ( 1-\frac{1}{K} \right )^{-\gamma} \right)\right| , \right.\left.
                    \left| \left(1-p_c \right)\left(1-\left ( 1-\frac{1}{K} \right )^{\gamma} \right)\right|\right \}\end{equation}
for some $\gamma<1/2$ then we can have $\hat{N} = N$. The number of observations $S_2$ required to compute $\hat{p_c}$
such that $|\hat{p}_c - p_c| \leq \epsilon_1$  for all SUs is given by Hoeffding's inequality. 
 \[Pr(|\hat{p}_c - p_c| >\epsilon_1) \leq 2\exp(-2\cdot S_2\cdot\epsilon^2_1) < \frac{\delta_1}{4K} \]
\[\implies S_2 > \frac{1}{2\epsilon^2_1}\ln \left(\frac{8K}{\delta_1}\right) \]

We can further simplify our expression for $\epsilon_1$. Substituting the value of $p_c$, we have 
 \begin{equation}\epsilon_1 = min \left \{\left|\left(1-\frac{1}{K}\right)^{N-1} \left(1-\frac{J}{K} \right) \left(1-\left ( 1-\frac{1}{K} \right )^{-\gamma} \right)\right| ,\right.  \left.
                    \left| \left(1-\frac{1}{K}\right)^{N-1} \left(1-\frac{J}{K} \right) \left(1-\left ( 1-\frac{1}{K} \right )^{\gamma} \right)\right|\right \}\end{equation}
Using the results in \cite{MC} and  $\left(1-\frac{J}{K} \right)\geq \frac{1}{2}$  as $K>N>J$ ,we have 
\[\left|\left(1-\frac{1}{K}\right)^{N-1} \left(1-\frac{J}{K} \right)\left(1-\left ( 1-\frac{1}{K} \right )^{-\gamma} \right)\right| \geq \frac{\gamma}{2\exp(1)K}\]
\[\left|\left(1-\frac{1}{K}\right)^{N-1}\left(1-\frac{J}{K} \right) \left(1-\left ( 1-\frac{1}{K} \right )^{\gamma} \right)\right| \geq \frac{\gamma}{2\exp(1)K}\]
Therefore, combining above two, we have 

\[\epsilon_1 \geq \frac{\gamma}{2\exp(1).K}\]

If all SUs have $S_2>\frac{1}{2\epsilon^2_1}\ln \left(\frac{8K}{\delta_1}\right)$ observations that the selected channel was free, where $\epsilon_1 =\frac{\gamma}{2\exp(1).K}$ then with 
probability at least $1-\frac{\delta_1}{4}$, we have  $\hat{N} = N$ for all SUs given $J$.
\\

We want to show that there exists a $T_F$ large enough such that all SUs have greater than $S_2$ observations of selected channel were 
free. 
Let $A_{I_t}(t)$ be a random variable which indicates if selected channel $k$ is free at round $t$. We have

                 \begin{equation}\label{eq:ct}
                 A_{I_t}(t)= \left\{ \begin{array}{cl}
                 1& \mbox{ if  $I_t$  is free}\\
                 0&   \mbox{otherwise} \end{array} \right . 
                 \end{equation}

\begin{align*}
E[A_{I_t}(t)] &= \sum\limits_{k=1}^{K}Pr(I_t=k) Pr(A_{I_t}(t)=1/I_t = k) \\
	&= \sum\limits_{k=1}^{K}\frac{1}{K} (1-p_k)\\
	&\geq \theta 
\end{align*}
We have
\begin{align*}
Pr &\left(A \ SU \ has < \frac{1}{2}\cdot T_F\cdot E[A_{I_t}(t)] observations\right)  \\ 
 &= Pr\left(\sum\limits_{t=1}^{T_F} A_{I_t}(t) < \frac{1}{2}\cdot T_F\cdot E[A_{I_t}(t)]\right)\\
 &\leq \exp \left(-\frac{1}{8}\cdot T_F\cdot E[A_{I_t}(t)] \right) \tag{Using Chernoff's Bound}
\end{align*}

We set,
\begin{align*}
&Pr \left(\mbox{SUs do not have} > \frac{1}{2} T_{F}E[A_{I_t}(t)] \mbox{observations}\right)< \frac{\delta_1}{4} \\
&\implies \exp \left(-\frac{1}{8} T_{F} E[A_{I_t}(t)] \right) < \frac{\delta_1}{4K} \\
&\implies T_{F} > 8  \ln \left(\frac{4K}{\delta_1}\right)  \frac{1}{E[A_{I_t}(t)] } \\ 
&\implies T_{F} >\frac{8}{\theta}  \ln \left(\frac{4K}{\delta_1}\right),  
\end{align*}


We have shown that for $T_F > \frac{8}{\theta}\cdot \ln \left(\frac{4K}{\delta_1}\right)$ , the number of observations of selected channels 
being free for all SUs, $\sum\limits_{t=1}^{T_F} A_{I_t}(t)$ is greater than $ \frac{1}{2}\cdot T_F\cdot E[A_{I_t}(t)]$ with probability at least
$1-\frac{\delta_1}{4} $.
We also want that total number of observations of selected channels being free to be at least $S_2$.
Therefore, \\ 
\[ \sum\limits_{t=1}^{T_{S_2}} A_{I_t}(t) \geq \frac{1}{2}\cdot T_{S_2}\cdot E[A_{I_t}(t)] \geq S_2\]
\[\implies \frac{1}{2}\cdot T_{S_2}\cdot E[A_{I_t}(t)] \geq \frac{1}{2\epsilon^2_1}\ln \left(\frac{8K}{\delta_1}\right) \]
\[\implies  T_{S_2} \geq \frac{1}{\epsilon^2_1 E[A_{I_t}(t)]}\ln \left(\frac{8K}{\delta_1}\right) \]
\[\implies  T_{S_2} \geq \frac{1}{\epsilon^2_1 \theta}\ln \left(\frac{8K}{\delta_1}\right) \]
\\
We define two more events:
$E$ - event that all SUs have observed at least $S_2$ times the selected channel was free \\ 
$D$ - event that all SUs have correct estimate of $N$ given $J$ \\
We have

\begin{align*}
Pr(D) &= 1 - Pr(\overline{D})  \\
     &= 1 - Pr(\overline{D}/E)Pr(E) -  Pr(\overline{D}/\overline{E})Pr(\overline{E}) \\
     &\geq 1 - Pr(\overline{D}/E) - Pr(\overline{E}) \\ 
     &\geq 1 - \frac{\delta_1}{4}- \frac{\delta_1}{4}\\
  	&\geq 1 - \frac{\delta_1}{2} 
\end{align*}

If $T_{NE} = max\left(\frac{8}{\theta}\ln \left(\frac{4K}{\delta_1}\right),\frac{1}{\epsilon^2_1 \theta}\ln \left(\frac{8K}{\delta_1}\right)\right)$ then for $t \geq T_{NE}$, all SUs would have correct estimate of $N$ given $J$ with probability atleast 
$1-\frac{\delta_1}{2}$ 
\newline
For $A$ and $D$ as defined previously,
\begin{align*}
Pr(AD)&= 1-Pr(\overline{AD}) \\
      &\geq 1 - Pr(\overline{A}) - Pr(\overline{D})\\
      &\geq 1 -  \frac{\delta_1}{2}-\frac{\delta_1}{2}\\
      &\geq 1-\delta_1 
\end{align*}
If $T_C = max (T_{NE},T_{R})$, then with probability $1-\delta_1$, all SUs would have $\epsilon$-correct channel ranking as well as correct estimate of $N$ given $J$.
\begin{lemma} \label{thm:le3}
For any $\delta>0$, all SUs will find non-overlapping channels with probability at least $1-\delta/3$ if OR subroutine is run for $T_O$ number of time slots.
\end{lemma}
\begin{lemma} \label{thm:le4}
For any $\delta>0$, all SUs will estimate $\hat{J}=J$ with probability at least $1-\delta/3$ if JE subroutine is run for $T_J$ number of time slots.
\end{lemma}
\noindent \textbf{Proof:} Given that SUs are orthogonalized on $K$ channels we want to compute number of slots required to estimate $J$ correctly with high probability for all SUs. 
We will compute collision probability only when the channel is free.
Let $S_3$ be the number of times we observe that selected channel is free after orthogonalization.
Given $S_3$, we want to ensure correct estimation of $J$ with high probability.
We also assume that jammer is hitting top $N$ channels.
\begin{align*} {\label{eq:Pc4}}
Pr(collision) = p_c &= \sum\limits_{i=1}^{N} \frac{1}{K} \left(1-1-\frac{J}{N}\right) \\
 p_c &= \sum\limits_{i=1}^{N} \frac{1}{K} \frac{J}{N} \\
p_c = \frac{J}{K} \\
\implies J = Kp_c
\end{align*}

Let $\hat{p}_c$ be the estimate of $p_c$. Then estimate of $J$ is given by, \\
\[\hat{J}=K\hat{p}_c \]

We want to ensure $\hat{J}$ is close to $J$. Therefore, for some $\gamma < 1/2$ 
\begin{align*} 
\left| \hat{J}- J \right| &\leq \gamma \\
\left| \hat{p}_c - p_c \right| &\leq \frac{\gamma}{K} \\
\end{align*}

The number of observations $S_3$ required to compute $\hat{p_c}$
such that $|\hat{p}_c - p_c| \leq \epsilon_2$, where $\epsilon_2= \frac{\gamma}{K}$ for all SUs is given by Hoeffding's inequality. 
 \[Pr(|\hat{p}_c - p_c| >\epsilon_2) \leq 2\exp(-2\cdot S_3\cdot\epsilon^2_2) < \frac{\delta_3}{2K} \]
\[\implies S_3 > \frac{1}{2\epsilon^2_2}\ln \left(\frac{4K}{\delta_3}\right) \]

To ensure we have $S_3$ observations of selected channels being free for all users, we define $A_{I_t}(t)$ and $E[A_{I_t}(t)] \geq \theta$ 
as defined in proof of Lemma \ref{thm:le1} used for $N$ and $J$ estimation.

We set,
\begin{align*}
&Pr \left(\mbox{SUs do not have} > \frac{1}{2} T_{J}E[A_{I_t}(t)] \mbox{observations}\right)< \frac{\delta_3}{2} \\
&\implies \exp \left(-\frac{1}{8} T_{J} E[A_{I_t}(t)] \right) < \frac{\delta_3}{2K} \\
&\implies T_{J} > 8  \ln \left(\frac{2K}{\delta_3}\right)  \frac{1}{E[A_{I_t}(t)] } \\ 
&\implies T_{J} >\frac{8}{\theta}  \ln \left(\frac{2K}{\delta_3}\right)  
\end{align*}

where $T_J$ is the number of slots for Jammer Estimation.
We also want that total number of observations of selected channels being free to be at least $S_3$.
Therefore, \\ 
\[ \sum\limits_{t=1}^{T_J} A_{I_t}(t) \geq \frac{1}{2}\cdot T_J\cdot E[A_{I_t}(t)] \geq S_3\]
\[\implies \frac{1}{2}\cdot T_J\cdot E[A_{I_t}(t)] > \frac{1}{2\epsilon^2_2}\ln \left(\frac{4K}{\delta_3}\right)\]
\[\implies  T_J > \frac{1}{\epsilon^2_2 E[A_{I_t}(t)]}\ln \left(\frac{4K}{\delta_3}\right) \]
\[\implies  T_J > \frac{1}{\epsilon^2_2 \theta}\ln \left(\frac{4K}{\delta_3}\right) \]
\[T_J = max\left(\frac{8}{\theta }  \cdot \ln \left(\frac{2K}{\delta_3}\right),\frac{1}{\epsilon^2_2 \theta}\ln \left(\frac{4K}{\delta_3}\right)\right) \]

We define two more events:\\
$G$ - event that all SUs have observed at least $S_3$ times the selected channel was free \\ 
$H$ - event that all SUs have correct estimate of $J$\\

We have

\begin{align*}
Pr(H) &= 1 - Pr(\overline{H})  \\
     &= 1 - Pr(\overline{H}/G)Pr(G) -  Pr(\overline{H}/\overline{E})Pr(\overline{E}) \\
     &\geq 1 - Pr(\overline{H}/G) - Pr(\overline{G}) \\ 
     &\geq 1 - \frac{\delta_3}{2}- \frac{\delta_3}{2}\\
  	&\geq 1 - \delta_3 \\
\end{align*}

The following theorem states the expected regret of the CNJ against the jammers who employs the jamming strategy as in CDJ. Again the expectation is over the randomness of the algorithm.
\begin{thm}\label{thm:t2}
For all 
	$\Delta > \epsilon, \gamma \in (0,0.5)$ and $0 < \delta\leq 1$, with probability at least $1-\delta$, the expected regret of SUs using CNJ in the presence of $J$ jammers after $T_L=T_C+T_O+T_J$ slots of learning is upper bounded by $NT_L + N^2exp(1)$, where
	\begin{eqnarray}
	\label{eq:TC2}
	T_{C}&=&\mbox{round} \bigg (\max\bigg(\frac{8}{\theta}\ln \left(\frac{12K}{\delta}\right),\frac{1}{\epsilon^2_1 \theta}\ln \left(\frac{24K}{\delta}\right) , \ln \left(\frac{12K^2}{\delta}\right), \frac{4K}{\epsilon^2}\ln \left ( \frac{24K^2}{\delta} \right) \bigg)\bigg)\\
		\label{eq:TO2}
	T_O &=&\mbox{round}\bigg(\frac{\log \left(\frac{\delta}{3K}\right)}{\log \left(1-\frac{\theta}{K}\left(1-\frac{1}{K} \right) ^	{K-1} \right)} \bigg )\\
		\label{eq:TJ2}
	T_J &=& \hspace{-.2cm}\mbox{round} \bigg (max\left(\frac{8}{\theta }  \ln \left(\frac{6K}{\delta}\right),\frac{1}{\epsilon^2_2 \theta}\ln \left(\frac{12K}{\delta}\right)\right)\bigg)
	\end{eqnarray}
with $\epsilon_1=\frac{\gamma}{2\exp(1).K}$ and $\epsilon_2= \frac
{\gamma}{K}$.
\end{thm}
\noindent
The terms $NT_C, NT_O, NT_J$ correspond to the total regret incurred by all the players in the first three subroutines, i.e., CR2, OR and JE respectively. The last term corresponds to the OR subroutine that runs till the end of time horizon.

\noindent \textbf{Proof:} Let Y be an event that all SUs have correct estimate of $N$ and $J$ as well as they are orthogonalized and have correct channel ranking after $T_L$ learning rounds.
Let $M$ be an event that all SUs have correct estimate of $N$.
A,D,F,H are as defined in proof of lemmas \ref{thm:le2}, \ref{thm:le3}, \ref{thm:le4}
\begin{align*}
Pr(Y) &\geq Pr(Y/AMFH)\cdot Pr(AMFH)  \\
   &\geq Pr(Y/AMFH)\cdot Pr(AM/FH)\cdot Pr(FH) \\ 
   &\geq Pr(Y/AMFH)\cdot Pr(AD)\cdot PR(H/F)\cdot Pr(F) \\  \tag{event $M$ given $H$ is $D$}
\end{align*}

Using Lemmas \ref{thm:le2},\ref{thm:le3},\ref{thm:le4} , we have, 
\begin{equation*}
Pr(Y)  \geq Pr(Y/AMFH)Pr(AD)PR(H/F)Pr(F)
\end{equation*}
\begin{equation*}
Pr(Y)  \geq (1-\delta_1)(1-\delta_3)(1-\delta_2) 
\end{equation*}
Setting $ \delta_1 = \delta_2=\delta_3=\frac{\delta}{3}$ 
\begin{equation*}
Pr(Y)  \geq \left(1-\frac{\delta}{3}\right)^3 
\end{equation*}
\begin{equation*}
Pr(Y)  \geq 1-3\frac{\delta}{3}
\end{equation*}
\begin{equation*}
Pr(Y)  \geq 1-\delta
\end{equation*}

Setting $T_L = T_C + T_O+T_J$, with probability at least $1-\delta$, all SUs have correct estimate of $N$ and $J$ as well as they are orthogonalized and have correct channel ranking.

Substituting value of $\delta_1,\delta_2,\delta_3$, we have,
\begin{equation*}
T_C= max \left( \frac{8}{\theta}\ln \left(\frac{12K}{\delta}\right), \frac{1}{\epsilon^2_1 \theta}\ln \left(\frac{24K}{\delta}\right),\right.  \left. 8K\ln \left(\frac{12K^2}{\delta}\right),\frac{4K}{\epsilon^2}\ln \left ( \frac{24 \cdot K^2}{\delta} \right) \right)
\end{equation*}

\[ T_O=\frac{\log \left(\frac{\delta}{3K}\right)}{\log \left(1-\frac{\theta}{k}\left(1-\frac{1}{K} \right) ^ {K-1} \right)} \]
\[T_J=max\left(\frac{8}{\theta }  \cdot \ln \left(\frac{6K}{\delta}\right),\frac{1}{\epsilon^2_2 \theta}\ln \left(\frac{12K}{\delta}\right) \right)\]

Regret bounds proof is similar to proof of 
Theorem $1$. 
\section{Analysis of CUJ Algorithm}\label{UJ}

The regret of the CUJ algorithm is same as the CNJ algorithm.

\begin{thm}\label{thm:t3}
For all  $\Delta > \epsilon$ and $0 \leq\delta\leq 1$, with probability at least $1-\delta$, the expected regret of $N$ SUs using the CUJ algorithm in the presence of $J$ Jammers after $T_L=T_C+T_O+T_J$ time slots of learning is upper bounded by $NT_L + N^2exp(1)$, where the values of $T_C, T_O,$ and $T_J$ are given in  (\ref{eq:TC2}), (\ref{eq:TO2}), and (\ref{eq:TJ2}), respectively with $\epsilon_1=\epsilon_2=\frac{\gamma}{\exp(1).K}$.
\end{thm}
\noindent \textbf{Proof:} \textbf{Channel Ranking Estimation} This part is similar to channel ranking estimation of Section~\ref{thm:le2}. We have,
If \[ T_{R} = max \left(8K\ln \left(\frac{4K}{\delta_1}\right) ,\frac{4K}{\epsilon^2 }\ln \left ( \frac{8 \cdot K^2}{\delta_1} \right)  \right),\] then for $t\geq T_{R}$ all SUs will have $\epsilon$-correct ranking of channels with probability at least 
 $1- \frac{\delta_1}{2}.$ 

\noindent \textbf{Estimation of $N+J$:} Now, we compute number of slots required to estimate $N+J$ with high probability. We define the following events: \\
 $W_k$ = event that channel $k$ is selected by a SU \\
$X_k$ = event that none of the other SUs/Jammer selects $k$ \\
$U_k$ = event that channel $k$ is free \\
Let $S_2$ be the number of times we observe that selected channel is free.  Probability of collision, denoted $p_c$, for any user conditioned on the events that the channel selected is free is given by
\begin{align*}
Pr(collision) &= \sum\limits_{i=1}^{K}  Pr(W_k/U_k)Pr(X_k/U_k)  \\
p_c	&= \sum\limits_{i=1}^{K} \frac{1}{K} \left(1-\left(1-\frac{1}{K}\right)^{N+J-1} \right)\\
p_c	&= 1-\left(1-\frac{1}{K}\right)^{N+J-1}  \\
\end{align*}

Solving for $N+J$, we get
 
\[N+J = 1 + \frac{\log \left (1-p_c\right)} {\log \left ( 1-\frac{1}{K} \right )}\] 
Let $\hat{p_c}$  be the estimate of $p_c$. 
\[\overline{N+J} = 1 + \frac{\log \left (1-\hat{p}_c\right)} {\log \left ( 1-\frac{1}{K} \right )}\] 

We want to make $\widehat{N+J}$ as close to $N+J$. Therefore, for $\gamma < \frac{1}{2}$
\[ \left | \widehat{N+J} - N+J \right| < \gamma \] 

\begin{align*}
\left| \frac{\log \left (1-\hat{p}_c\right)} {\log \left ( 1-\frac{1}{K} \right )}-\frac{\log \left (1-p_c\right)} {\log \left ( 1-\frac{1}{K} \right )} \right|  &< \gamma  \\
\left| \frac{\log \left(\frac{1-\hat{p}_c}{1-p_c}\right)}{\log \left ( 1-\frac{1}{K} \right )}\right| &< \gamma  \\
\end{align*}

Let $\alpha$ be the difference between $\hat{p}_c$ and $p_c$. Substituting,
 $\hat{p}_c = p_c +\alpha $.  

\begin{align*}
\left| \frac{\log \left(\frac{1- p_c -\alpha}{1-p_c}\right)}{\log \left ( 1-\frac{1}{K} \right )}\right| &< \gamma  \\
 -\gamma\log \left ( 1-\frac{1}{K} \right )&\leq \log \left(\frac{1-p_c - \alpha}{ 1-p_c}  \right) \leq  \gamma\log \left ( 1-\frac{1}{K} \right ) \\
\log \left ( 1-\frac{1}{K} \right )^{-\gamma}&\leq \log \left(\frac{1-p_c - \alpha}{ 1-p_c}  \right) \leq  \log \left ( 1-\frac{1}{K} \right )^\gamma \\
 \left ( 1-\frac{1}{K} \right )^{\gamma}&\leq  \left(\frac{1-p_c - \alpha}{ 1-p_c}  \right) \leq   \left ( 1-\frac{1}{K} \right )^{-\gamma} \\
 \left ( 1-\frac{1}{K} \right )^{\gamma} \left(1-p_c \right)&\leq  \left(1-p_c - \alpha  \right) \leq   \left ( 1-\frac{1}{K} \right )^{-\gamma} \left(1-p_c \right) \\
\left(1-p_c \right)\left(1-\left ( 1-\frac{1}{K} \right )^{-\gamma} \right) &\leq \alpha \leq \left(1-p_c \right)\left(1-\left ( 1-\frac{1}{K} \right )^{\gamma} \right)
\end{align*}

If we ensure $|\hat{p}_c - p_c| < \epsilon_1$ where
\begin{equation}\epsilon_1 = min \left \{\left| \left(1-p_c \right)\left(1-\left ( 1-\frac{1}{K} \right )^{-\gamma} \right)\right| \right.,  \left.
                    \left| \left(1-p_c \right)\left(1-\left ( 1-\frac{1}{K} \right )^{\gamma} \right)\right|\right \}\end{equation}
for some $\gamma<1/2$ then we can have $\widehat{N+J} = N+J$. The number of observations $S_2$ required to compute $\hat{p_c}$
such that $|\hat{p}_c - p_c| < \epsilon_1$ is given by Hoeffding's inequality. 
 \[Pr(|\hat{p}_c - p_c| >\epsilon_1) \leq 2\exp(-2\cdot S_2\cdot\epsilon^2_1) < \frac{\delta_1}{4K} \]
\[\implies S_2 > \frac{1}{2\epsilon^2_1}\ln \left(\frac{8K}{\delta_1}\right) \]

We can further simplify our expression for $\epsilon_1$. Substituing the value of $p_c$, we have 
 \begin{equation}\epsilon_1 = min \left \{\left|\left(1-\frac{1}{K}\right)^{N+J-1} \left(1-\left ( 1-\frac{1}{K} \right )^{-\gamma} \right)\right| , \right. \left.
                    \left| \left(1-\frac{1}{K}\right)^{N+J-1} \left(1-\left ( 1-\frac{1}{K} \right )^{\gamma} \right)\right|\right \}
\end{equation}
Using the results in \cite{MC}, we have 
\[\left|\left(1-\frac{1}{K}\right)^{N+J-1} \left(1-\left ( 1-\frac{1}{K} \right )^{-\gamma} \right)\right| \geq \frac{\gamma}{\exp(1).K}\]
\[\left|\left(1-\frac{1}{K}\right)^{N+J-1} \left(1-\left ( 1-\frac{1}{K} \right )^{\gamma} \right)\right| \geq \frac{\gamma}{\exp(1).K}\]

Therefore, combining above two, we have 

\[\epsilon_1 \geq \frac{\gamma}{\exp(1).K}\]

If all SUs have $S_2> \frac{1}{2\epsilon^2_1}\ln \left(\frac{8K}{\delta_1}\right) $ observations that the selected channel was free, where $\epsilon_1 =\frac{\gamma}{\exp(1).K}$ then with 
probability at least $1-\frac{\delta_1}{4}$, we have  $\widehat{N+J} = N+J$ for all SUs.
We want to show that there exists a $T_F$ large enough such that all SUs have greater than $S_2$ observations of selected channel were 
free. 
Let $A_{I_t}(t)$ be a random variable which indicates if selected channel $k$ is free at round $t$. We have
\begin{equation}\label{eq:ct}
A_{I_t}(t)= \left\{ \begin{array}{cl}
1& \mbox{ if  $I_t$  is free}\\
0&   \mbox{otherwise} \end{array} \right . 
\end{equation}

\begin{align*}
E[A_{I_t}(t)] &= \sum\limits_{k=1}^{K}Pr(I_t=k) Pr(A_{I_t}(t)=1/I_t = k) \\
	&= \sum\limits_{k=1}^{K}\frac{1}{K} (1-p_k)\\
	&\geq \theta 
\end{align*}
\begin{align*}
Pr &\left(A \ SU \ has < \frac{1}{2}\cdot T_F\cdot E[A_{I_t}(t)] observations\right)  \\ 
 &= Pr\left(\sum\limits_{t=1}^{T_F} A_{I_t}(t) < \frac{1}{2}\cdot T_F\cdot E[A_{I_t}(t)]\right)\\
 &\leq \exp \left(-\frac{1}{8}\cdot T_F\cdot E[A_{I_t}(t)] \right) \tag{Using Chernoff's Bound}
\end{align*}

We set,
\begin{align*}
&Pr \left(\mbox{SUs do not have} > \frac{1}{2} T_{F}E[A_{I_t}(t)] \mbox{observations}\right)< \frac{\delta_1}{4} \\
&\implies \exp \left(-\frac{1}{8} T_{F} E[A_{I_t}(t)] \right) < \frac{\delta_1}{4K} \\
&\implies T_{F} > 8  \ln \left(\frac{4K}{\delta_1}\right)  \frac{1}{E[A_{I_t}(t)] } \\ 
&\implies T_{F} >\frac{8}{\theta}  \ln \left(\frac{4K}{\delta_1}\right),  
\end{align*}


We have shown that for $T_F > \frac{8}{\theta}\cdot \ln \left(\frac{4K}{\delta_1}\right)$ , the number of observations of selected channels 
being free for all SUs, $\sum\limits_{t=1}^{T_F} A_{I_t}(t)$ is greater than $ \frac{1}{2}\cdot T_F\cdot E[A_{I_t}(t)]$ with probability at least
$1-\frac{\delta_1}{4} $

We also want that total number of observations of selected channels being free to be at least $S_2$.
Therefore, \\ 
\[ \sum\limits_{t=1}^{T_F} A_{I_t}(t) \geq \frac{1}{2}\cdot T_F\cdot E[A_{I_t}(t)] \geq S_2\]
\[\implies \frac{1}{2}\cdot T_F\cdot E[A_{I_t}(t)] > \frac{1}{2\epsilon^2_1}\ln \left(\frac{8K}{\delta_1}\right) \]
\[\implies  T_F > \frac{1}{\epsilon^2_1 E[A_{I_t}(t)]}\ln \left(\frac{8K}{\delta_1}\right) \]
\[\implies  T_F > \frac{1}{\epsilon^2_1 \theta}\ln \left(\frac{8K}{\delta_1}\right) \]
\\
We define two more events:
$E$ - event that all SUs have observed at least $S_2$ times the selected channel was free \\ 
$D$ - event that all SUs have correct estimate of $N$ given $J$ \\

We have
\begin{align*}
Pr(D) &= 1 - Pr(\overline{D})  \\
     &= 1 - Pr(\overline{D}/E)Pr(E) -  Pr(\overline{D}/\overline{E})Pr(\overline{E}) \\
     &\geq 1 - Pr(\overline{D}/E) - Pr(\overline{E}) \\ 
     &\geq 1 - \frac{\delta_1}{4}- \frac{\delta_1}{4}\\
  	&\geq 1 - \frac{\delta_1}{2} 
\end{align*}

If $T_{E} = max\left(\frac{8}{\theta}\ln \left(\frac{4K}{\delta_1}\right),\frac{1}{\epsilon^2_1 \theta}\ln \left(\frac{8K}{\delta_1}\right)\right)$ then for $t \geq T_{E}$, all SUs would have correct estimate of $N+J$ with probability at least 
$1-\frac{\delta_1}{2}$ 
\newline
For events $A$ and $D$ as defined previously, we have
\begin{align*}
Pr(AD)&= 1-Pr(\overline{AD}) \\
      &\geq 1 - Pr(\overline{A}) - Pr(\overline{D})\\
      &\geq 1 -  \frac{\delta_1}{2}-\frac{\delta_1}{2}\\
      &\geq 1-\delta_1 
\end{align*}
If $T_C = max (T_{E},T_{R})$, then with probability $1-\delta_1$, all SUs would have $\epsilon$-correct channel ranking as well as correct estimate of $N$ given $J$.


\noindent \textbf{Orthogonalization}
We want to compute $T_O$ such that all SUs are orthogonalised on $N+J$ channels with high probability within $T_O$.
If $p_c$ is the collision probability of a SU at any time slot $t$, and if none of the other SUs are orthogonalized (worst-case) 
then probability that a SUs will find a an orthogonal channel within $T_O$ is given by 
   \begin{equation*} \label{eq:2} \sum\limits_{t=1}^{T_O} p_c^{t-1}\cdot (1-p_c) \end{equation*}
We want this probability to be at least $1-\frac{\delta_2}{K}$ to ensure all SUs are orthogonalised within $T_O$.
\begin{align*}
\sum\limits_{t=1}^{T_O} p_c^{t-1}\cdot (1-p_c) &\geq 1-\frac{\delta_2}{K} \\
       1-p_c^{T_O} &\geq 1-\frac{\delta_2}{K}\\
	T_O \log p_c &\leq \log \left(\frac{\delta_2}{K}\right) \\
 T_O &\geq \frac{\log \left(\frac{\delta_2}{K}\right)}{\log p_c} \\
\end{align*}
Consider $p_{ns}$ be the probability of no collision due to non-settled SUs and $p_{s}$ be the probability of no collision due to settled SUs. Then,
\begin{align*}
Pr(no \ collision)
	&= \sum\limits_{k=1}^{N+J}\frac{1}{N+J} (1-p_k) \left(p_{ns}+p_{s}\right)\left(1-\frac{1}{N+J}\right)^{J} \\ 
   &+ \sum\limits_{k=1}^{N+J}\frac{p_k}{N+J}\\
	 p_{nc}&\geq \sum\limits_{k=1}^{N+J}\frac{1}{N+J} (1-p_k) \left(p_{ns}+p_{s}\right)\\
	 &\left(1-\frac{1}{N+J}\right)^{J}\\
    &\geq \sum\limits_{k=1}^{N+J}\frac{1}{N+J} (1-p_k) p_{ns}\left(1-\frac{1}{N+J}\right)^{J}\\
    &\geq \sum\limits_{k=1}^{N+J}\frac{1}{N+J} (1-p_k) \left(1-\frac{1}{N+J} \right)^{N+J-1} \\
    &\geq \theta\left(1-\frac{1}{N+J} \right)^{N+J-1} \\ 
  &\geq \theta\left(1-\frac{1}{K} \right)^{K-1}\\
  p_c  &\leq 1-\theta\left(1-\frac{1}{K} \right)^{K-1}\\ 
\implies T_O &\geq \frac{\log \left(\frac{\delta_2}{K}\right)}{\log \left(1-\theta\left(1-\frac{1}{K} \right) ^ {K-1} \right)} \\
\end{align*}
%
We define an event $F$ such that all SUs are orthogonalised given all SUs have correct estimate of $N+J$.
We just showed that $Pr(F) \geq  1-\delta_2.$

\noindent \textbf{Jammer Estimation}
The SUs are orthogonalised and locked themselves on top $N+J$ channels.
We want to compute $T_J$, slots required to estimate $J$ correctly with high probability.
Let $S_3$ be the number of observations that the channel was free. Then, SUs will estimate $J$ by computing collision probabilities
only when the channel is free.

\begin{align*}
Pr(collision)=p_c &= 1-\left(1-\frac{1}{N+J}\right)^J \\
\end{align*} 	
Solving for $J$, we get
\[J = \frac{\log \left (1-p_c\right)} {\log \left ( 1-\frac{1}{N+J} \right )}\] 
Let $\hat{p_c}$  be the estimate of $p_c$. 
\[\hat{J} =  \frac{\log \left (1-\hat{p}_c\right)} {\log \left ( 1-\frac{1}{N+J} \right )}\] 

We want to make $\hat{J}$ as close to $J$. Therefore, for $\gamma < \frac{1}{2}$
\[ \left | \hat{J} - J \right| < \gamma \] 

\begin{align*}
\left| \frac{\log \left (1-\hat{p}_c\right)} {\log \left ( 1-\frac{1}{N+J} \right )}-\frac{\log \left (1-p_c\right)} {\log \left ( 1-\frac{1}{N+J} \right )} \right|  &< \gamma  \\
\left| \frac{\log \left(\frac{1-\hat{p}_c}{1-p_c}\right)}{\log \left ( 1-\frac{1}{N+J} \right )}\right| &< \gamma  \\
\end{align*}

Let $\alpha$ be the difference between $\hat{p}_c$ and $p_c$. Substituting,
 $\hat{p}_c = p_c +\alpha $.  
\[\log \left ( 1-\frac{1}{N+J} \right )^{-\gamma}\leq \log \left(\frac{1-p_c - \alpha}{ 1-p_c}  \right) \leq  \log \left ( 1-\frac{1}{N+J} \right )^\gamma \]
\[\left ( 1-\frac{1}{N+J} \right )^{\gamma}\leq  \left(\frac{1-p_c - \alpha}{ 1-p_c}  \right) \leq   \left ( 1-\frac{1}{N+J} \right )^{-\gamma}\]
\[\left ( 1-\frac{1}{N+J} \right )^{\gamma} \left(1-p_c \right)\leq  \left(1-p_c - \alpha  \right) \leq   \left ( 1-\frac{1}{N+J} \right )^{-\gamma} \left(1-p_c \right)\]
\begin{equation}\left(1-p_c \right)\left(1-\left ( 1-\frac{1}{N+J} \right )^{-\gamma} \right) \leq \alpha \leq \left(1-p_c \right)\left(1-\left ( 1-\frac{1}{N+J} \right )^{\gamma} \right)\end{equation}

If we ensure $|\hat{p}_c - p_c| < \epsilon_2$ where
\begin{equation}\epsilon_2= min \left \{\left| \left(1-p_c \right)\left(1-\left ( 1-\frac{1}{N+J} \right )^{-\gamma} \right)\right| , 
            \right.   \left.      \left| \left(1-p_c \right)\left(1-\left ( 1-\frac{1}{N+J} \right )^{\gamma} \right)\right|\right \}
\end{equation}
for some $\gamma<1/2$ then we can have $\hat{J} = J$. The number of observations $S_3$ required to compute $\hat{p_c}$
such that $|\hat{p}_c - p_c| \leq \epsilon_2$ is given by Hoeffding's inequality. 
 \[Pr(|\hat{p}_c - p_c| >\epsilon_2) \leq 2\exp(-2\cdot S_3\cdot\epsilon^2_2) < \frac{\delta_3}{2K} \]
\[\implies S_3 > \frac{1}{2\epsilon^2_2}\ln \left(\frac{4K}{\delta_3}\right) \]

We can further simplify our expression for $\epsilon_1$. Substituing the value of $p_c$, we have 
 \begin{equation}\epsilon_2 = min \left \{\left|\left(1-\frac{1}{N+J}\right)^{J} \left(1-\left ( 1-\frac{1}{N+J} \right )^{-\gamma} \right)\right| , 
         \right.    \left.       \left| \left(1-\frac{1}{N+J}\right)^{J} \left(1-\left ( 1-\frac{1}{N+J} \right )^{\gamma} \right)\right|\right \}
\end{equation}
Using the results in \cite{MC}, we have 
\[\left|\left(1-\frac{1}{N+J}\right)^{J} \left(1-\left ( 1-\frac{1}{N+J} \right )^{-\gamma} \right)\right| \geq \frac{\gamma}{\exp(1).K}\]
\[\left|\left(1-\frac{1}{N+J}\right)^{J} \left(1-\left ( 1-\frac{1}{N+J} \right )^{\gamma} \right)\right| \geq \frac{\gamma}{\exp(1).K}\]

Therefore, combining above two, we have 
\[
\epsilon_2 \geq \frac{\gamma}{\exp(1).K}\]

If SUs  have $S_3> \frac{1}{2\epsilon^2_2}\ln \left(\frac{4K}{\delta_3}\right)  $ observations that the selected channel was free, where $\epsilon_2 =\frac{\gamma}{\exp(1).K}$ then for all SUs with 
probability at least $1-\frac{\delta_3}{2}$, we have  $\hat{J} = J$

To ensure we have $S_3$ observations of selected channels being free for all SUs, using similar approach as in proof of Lemma \ref{thm:le1} used for $N+J$ estimation, we can say ,
   
\begin{align*}
Pr \left(all \ SUs \ has < \frac{1}{2}\cdot T_J\cdot E[A_{I_t}(t)] observations\right)< \frac{\delta_3}{2} \\
\implies \exp \left(-\frac{1}{8}\cdot T_J\cdot E[A_{I_t}(t)] \right) < \frac{\delta_3}{2K} \\
\implies T_J > 8 \cdot \ln \left(\frac{2K}{\delta_3}\right) \cdot \frac{1}{E[A_{I_t}(t)] }  
\end{align*}

where $T_J$ is the number of slots for Jammer Estimation. In this case, $E[A_k(t)] = 1-p_k$.
We also want that total number of observations of selected channels being free to be at least $S_3$.
Therefore, \\ 
\[ \sum\limits_{t=1}^{T_J} A_{I_t}(t) \geq \frac{1}{2}\cdot T_J\cdot E[A_{I_t}(t)] \geq S_3\]
\[\implies \frac{1}{2}\cdot T_J\cdot E[A_{I_t}(t)] > \frac{1}{2\epsilon^2_2}\ln \left(\frac{4K}{\delta_3}\right)\]
\[\implies  T_J > \frac{1}{\epsilon^2_2 E[A_{I_t}(t)]}\ln \left(\frac{4K}{\delta_3}\right) \]

We now define following events:\\
\noindent $G$ - event that all SUs have observed at least $S_3$ times the selected channel was free \\ 
$G_n$ - event that $n^{th}$ SU has observed at least $S_3$ times the selected channel was free \\
$H$ - event that all SUs have correct estimate of $J$ given all SUs are orthogonalised when selected channel was free \\
$H_n$ - event that $n^{th}$ SU has correct estimate of $N$ given all SUs are orthogonalised when selected channel was free \\

\begin{align*}
Pr(H) &= 1 - Pr(\overline{H})  \\
     &= 1 - Pr(\overline{H}/G)Pr(G) -  Pr(\overline{H}/\overline{E})Pr(\overline{E}) \\
     &\geq 1 - Pr(\overline{H}/G) - Pr(\overline{G}) \\ 
     &\geq 1 - \sum\limits_{n=1}^{N} Pr(\overline{H_n}/G_n) - \sum\limits_{n=1}^{N} Pr(\overline{G_n})  \tag{Union Bound}\\  	
     &\geq 1 - 	\sum\limits_{n=1}^{N} \frac{\delta_3}{2K}-\sum\limits_{n=1}^{N} \frac{\delta_3}{2K}\\
     &\geq 1 - \frac{\delta_3}{K}N \\
 	&\geq 1 - \frac{\delta_1}{K}K \\
  	&\geq 1 - \delta_3 \\
\end{align*}

\noindent Let Y be an event that all SUs have correct estimate of $N$ and $J$ as well as they are orthogonalised and have correct channel ranking.
\begin{align*}
Pr(Y) &\geq Pr(Y/ADFH)\cdot Pr(ADFH)  \\
   &\geq Pr(Y/ADFH)\cdot Pr(H/ADF)\cdot Pr(ADF) \\ 
   &\geq Pr(Y/ADFH)\cdot Pr(H/ADF)\cdot Pr(F/AD) \cdot Pr(AD) \\ 
\end{align*}

 We have, 
\begin{equation*}
   \geq Pr(Y/ADFH)\cdot Pr(H/ADF)\cdot Pr(F/AD) \cdot Pr(AD) 
\end{equation*}
   \begin{equation*}
Pr(Y)  \geq (1-\delta_3)(1-\delta_2)(1-\delta_1) 
\end{equation*}

Setting $ \delta_1 = \delta_2=\delta_3=\frac{\delta}{3}$ 

\begin{equation*}
Pr(Y)  \geq \left(1-\frac{\delta}{3}\right)^3 
\end{equation*}
\begin{equation*}
Pr(Y)  \geq 1-3\frac{\delta}{3}
\end{equation*}
\begin{equation*}
Pr(Y)  \geq 1-\delta
\end{equation*}

After estimating $N,J$ and channel ranking, all SUs run the OR subroutine for the second time. We will compute the maximum slots required such that all SUs get orthogonalized and start sequential hopping. Our approach to bound regret is similar to approach in \cite{MC}. We next derive the probability of collision for any SU at any round $t$.

\noindent Now, to bound regret consider following:
if $(m<=J)$: \\
\begin{align*}
Pr(no \ collision)
	&= \sum\limits_{k=1}^{N+m}\frac{(1-p_k)}{N+m}  \left(p_{ns}+p_{s}\right)\left(1-\frac{1}{N+J}\right)^{J} \\
	&+\sum\limits_{k=1}^{N+m}\frac{p_k}{N+m}\\
p_{nc}&\geq \sum\limits_{k=1}^{N+m}\frac{(1-p_k)}{N+m}  \left(p_{ns}+p_{s}\right)\left(1-\frac{1}{N+J}\right)^{J}\\
    &\geq \sum\limits_{k=1}^{N+m}\frac{(1-p_k)}{N+m}  p_{ns}\left(1-\frac{1}{N+J}\right)^{J}\\
    &\geq \sum\limits_{k=1}^{N+m}\frac{(1-p_k)}{N+m}  \left(1-\frac{1}{N+m} \right)^{N-1}\left(1-\frac{1}{N+J} \right)^{J} \\
    &\geq \theta\left(1-\frac{1}{N+J} \right)^{N+J-1} \\ 
  &\geq \theta\left(1-\frac{1}{K} \right)^{K-1}\\
\end{align*}

if $(m>J)$: \\
\begin{align*}
Pr(no \ collision)
	&= \sum\limits_{k=1}^{N+J}\frac{(1-p_k)}{N+m}  \left(p_{ns}+p_{s}\right)\left(1-\frac{1}{N+J}\right)^{J} \\ 
    &+\sum\limits_{k=N+J+1}^{N+m}\frac{(1-p_k)}{N+m}  \left(p_{ns}+p_{s}\right)\\
    &+ \sum\limits_{k=1}^{N+m}\frac{p_k}{N+m}\\
p_{nc}&\geq \sum\limits_{k=1}^{N+m}\frac{(1-p_k)}{N+m}  \left(p_{ns}+p_{s}\right)\left(1-\frac{1}{N+J}\right)^{J}\\
    &\geq \sum\limits_{k=1}^{N+m}\frac{(1-p_k)}{N+m}  p_{ns}\left(1-\frac{1}{N+J}\right)^{J}\\
    &\geq \sum\limits_{k=1}^{N+m}\frac{(1-p_k)}{N+m}  \left(1-\frac{1}{N+m} \right)^{N-1}\\
    &\left(1-\frac{1}{N+J} \right)^{J} \\
    &\geq \theta\left(1-\frac{1}{N+m} \right)^{N+J-1} \\ 
  &\geq \theta\left(1-\frac{1}{K} \right)^{K-1}\\
    &\geq \frac{\theta}{exp(1)} 
\end{align*}
Combining above two cases, we can say 
\[p_{nc} \geq \frac{\theta}{exp(1)}\]
Since the number of slots required to find a collision free transmission and start sequential hopping is geometric distributed, the expected number of slots for first success, denoted by $T_{SH}$, is bounded by :
\[T_{SH} < \frac{1}{p_{nc}} \implies T_{{SH}} < \frac{Nexp(1)}{\theta} .\]  

Regret will again be bounded as $NT_{SH}$, i.e., 
$\frac{N^2exp(1)}{\theta}$. Hence proving theorem \ref{thm:t3}

\section{Additional Simulation Results}\label{SF}
In Figures 1-5, we  compare the performance of proposed  CDJ, CNJ and CUJ algorithms with the respective myopic algorithm, $\rho^{rand}$ \cite{prand} and MC \cite{MC} algorithms in terms of regret and throughput for various combinations of $K, N$, $J$ and $\{p_i\}$. It can be observed that proposed algorithms offer lower regret and higher throughput in all cases. Furthermore, the difference between the throughput of the proposed algorithms and other algorithms increases as $T$ increases. Few observations which support simulation results in main papers are: 

\begin{enumerate}
	\item For a given values of $K$ and $J$ (as shown in Fig. \ref{cc} and \ref{ee} where $K = 16$ and $J = 4$ ), the average regret increases with increase in $N$ from 8 to 10 as discussed in Fig. 4 of the main manuscript. 
	\item From  Fig. \ref{cc} and \ref{dd} where $K = 16$ and $N = 8$, the average regret decreases with the increase in the value of $J$ from 4 to 6 as discussed in Fig. 6 of the main manuscript.
	\item With increase in value of $K$ for a given $N$ and $J$, the average regret increases as depicted in Fig. \ref{bb} and \ref{cc}.
	
\end{enumerate}


%
 \par


\begin{figure*}[!htb]
	\centering
	\subfloat[]{\includegraphics[scale=0.35]{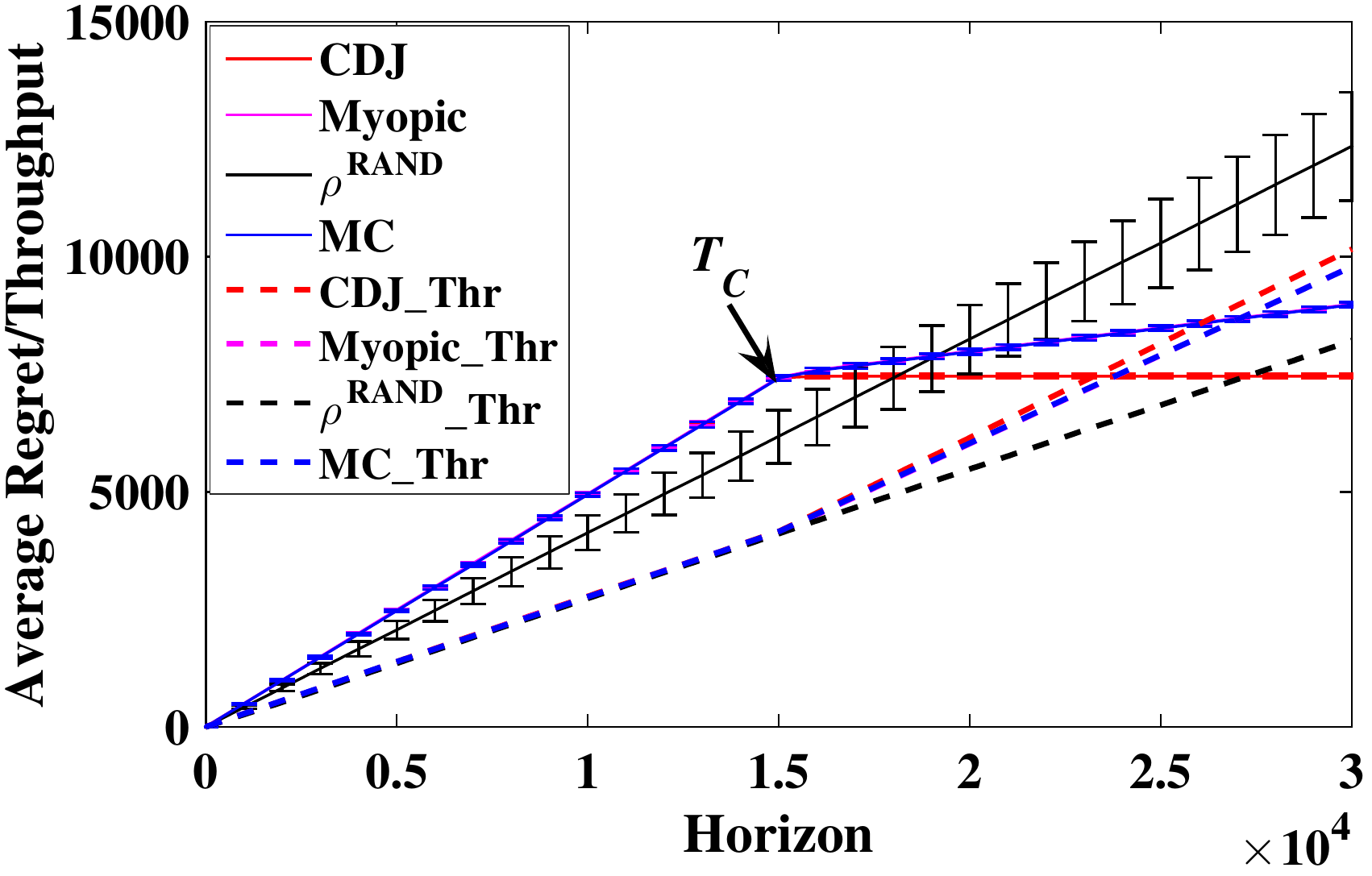}%
		\label{aa1}}
		\hspace{0.1cm}
	\subfloat[]{\includegraphics[scale=0.35]{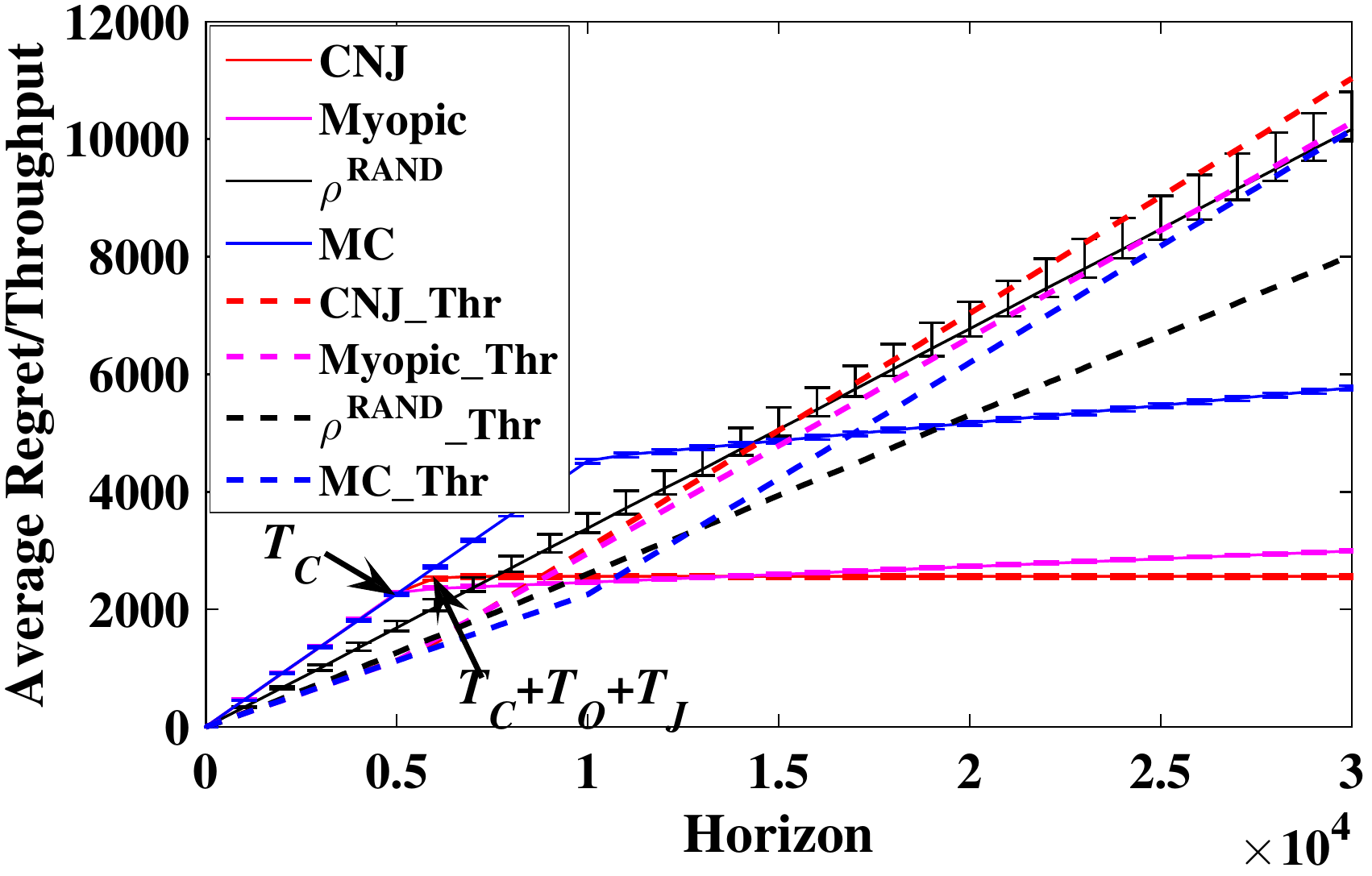}%
		\label{aa2}}
	\hspace{0.1cm}
		\subfloat[]{\includegraphics[scale=0.35]{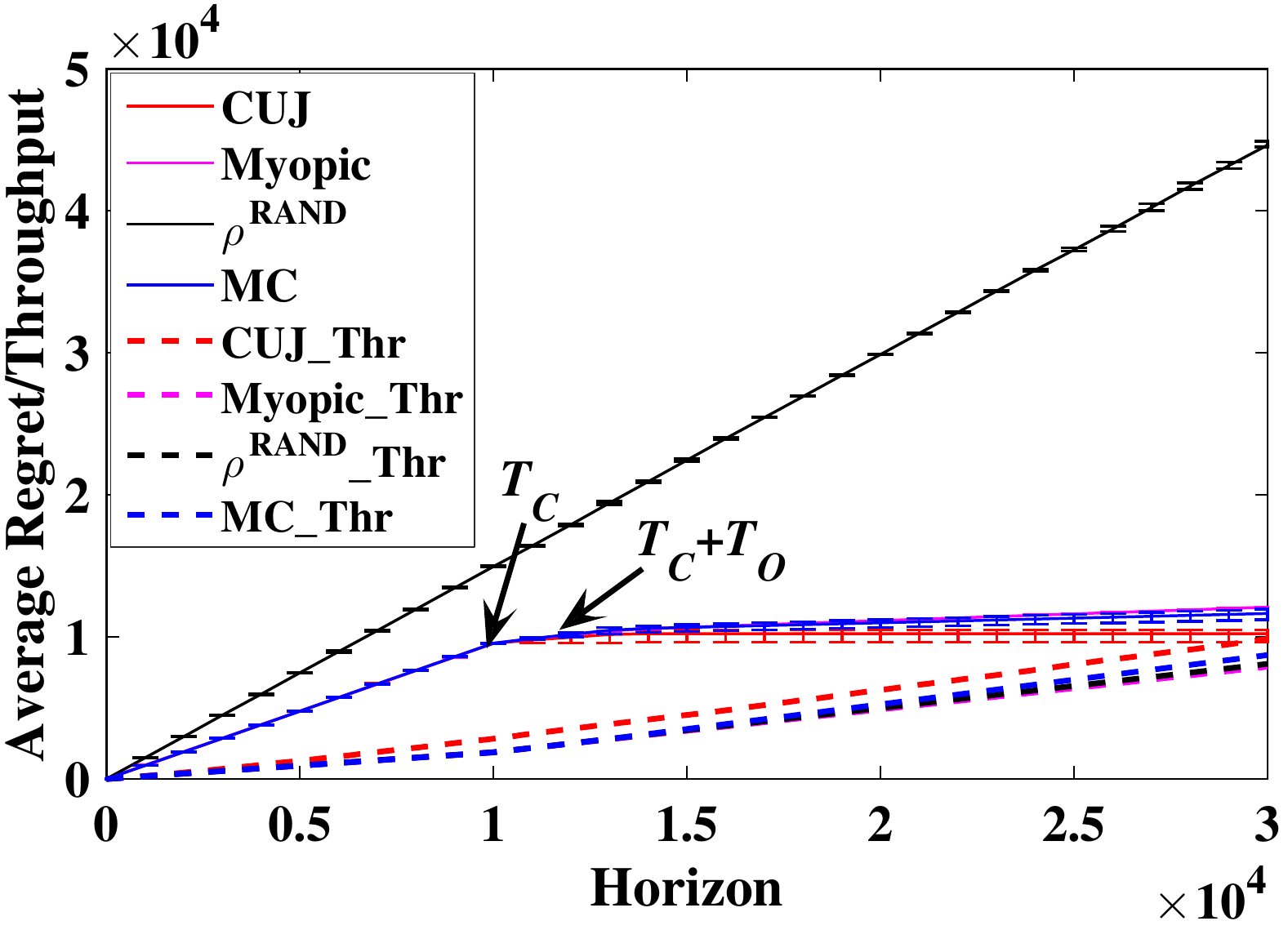}%
		\label{aa3}}	
\caption {The comparison of average regret of the Myopic algorithm with (a) CDJ algorithm, (b) CNJ algorithm and (c) CUJ algorithm at different instants of the horizon. Here, we fix  $K=8$, $N=4$, $J=2$ with $p_i = \{0.2,0.3,0.4,...,0.9\}$.}
\label{aa}
\end{figure*}

\begin{figure*}[!htb]
	\centering
	\subfloat[]{\includegraphics[scale=0.35]{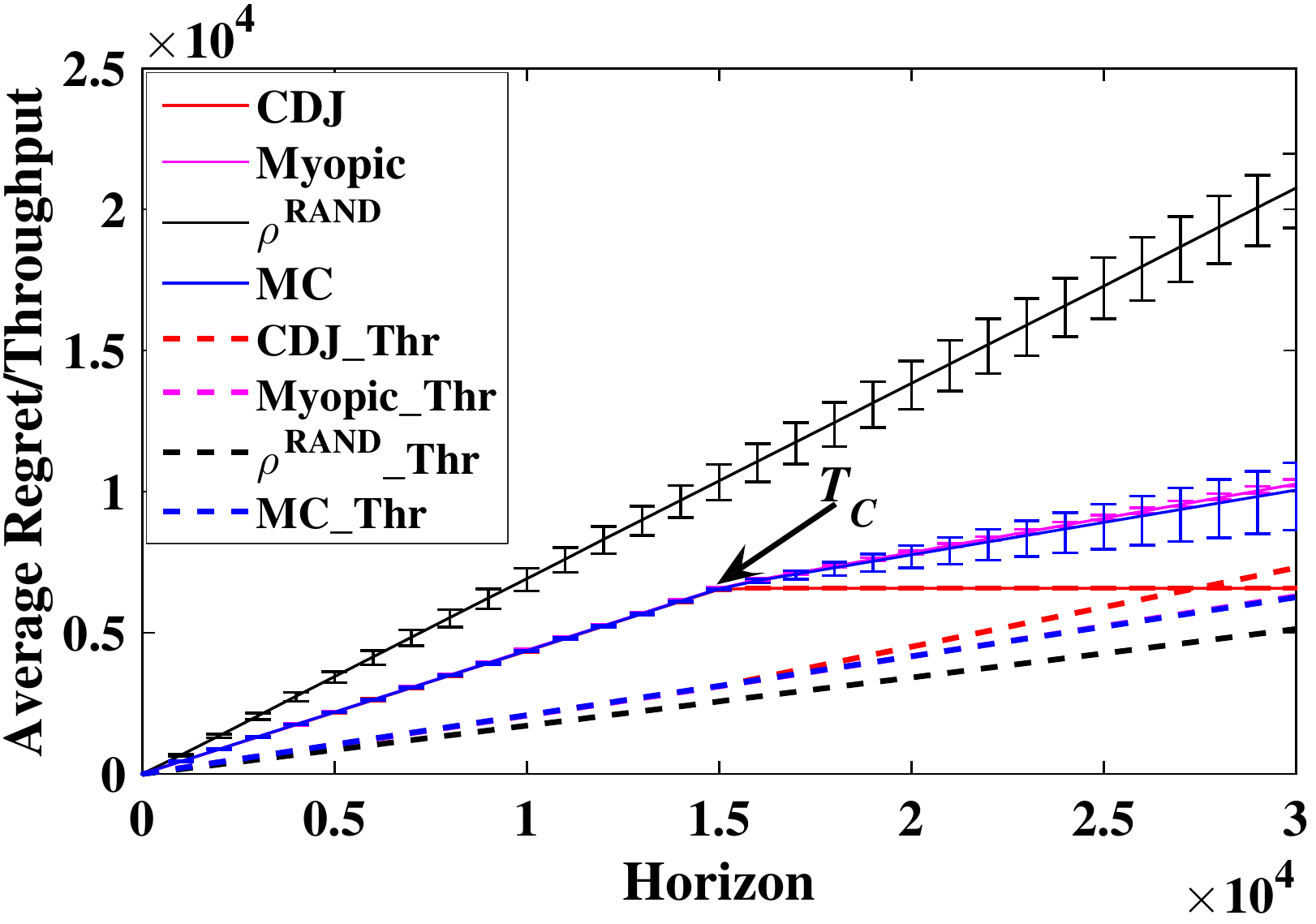}%
		\label{bb1}}
		\hspace{0.1cm}
	\subfloat[]{\includegraphics[scale=0.35]{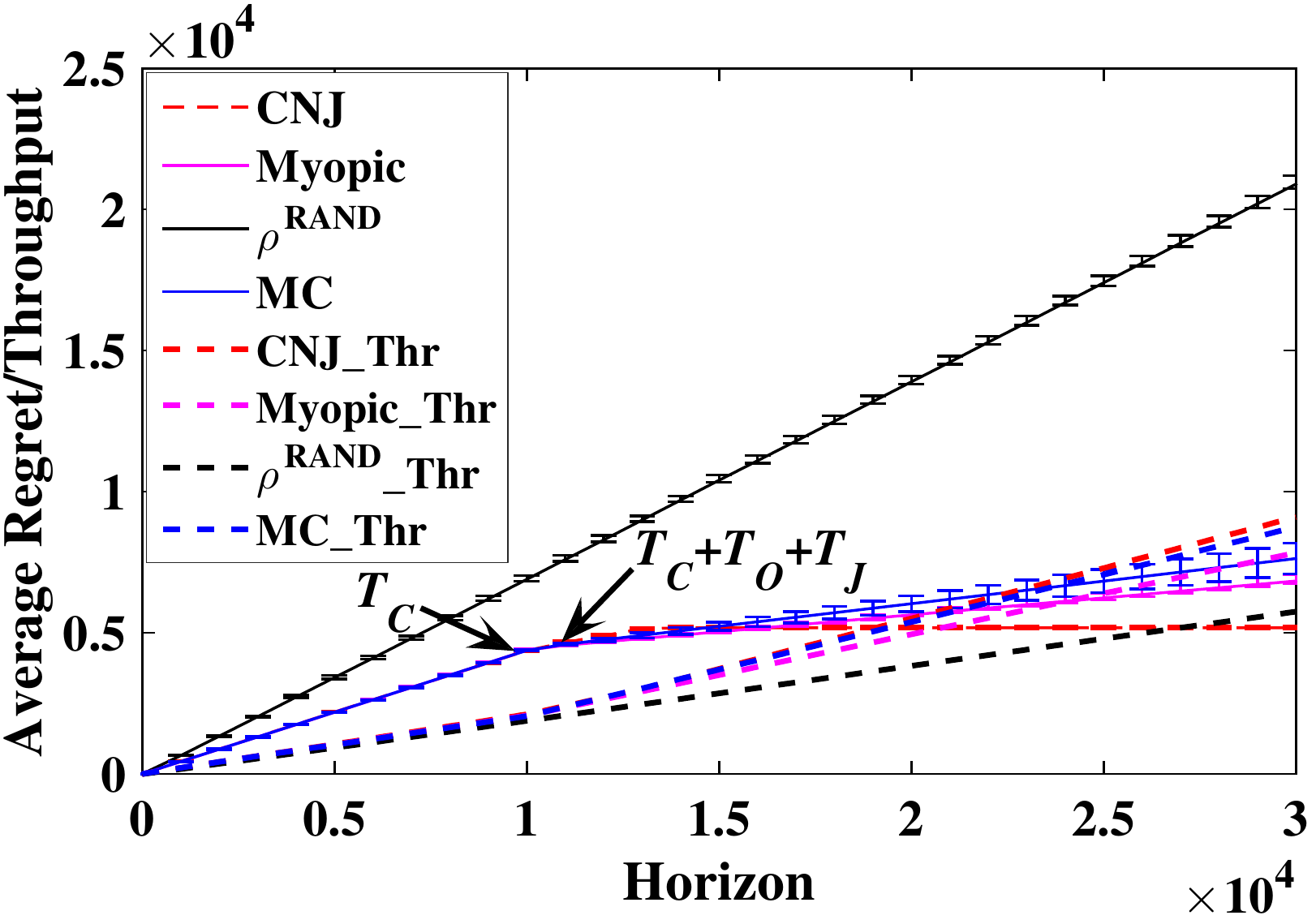}%
		\label{bb2}}
	\hspace{0.1cm}
		\subfloat[]{\includegraphics[scale=0.35]{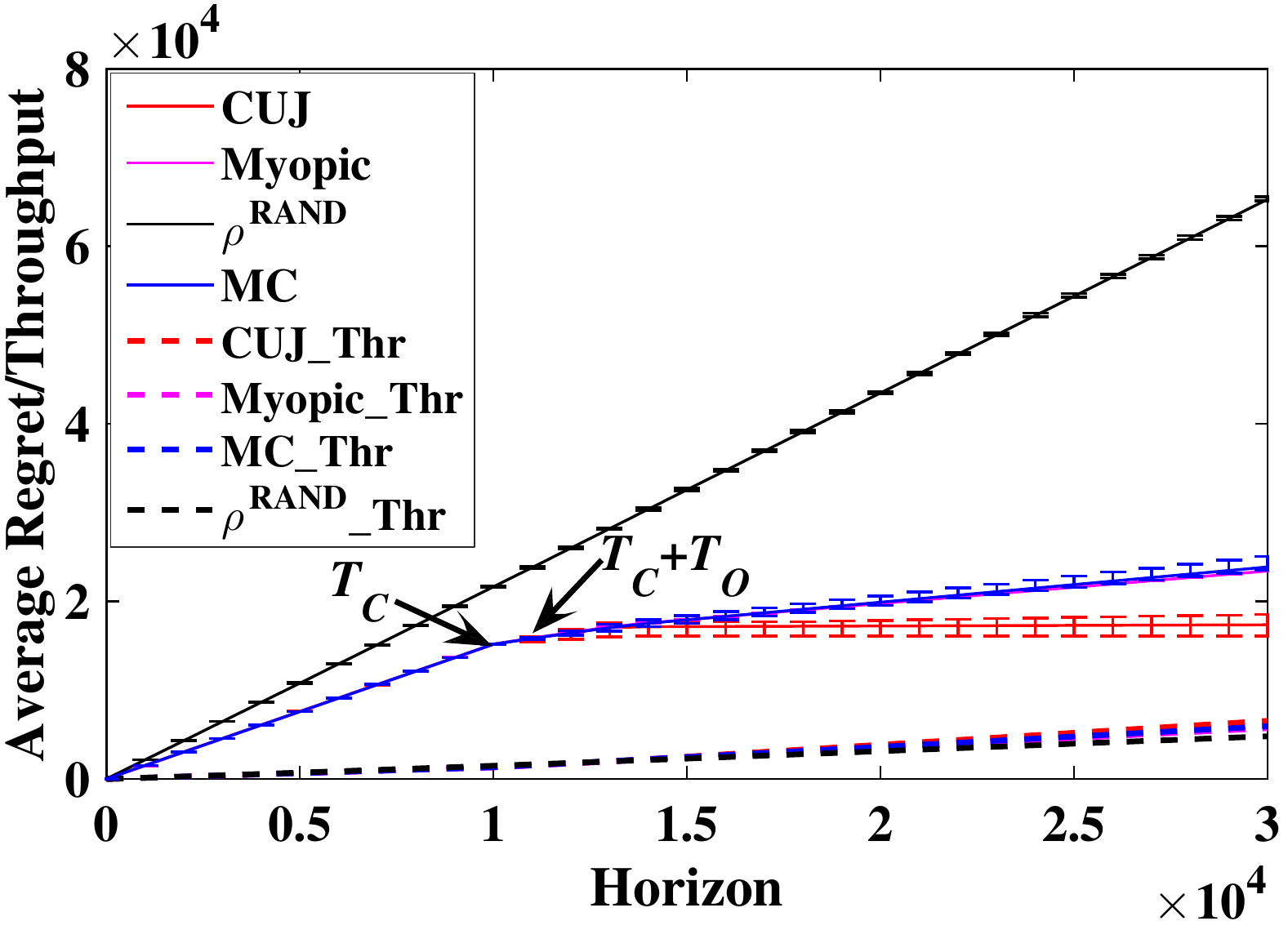}%
		\label{bb3}}	
\caption {The comparison of average regret of the Myopic algorithm with (a) CDJ algorithm, (b) CNJ algorithm and (c) CUJ algorithm at different instants of the horizon. Here, we fix  $K=12$, $N=8$, $J=4$ with $p_i = \{0.8,0.16,0.24,...,0.96\}$.}
\label{bb}
\end{figure*}

\begin{figure*}[!htb]
	\centering
	\subfloat[]{\includegraphics[scale=0.35]{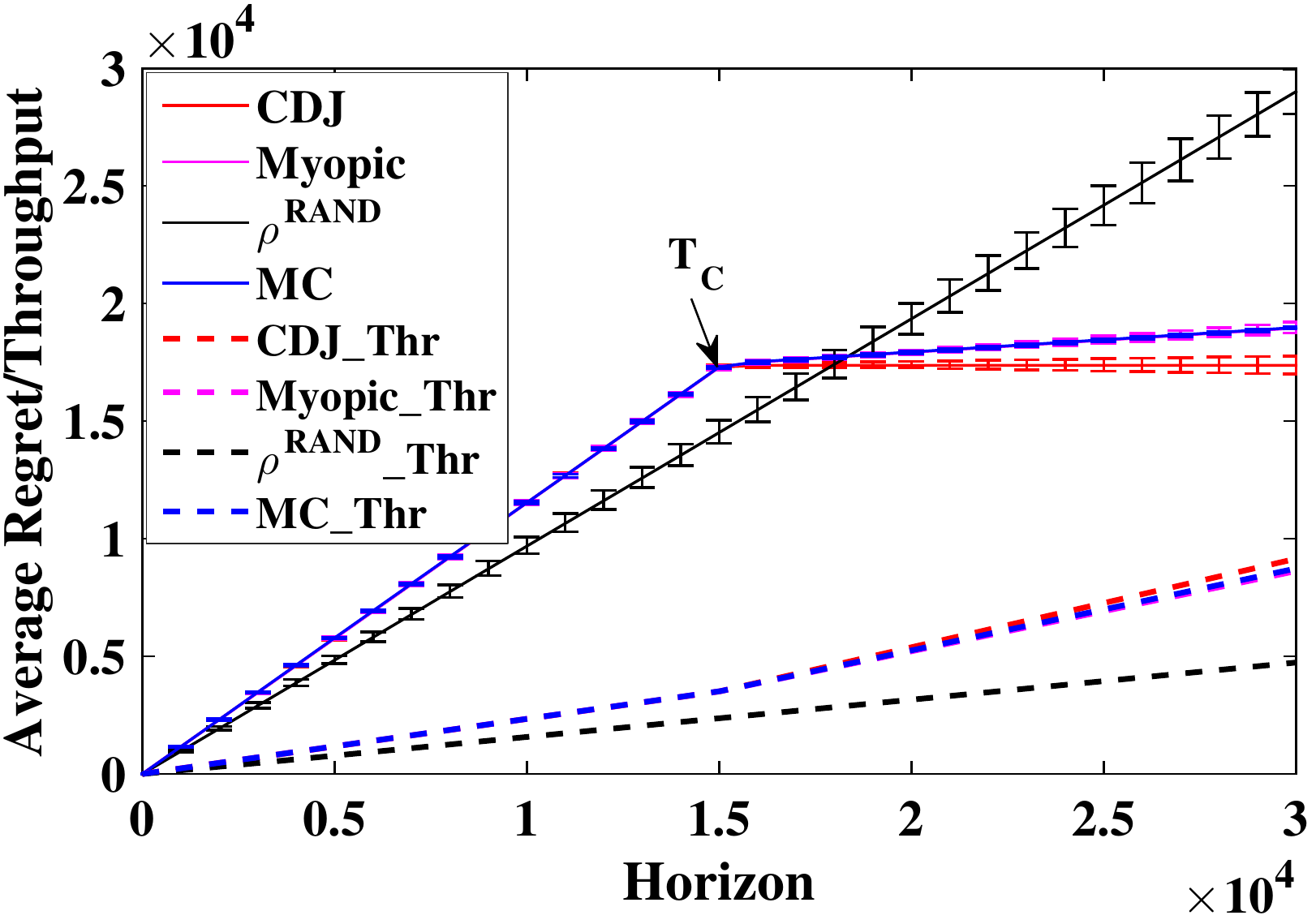}%
		\label{cc1}}
		\hspace{0.1cm}
	\subfloat[]{\includegraphics[scale=0.35]{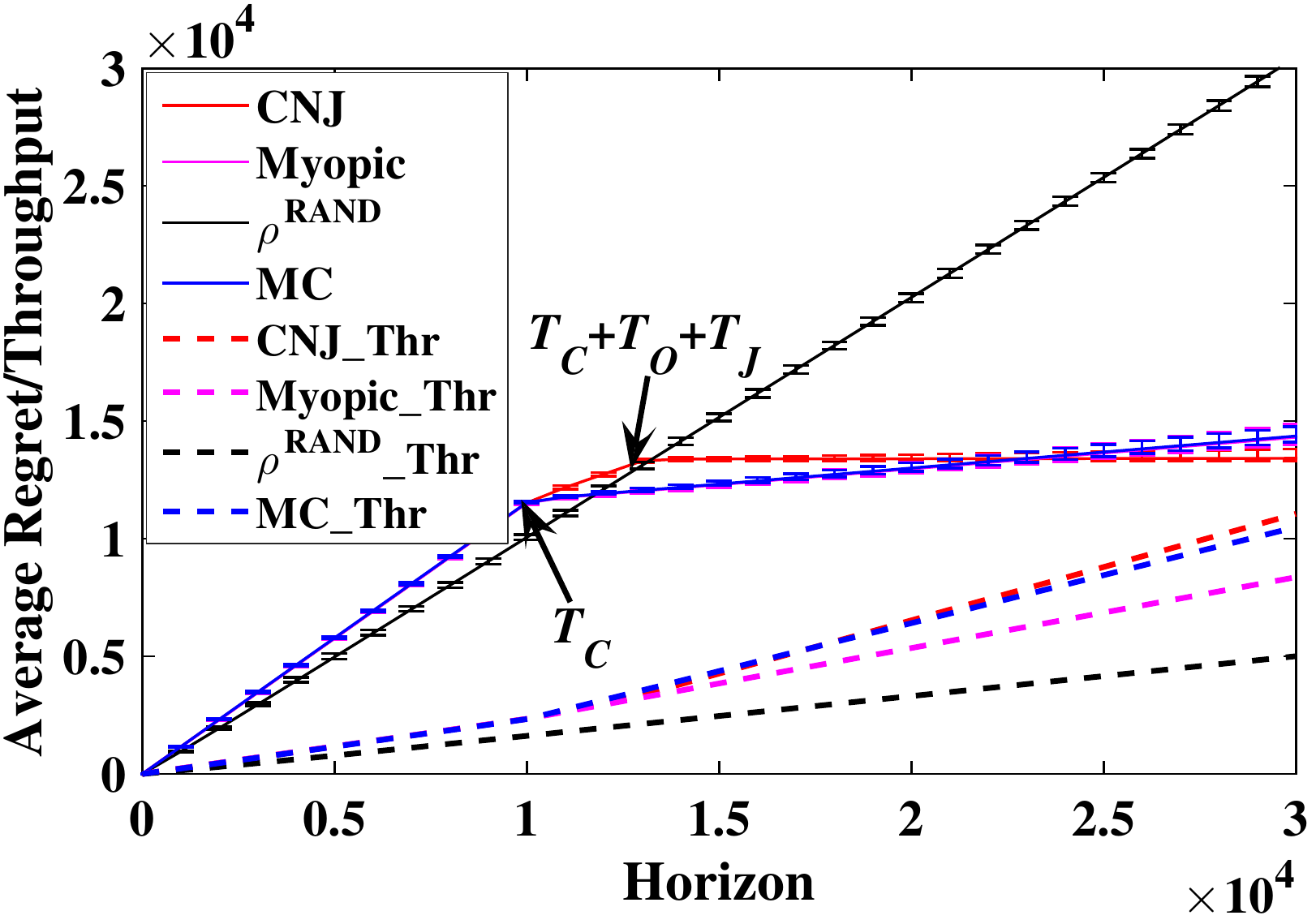}%
		\label{cc2}}
	\hspace{0.1cm}
		\subfloat[]{\includegraphics[scale=0.35]{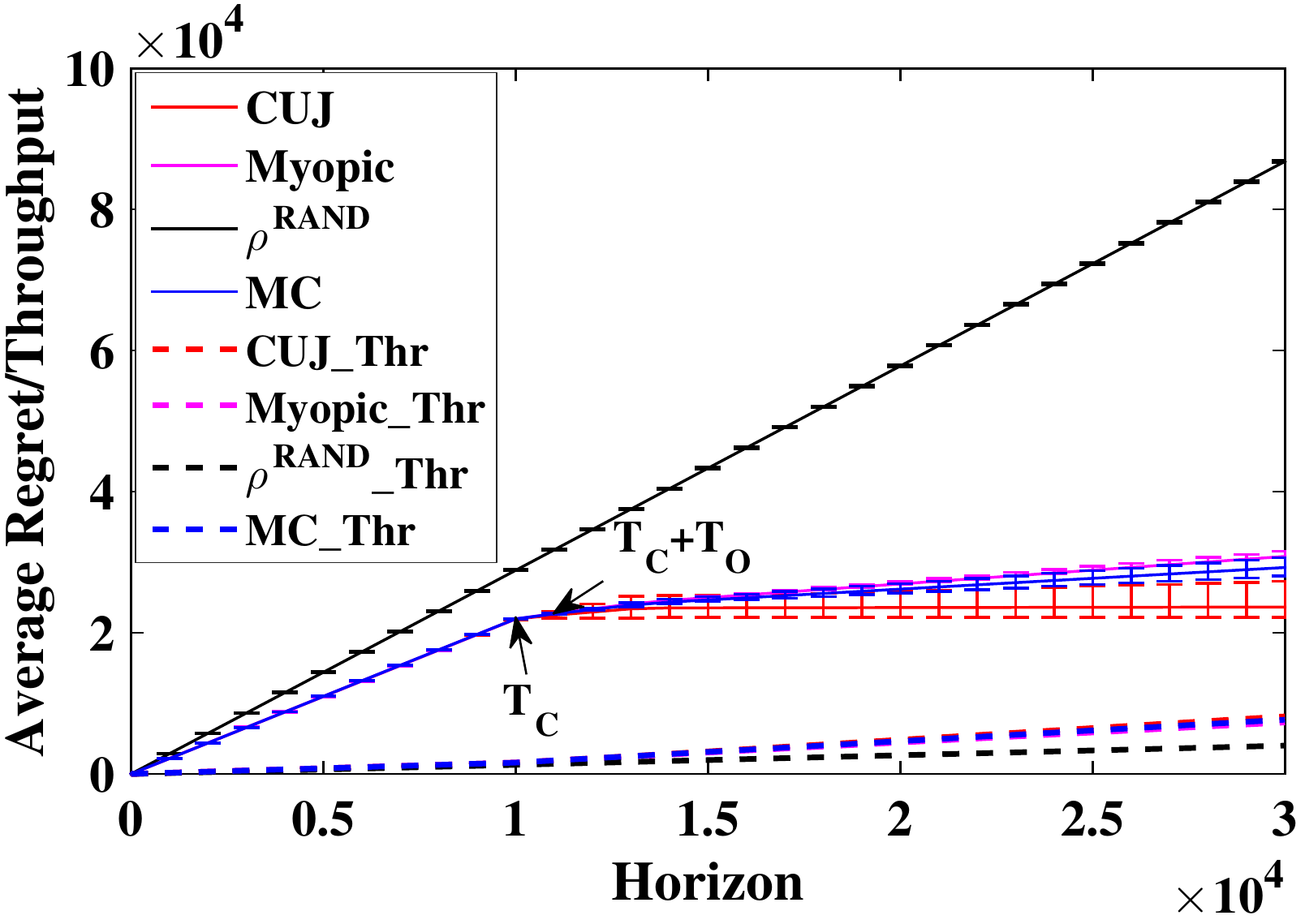}%
		\label{cc3}}	
\caption {The comparison of average regret of the Myopic algorithm with (a) CDJ algorithm, (b) CNJ algorithm and (c) CUJ algorithm at different instants of the horizon. Here, we fix  $K=16$, $N=8$, $J=4$ with $p_i = \{0.06,0.12,0.18,...,0.96\}$.}
\label{cc}
\end{figure*}

\begin{figure*}[!htb]
	\centering
	\subfloat[]{\includegraphics[scale=0.35]{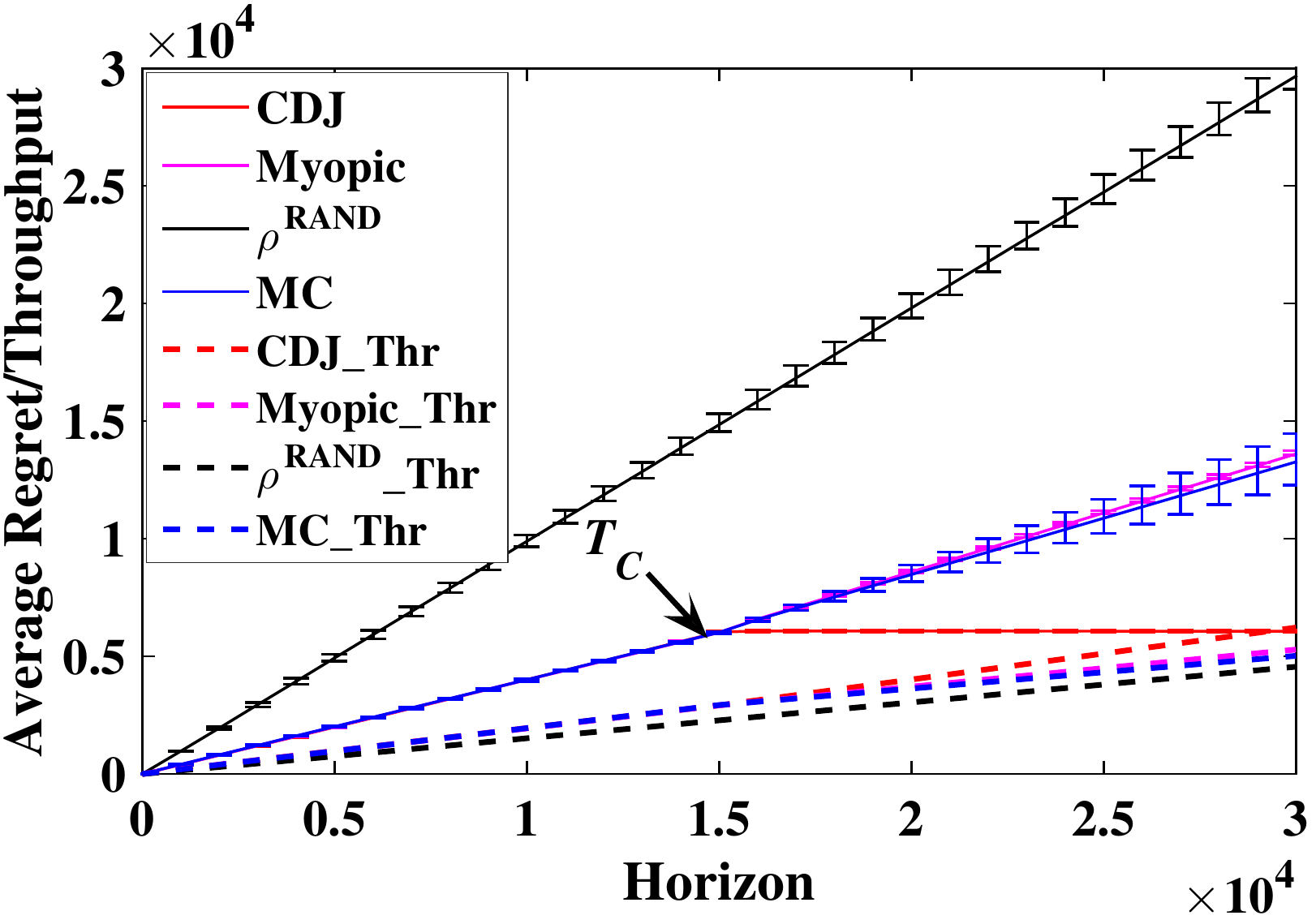}%
		\label{dd1}}
		\hspace{0.1cm}
	\subfloat[]{\includegraphics[scale=0.35]{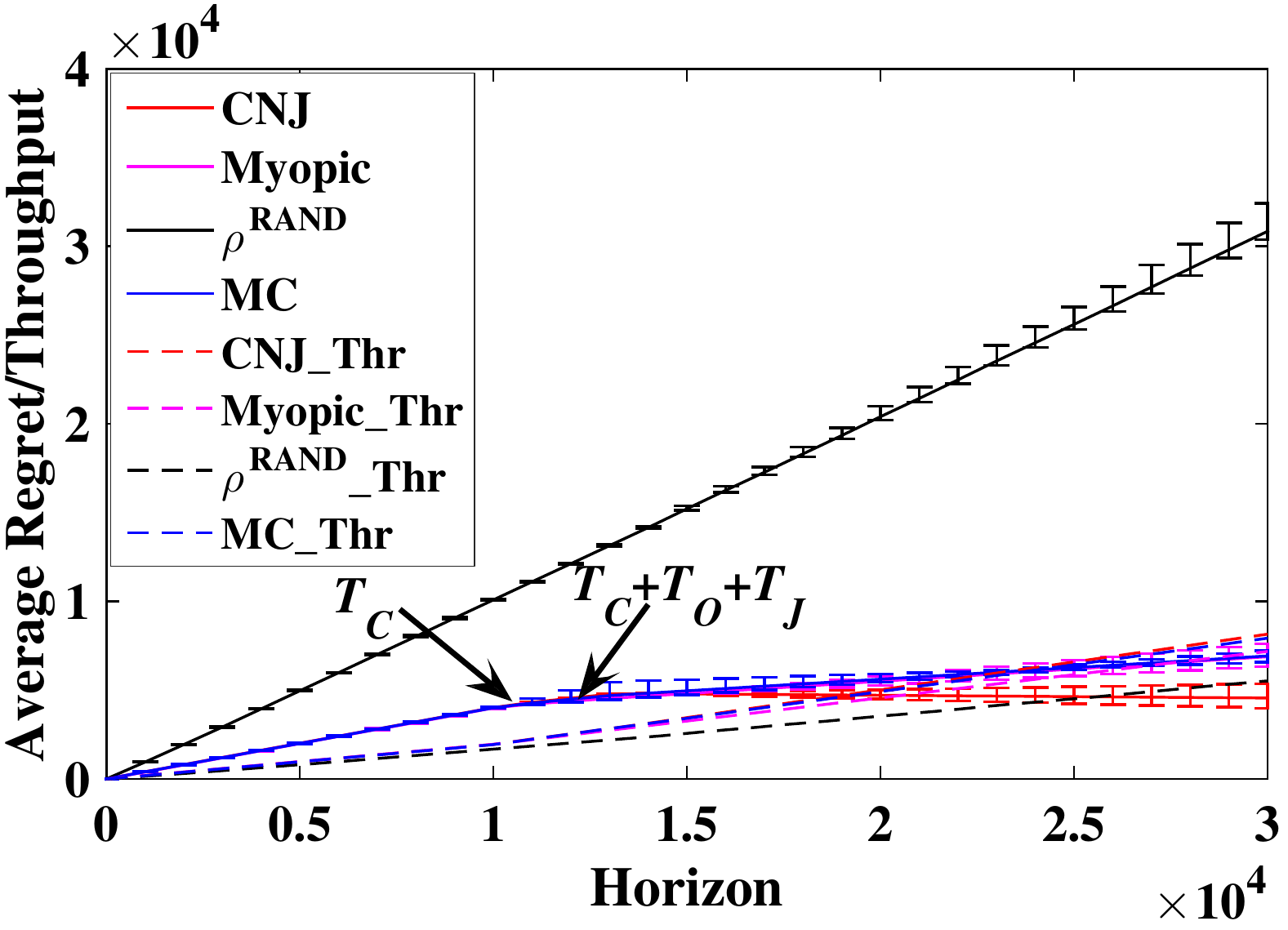}%
		\label{dd2}}
	\hspace{0.1cm}
		\subfloat[]{\includegraphics[scale=0.35]{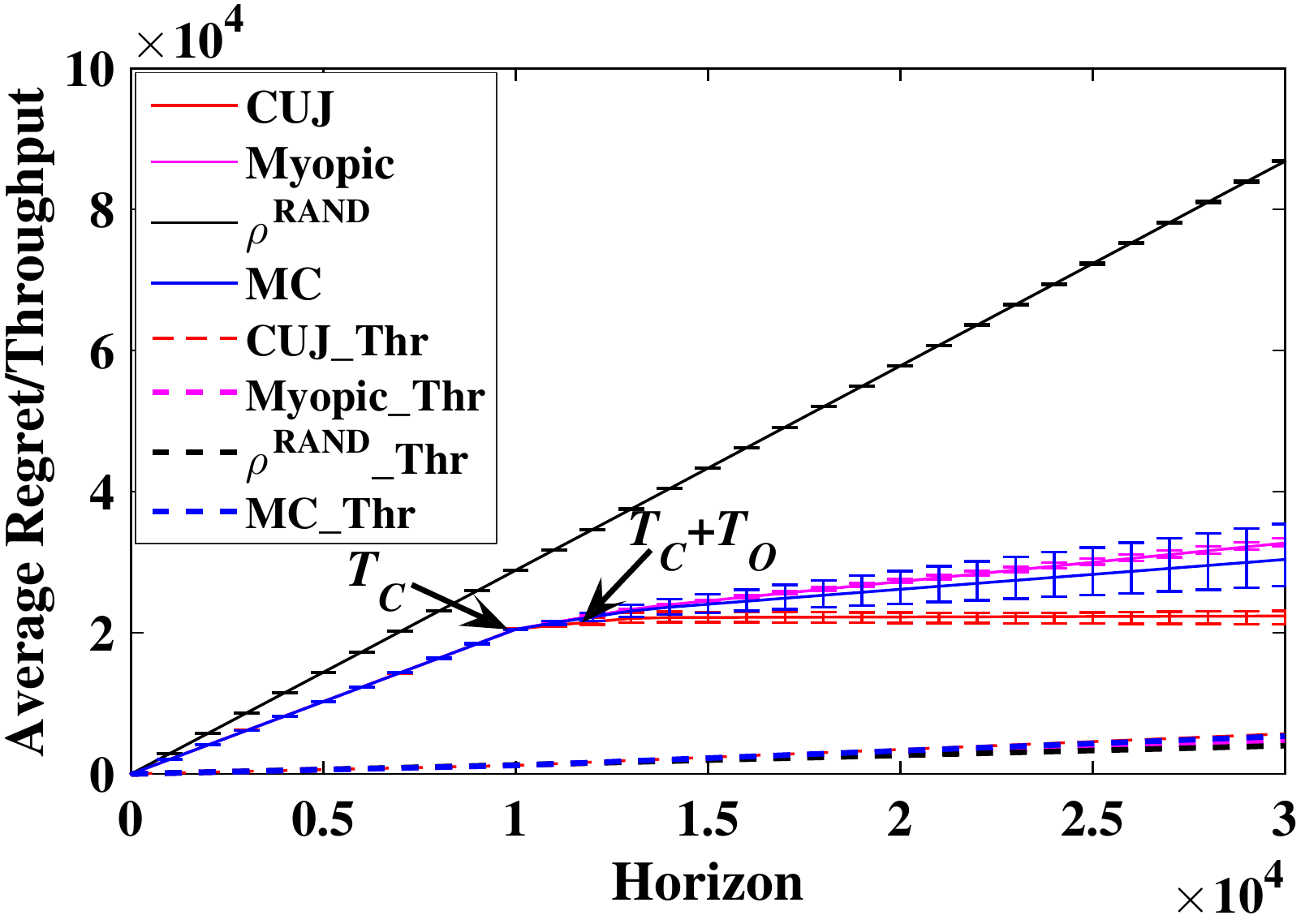}%
		\label{dd3}}	
\caption {The comparison of average regret of the Myopic algorithm with (a) CDJ algorithm, (b) CNJ algorithm and (c) CUJ algorithm at different instants of the horizon. Here, we fix  $K=16$, $N=8$, $J=6$ with $p_i = \{0.06,0.12,0.18,...,0.96\}$.}
\label{dd}
\end{figure*}

\begin{figure*}[!htb]
	\centering
	\subfloat[]{\includegraphics[scale=0.35]{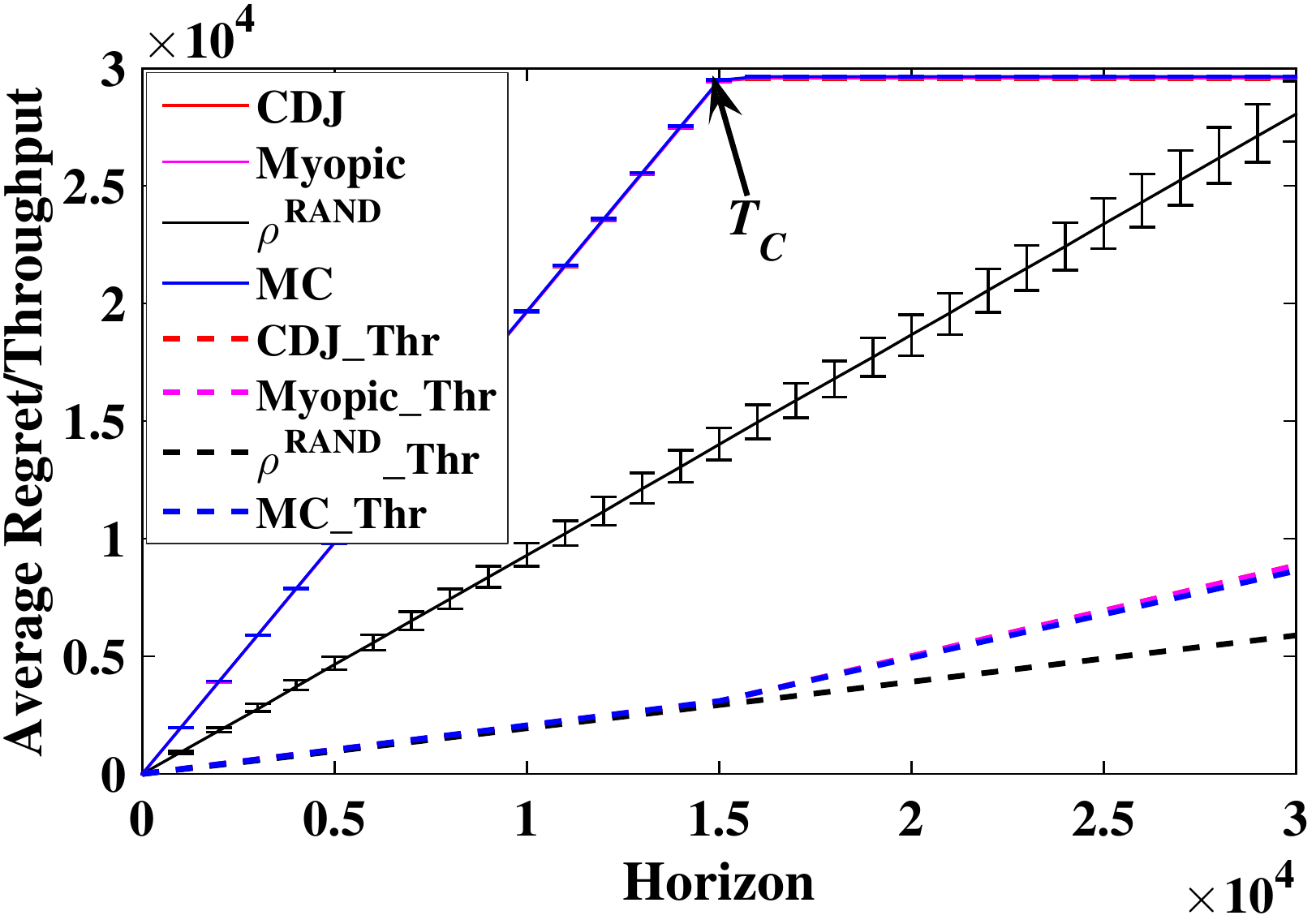}%
		\label{ee1}}
		\hspace{0.1cm}
	\subfloat[]{\includegraphics[scale=0.35]{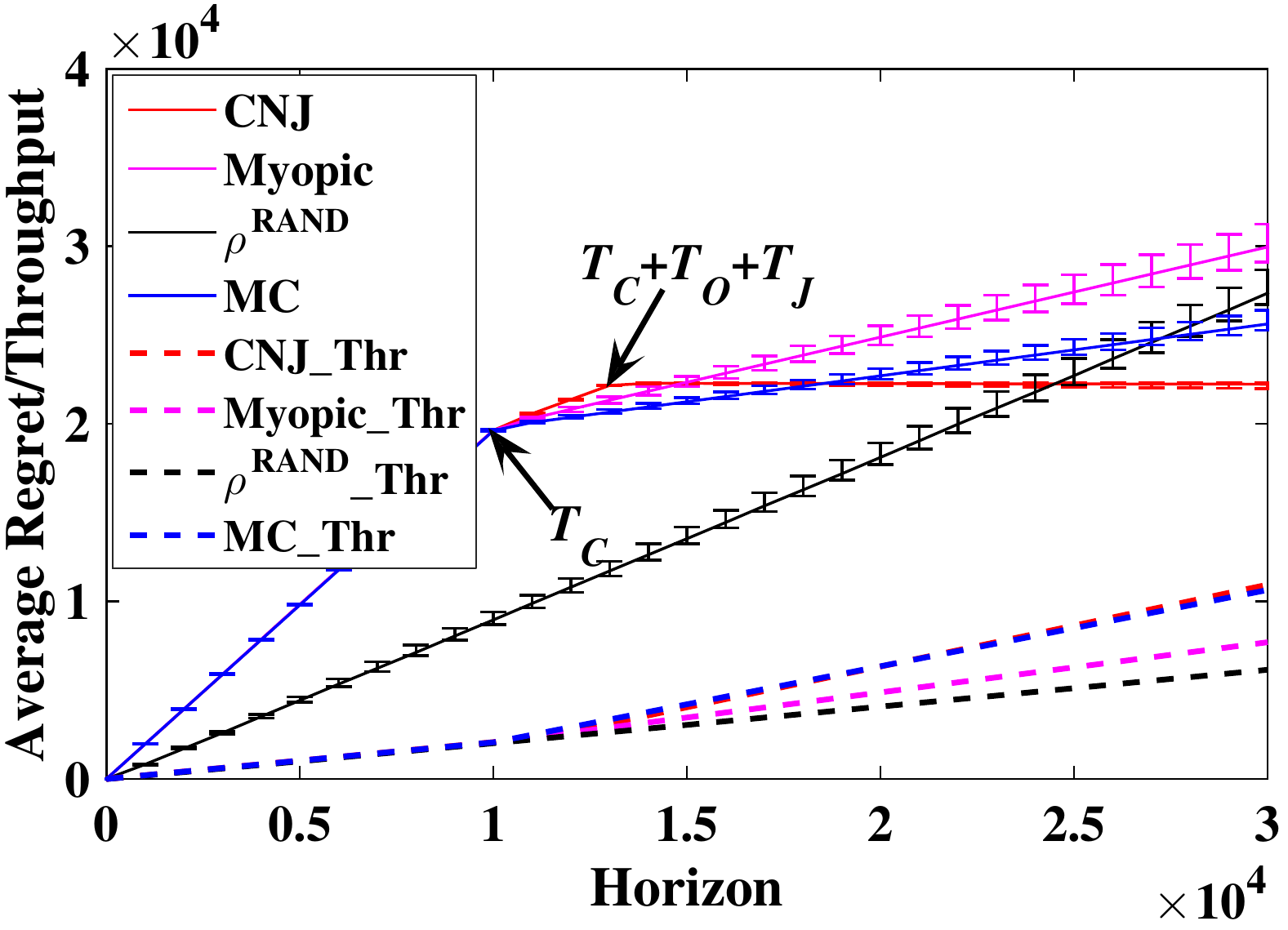}%
		\label{ee2}}
	\hspace{0.1cm}
		\subfloat[]{\includegraphics[scale=0.35]{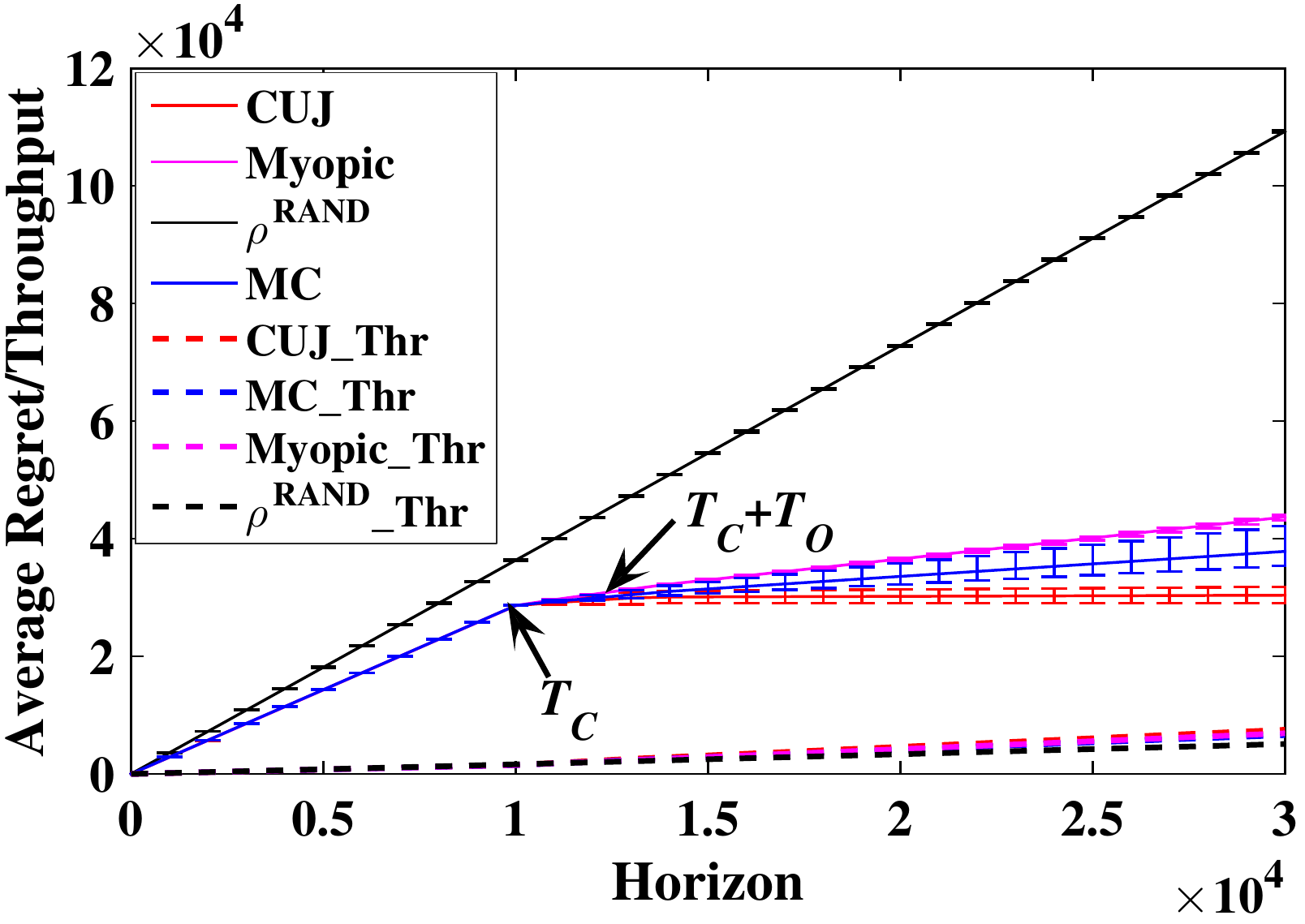}%
		\label{ee3}}	
\caption {The comparison of average regret of the Myopic algorithm with (a) CDJ algorithm, (b) CNJ algorithm and (c) CUJ algorithm at different instants of the horizon. Here, we fix  $K=16$, $N=10$, $J=4$ with $p_i = \{0.06,0.12,0.18,...,0.96\}$.}
\label{ee}
\end{figure*}

\end{document}